\documentclass[twocolumn, times, tighten]{aastex62}
\submitjournal{The Astrophysical Journal}
\accepted{10/8/2019}
\usepackage{amssymb}
\usepackage{amsmath}
\usepackage{natbib}

\defcitealias{Cardelli89}{CCM} 
\defcitealias{Massa00}{M00} 
\defcitealias{Fitzpatrick07}{F07} 
\defcitealias{Fitzpatrick05a}{F05} 
\defcitealias{Bastiaansen92}{B92} 
\defcitealias{Fitzpatrick99review}{F99}

\begin{document}

\newcommand{\atlas}{{ATLAS9}}
\newcommand{\tlusty}{{TLUSTY}}
\newcommand{\synspec}{{SYNSPEC}}
\newcommand{\spectrum}{{SPECTRUM}}

\newcommand{\iras}{{\it IRAS}}
\newcommand{\hip}{{\it Hipparcos}}
\newcommand{\cop}{{\it Copernicus}}
\newcommand{\ans}{{\it ANS}}
\newcommand{\hst}{{\it HST}}
\newcommand{\iue}{{\it IUE}}
\newcommand{\oao}{{\it OAO-2}}
\newcommand{\tmass}{{2MASS}} 
\newcommand{\ghrs}{{GHRS}}

\newcommand{\jhk}{{\it JHK}}  
\newcommand{\ubv}{{\it UBV}}
\newcommand{\uvby}{{\it uvby}}
\newcommand{\ugriz}{{\it ugriz}}

\newcommand{\ebv}{\mbox{$E(B\!-\!V)$}}
\newcommand{\eub}{\mbox{$E(U\!-\!B)$}}

\newcommand{\klam}{\mbox{$k(\lambda\!-\!V)$}}
\newcommand{\kfive}{\mbox{$k(\lambda\!-\!55)$}}
\newcommand{\elamV}{\mbox{$E(\lambda\!-\!V)$}}
\newcommand{\elamfive}{\mbox{$E(\lambda\!-\!55)$}}
\newcommand{\efourfive}{\mbox{$E(44\!-\!55)$}}

\newcommand{\mic}{\mbox{$\mu{\rm m}$}}
\newcommand{\invmic}{\mbox{$\mu{\rm m}^{-1}$}}
\newcommand{\invlam}{\mbox{$\lambda^{-1}$}}

\newcommand{\mast}{{MAST}}
\newcommand{\irsa}{{IRSA}}
\newcommand{\simbad}{{SIMBAD}}

\newcommand{\teff}{\mbox{$T_{\rm eff}$}}
\newcommand{\logg}{{$\log g$}}
\newcommand{\abund}{[m/H]}
\newcommand{\vturb}{$v_{turb}$}
\newcommand{\vrot}{$v \sin i$}
\newcommand{\vrad}{$v_{rad}$}

\newcommand{\msun}{${\rm M}_\sun$}
\newcommand{\rsun}{${\rm R}_\sun$}

\newcommand{\kms}{km\,s$^{-1}$}

\newcommand{\bs}{\boldsymbol}
\newcommand{\invr}{$R(V)^{-1}$}
\newcommand{\etal}{\it et al.\rm}

\shortauthors{Fitzpatrick et al.}
\shorttitle{Optical Spectrophotometric Extinction}

\title{An Analysis of the Shapes of Interstellar Extinction Curves. VII. \\
Milky Way Spectrophotometric Optical-through-Ultraviolet Extinction and Its $R$-Dependence
\footnote{Based on observations made with the NASA/ESA Hubble Space Telescope, 
obtained at the Space Telescope Science Institute, which is operated by the 
Association of Universities for Research in Astronomy, Inc., under NASA 
contract NAS 5-26555. These observations are associated with program \# 
13760.}}

\correspondingauthor{Derck Massa}
\email{dmassa@spacescience.org}

\author[0000-0002-2371-5477]{E.\ L.\ Fitzpatrick}
\affiliation{Department of Astrophysics \& Planetary Science, Villanova 
University, 800 Lancaster Avenue, Villanova, PA 19085, USA}

\author[0000-0002-9139-2964]{Derck Massa}
\affil{Space Science Institute
4750 Walnut Street, Suite 205 
Boulder, Colorado 80301, USA} 

\author[0000-0001-5340-6774]{Karl D.\ Gordon}
\affiliation{Space Telescope Science Institute, 3700 San Martin
  Drive, Baltimore, MD, 21218, USA}
\affiliation{Sterrenkundig Observatorium, Universiteit Gent,
  Gent, Belgium}

\author{Ralph Bohlin}
\affiliation{Space Telescope Science Institute, 3700 San Martin
  Drive, Baltimore, MD, 21218, USA}
  
\author[0000-0002-0141-7436]{Geoffrey C. Clayton}
\affiliation{Department of Physics \& Astronomy, Louisiana State University, 
  Baton Rouge, LA 70803, USA}

\begin{abstract} 
We produce a set of 72 NIR through UV extinction curves by combining new 
\hst/STIS optical spectrophotometry with existing \iue\ 
spectrophotometry (yielding gapless coverage from 1150 -- 
10000~\AA) and NIR photometry.  These curves are used to determine a new, internally 
consistent, NIR through UV Milky Way Mean Curve and to 
characterize how the shapes of the extinction curves 
depend on $R(V)$.  We emphasize that while this dependence captures much of 
the curve variability, there remains considerable variation which is 
independent of $R(V)$.  We use the optical spectrophotometry to verify the 
presence of structure at intermediate wavelength scales in the curves.  The 
fact that the optical through UV portions of the curves are sampled at 
relatively high resolution makes them very useful for determining how 
extinction affects different broad band systems, and we provide several 
examples.  Finally, we compare our results to previous investigations. 
\end{abstract}

\keywords{ISM --- dust, extinction}

\section{Introduction}

Dust grains play a large number of roles in the interstellar medium (ISM), serving, for 
example, as the primary formation site for H$_2$, as an important heat source (via the 
photoelectric effect), and as a chemical reservoir for the heavy elements.  Arguably, 
however, their most significant effect is in modulating the flow of electromagnetic 
radiation through interstellar space via the combined effect of absorption and scattering.
The ability of grains to transmit, redirect, and transmute electromagnetic energy 
is enormously important to the physics of interstellar space and, perhaps most 
far-reaching, places profound limits on our ability to study the universe at optical and 
ultraviolet (UV) wavelengths.

Studies of the interstellar ``extinction curve,'' i.e., the wavelength dependence 
of the combined effect of absorption and scattering out of the line of sight, have 
shown it to be highly spatially variable. This poses significant problems since 
a detailed knowledge of the curve is essential in numerous astrophysical 
applications, such as reconstructing the intrinsic spectral energy distributions 
(SEDs) of objects ranging from nearby stars to the most distant galaxies or for 
understanding the deposition of electromagnetic energy in star forming regions.
On the other hand, these variations are a boon for the study of the dust itself, 
since they reflect general differences in the grain populations from sightline to 
sightline.  Understanding how the various spectral regions of the curves relate to 
each other and how they respond to changes in the interstellar environments can 
provide information critical for characterizing the size, composition, and 
structure of interstellar grains \citep[e.g.,][]{Weingartner01, Clayton03, Zubko04}.

Detailed investigations of the wavelength dependence of interstellar extinction began with 
measurements in the 
optical, generally based on broadband photometry. Early on, it was discovered that optical 
extinction curves may vary in shape from sightline to sightline, as exemplified by the 2.2 
$\mu m^{-1}$ ``knee'' reported by \citet{Whitford58}.  Higher resolution studied revealed the 
presence of the ``Very Broad Structure,'' 
first reported by \citet{Whiteoak66} and studied in detail by \citet{York71}.  
This structure is apparent in the ``continuum subtracted'' residuals 
presented most recently by \citet{MaizApellaniz14}. Equally early, it was recognized that some of
these shape variations were related to the ratio of total-to-selective extinction, 
${R(V) \equiv A(V)/E(B-V)}$ \citep{Johnson63}.
The $R(V)$-dependence of optical and, also, UV extinction was later quantified by 
\citet{Cardelli89}, and it has become standard practice to represent extinction curves 
as a family whose broad characteristics are dependent on $R(V)$ \citep{Fitzpatrick99review, 
Fitzpatrick04, Valencic04, Gordon09FUSE, MaizApellaniz14}.

It is somewhat remarkable that -- despite its observational accessibility -- the optical is not the 
best-characterized region of the interstellar extinction curve. The difficulty in obtaining ground-based 
spectrophotometry and the resultant paucity of spectroscopic resolution studies of optical extinction 
have prevented detailed investigations of the optical extinction features, their spatial variability, 
and their relationship to features at other wavelengths.  Remarkably, is it the UV, 
accessible only by space-based instruments, where extinction has been most extensively studied at 
spectroscopic resolution \citep{Witt84, Fitzpatrick86, Fitzpatrick90, Fitzpatrick88, Aiello88, 
Cardelli89, Valencic04, Gordon09FUSE}, primarily as a result of the large database of UV 
spectrophotometry obtained by the {\it International Ultraviolet 
Explorer} satellite (\iue).    

To fill in this gap in our knowledge of optical extinction, and to better characterize the 
relationship between extinction in the optical and that at other wavelengths, we initiated a Hubble 
Space Telescope (\hst) 
SNAP program using the low-resolution STIS optical and near-infrared (NIR) gratings to observe a 
carefully selected 
set of moderately-reddened early-type stars that were previously observed in the UV. These 
observations provide low-resolution spectrophotometry covering the spectral range from the NIR
to the near-UV region, i.e., between $\sim$1 $\mu$m and $\sim$3000 \AA. When coupled with the 
existing UV data, these yield well-calibrated, low resolution SEDs for the program stars from 
1150 -- 10000 \AA\, which can be used to construct detailed extinction curves. These data resolve most, 
if not all, structure in the curves shortward of 1 $\mu$m; characterize the detailed shapes of the 
extinction curves in largely unexplored spectral regions; reveal the degree of variability in NIR 
through near-UV curves; and allow us to relate this variability to that seen in the UV and infrared. 

The current paper utilizes \iue\ and STIS low resolution spectrophotometry along with broadband NIR 
photometry from the {\it Two Micron All-Sky Survey} (\tmass) to study the $R(V)$-dependence of extinction along lines of sight sampling a range of 
Milky Way environments. A detailed analysis and characterization of the structure observed in optical 
extinction curves is addressed in a companion paper \citep{Massa20}.  

Section 2 describes the selection of sightlines used in our analysis, as well as the processing of the STIS, \iue, and \tmass\/
data. Section 3 explains how the extinction curves were derived. Section 4 presents our results 
along with an $R(V) = 3.1$ curve (which is typical of the diffuse Milky Way environment) and an $R(V)$ 
parameterization which captures much of the sight line to sight line variation.  Section 5 discusses 
our results and compares them to previous work.  Finally, Section 6 summarizes our findings.

\section{Observations and Data Processing}

\subsection{The Sample} 
Our selection of program stars is based on the \hst\/ SNAP Program 13760.  The initial sample 
included 130 reddened, normal, Milky Way, near main sequence (luminosity class III -- V) O7 -- B9.5 
stars.  The techniques we use to determine extinction curves give excellent 
results for stars in this range (see \S\ref{secCURVES}). This sample was restricted to stars with 
existing UV spectrophotometry from \iue\/ and NIR broadband photometry from the \tmass\/ project and comprised two overlapping sub-groups: 1) stars 
with moderate to high extinction (\ebv\/ $\gtrsim 0.20$~mag) which sample sight lines with 
extreme values of $R(V)$ or interesting Milky Way environments, and 2) stars which sample dust 
within Milky Way open clusters and star-forming regions that have the potential to sample small 
scale structure in the dust distribution.  Examples of the first group can be found in 
\citet{Valencic04} and \citet[][hereafter F07]{Fitzpatrick07}, who summarize most of the 
previous \iue\/ work, including the \citet{Clayton00} data on low density sight lines. The 
second group includes stars in open clusters observed by \citet{Aiello82, Panek83, Clayton87, 
Boggs89, Fitzpatrick90, Hackwell91}.  

Of the 130 Milky Way targets submitted for our SNAP program, 77 were observed. Four stars 
(HD~73882, HD~99872, HD~281159 and Cl*~NGC~457~Pes~10) were eliminated from the sample because 
their STIS spectra show evidence for offset and overlapping spectra from one or more other stars lying in the dispersion direction.  
A fifth star, HD~164865, turned out to be a late B supergiant, B9~Iab, and could not be well 
fit by the models we use to produce our extinction curves.  That left the final sample of 72 
stars used in this paper.  These stars are listed in Table~1 along with some of 
their basic properties ({\it i.e.}, spectral type, V magnitude, \ebv, distance, and galactic coordinates).  
Membership in a 
stellar cluster is indicated by the cluster name in parenthesis in the first column of the table.  

\begin{figure*}
\epsscale{1.1}
\plotone{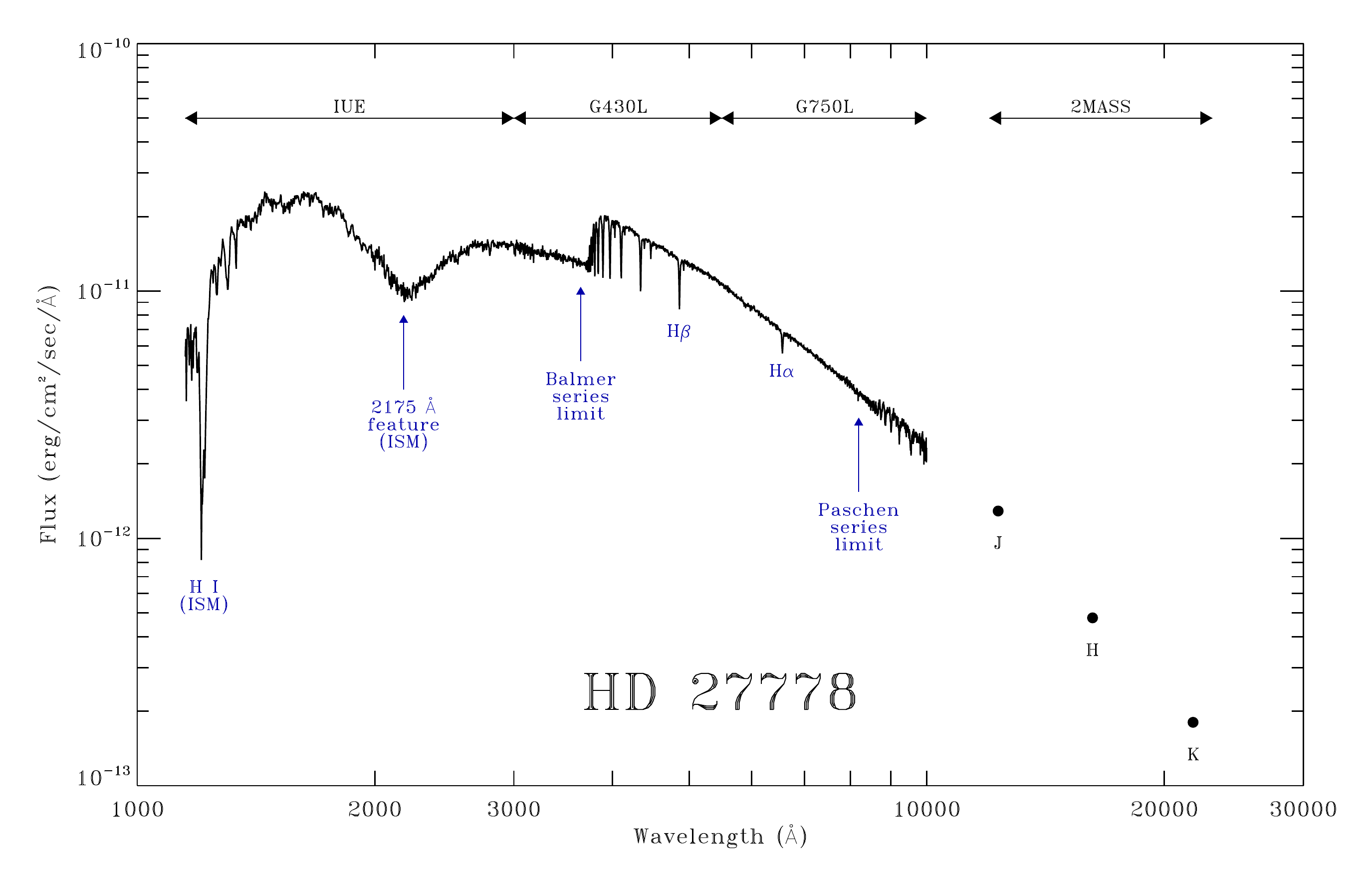}
\caption{Representative spectral energy distribution (SED) for our program stars, using HD~27778 as an 
example.  Sources and wavelength ranges of the various datasets are indicated at the top of the figure.  
{\it International Ultraviolet Explorer} satellite (\iue) data from the SWP, LWP, and LWR cameras, 
covering the range 1150-3000 \AA\/ was obtained from the {Mikulski Archive for Space Telescopes}.  
New {\it Hubble Space Telescope} spectrophotometric observations using the G430L and G750L gratings 
were acquired for this program, and cover the ranges 3000-5450 \AA\/ and 5450-10000 \AA, respectively.  
NIR \jhk\/ photometry from the {\it Two Micron All-Sky Survey} (\tmass) was obtained from the \tmass\/ 
archive maintained by the {Infrared Processing and Analysis Center}.  Prominent stellar and interstellar 
features are identified beneath the SED.
\label{figSED}}
\end{figure*}

Figure~\ref{figSED} shows an example of a typical program star SED, for the case of HD~27778 (B3 V). Prominent stellar and 
interstellar features are indicated beneath the SED. The sources of the data are indicated near the 
top of the figure.  All data used in this study were obtained from the \iue\/ satellite (UV 
spectrophotometry), the \hst\/ (``G430L'' and ``G750L'' optical spectrophotometry), and the 
\tmass\/ project (NIR \jhk\/ photometry).  In the subsections 
below, we describe the acquisition, processing, and calibration of the three datasets.

\startlongtable
\begin{deluxetable*}{lcrccrrc} 
\tabletypesize{\scriptsize}
\tablewidth{0pc} 
\tablecaption{Basic Data for Program Stars}
\tablehead{ 
\colhead{Star\tablenotemark{a}}         &
\colhead{Spectral}                      & 
\colhead{V}                             &
\colhead{$E(B-V)$\tablenotemark{c}}     &
\colhead{Distance\tablenotemark{d}}     &
\colhead{$l$}                &
\colhead{$b$}                &
\colhead{Reference}                     \\
\colhead{}                         &  
\colhead{Type\tablenotemark{b}}    & 
\colhead{(mag)}                    &
\colhead{(mag)}                    &
\colhead{(pc)}                     &
\colhead{($\degr$)}                & 
\colhead{($\degr$)}                & 
\colhead{}                         }
\startdata
BD+44 1080 &  B6 III & 9.12 & 0.85 & 521 & $162.484$ & $  1.51$ & 1  \\
BD+56 517 (NGC 869) &  B1.5 V               & 10.50 & 0.56 &     2079 & $134.612$ & $ -3.71$ & 2  \\
BD+56 518 (NGC 869) &  B1.5 V                & 10.60 & 0.59 &     2079 & $134.612$ & $ -3.71$ & 3  \\
BD+56 576 (NGC 884) &  B2 III                & 9.38 & 0.54 &     2345 & $135.047$ & $ -3.59$ & 4  \\
BD+69 1231 &  B9.5 V               & 9.27 & 0.23 & 389 & $110.255$ & $ 11.36$ & 5  \\
BD+71 92 (Cr 463) &  \nodata & 10.35 & 0.29 &      702 & $126.824$ & $  9.48$ & \nodata \\
CPD-41 7715 (NGC 6231) &  B2 IV-Vn              & 10.53 & 0.44 &     1243 & $343.413$ & $  1.18$ & 6  \\
CPD-57 3507 (NGC 3293) &  B1 V                 & 9.27 & 0.20 &     2910 & $285.832$ & $  0.08$ & 7  \\
CPD-57 3523 (NGC 3293) &  B1 II                & 8.02 & 0.27 &     2910 & $285.867$ & $  0.07$ & 7  \\
CPD-59 2591 (Tr 16) &  \nodata & 10.93 & 0.77 &     3200 & $287.591$ & $ -0.74$ & \nodata \\
CPD-59 2600 (Tr 16) &  O6 V((f))            & 8.61 & 0.51 &     3200 & $287.591$ & $ -0.74$ & 8  \\
CPD-59 2625 (Tr 16) &  B2 V                 & 11.58 & 0.44 &     3200 & $287.637$ & $ -0.68$ & 9  \\
GSC03712-01870 &  O5+ & 13.02 & 1.03 & 8200 & $137.859$ & $ -0.97$ & 10  \\
HD~13338 &  B1 V                  & 9.06 & 0.51 & 1506 & $133.513$ & $ -3.27$ & 11  \\
HD~14250 (NGC 884) &  B1 III                & 8.97 & 0.58 &     2345 & $134.802$ & $ -3.72$ & 3  \\
HD~14321 (NGC 884) &  B1 IV                & 9.22 & 0.53 &     2345 & $134.938$ & $ -3.85$ & 2  \\
HD~17443 &  B9 V                  & 8.74 & 0.38 & 284 & $133.924$ & $  7.55$ & 12  \\
HD~18352 &  B1 V                  & 6.83 & 0.47 & 566 & $137.715$ & $  2.16$ & 11  \\
HD~27778 &  B3 V                  & 6.36 & 0.36 & 262 & $172.759$ & $-17.38$ & 13  \\
HD~28475 &  B5 V                  & 6.79 & 0.27 & 254 & $185.139$ & $-25.09$ & 14  \\
HD~29647 &  B8 III                & 8.31 & 1.00 & 157 & $174.042$ & $-13.35$ & 15  \\
HD~30122 &  B5 III                & 6.34 & 0.26 & 371 & $176.616$ & $-14.03$ & 13  \\
HD~30675 &  B3 V                  & 7.53 & 0.52 & 355 & $173.597$ & $-10.20$ & 16  \\
HD~37061 (NGC 1978) &  B0.5 V & 6.83 & 0.53 &     2429 & $208.927$ & $-19.28$ & 17  \\
HD~38087 &  B5 V                 & 8.30 & 0.30 & 315 & $207.073$ & $-16.26$ & 17  \\
HD~40893 &  B0 IV:                & 8.90 & 0.44 & 2632 & $180.091$ & $  4.33$ & 11  \\
HD~46106 (NGC 2244) &  B1 V                  & 7.92 & 0.41 &     1670 & $206.188$ & $ -2.10$ & 18  \\
HD~46660 (NGC 2264) &  B1 V                  & 8.04 & 0.59 &      667 & $201.173$ & $  1.46$ & 11  \\
HD~54439 &  B2 IIIn               & 7.69 & 0.29 & 1295 & $225.402$ & $ -1.69$ & 11  \\
HD~62542 &  B5 V                  & 8.07 & 0.31 & 396 & $255.923$ & $ -9.25$ & 19  \\
HD~68633 &  B5 V                  & 8.00 & 0.48 & 274 & $266.224$ & $ -9.55$ & 20  \\
HD~70614 &  B6                   & 9.27 & 0.67 & \nodata & $259.807$ & $ -3.24$ & 20  \\
HD~91983 (NGC 3293) &  B1 III                & 8.58 & 0.28 &     2910 & $285.882$ & $  0.05$ & 21  \\
HD~92044 (NGC 3293) &  B0.5 II               & 8.25 & 0.40 &     2910 & $285.937$ & $  0.06$ & 21  \\
HD~93028 (Cr 228) &  O9 V                 & 8.37 & 0.21 &     2201 & $287.643$ & $ -1.20$ & 8  \\
HD~93222 (Cr 228) &  O7 III((f))          & 8.10 & 0.35 &     2201 & $287.726$ & $ -1.01$ & 8  \\
HD~104705 &  B0.5 III              & 7.79 & 0.27 & 2082 & $297.465$ & $ -0.34$ & 11  \\
HD~110336 &  B9 IV                 & 8.64 & 0.45 & 333 & $302.463$ & $-14.50$ & 22  \\
HD~110946 &  B1 V:                 & 9.14 & 0.50 & 1509 & $302.417$ & $ -2.06$ & 23  \\
HD~112607 &  B7/B8 III             & 8.06 & 0.31 & 531 & $303.784$ & $ -0.78$ & 22  \\
HD~142096 &  B2.5 V & 5.03 & 0.15 & 94 & $350.724$ & $ 25.38$ & 24  \\
HD~142165 &  B6 IVn & 5.37 & 0.11 & 128 & $347.514$ & $ 22.15$ & 24  \\
HD~146285 &  B8 V                  & 7.93 & 0.32 & 208 & $351.012$ & $ 18.20$ & 24  \\
HD~147196 &  B5 V                  & 7.04 & 0.26 & 272 & $352.811$ & $ 18.25$ & 25  \\
HD~147889 &  B2 V                  & 7.90 & 1.09 & 153 & $352.868$ & $ 17.04$ & 26  \\
HD~149452 &  O8 Vn((f))           & 9.06 & 0.87 & 1636 & $337.455$ & $  0.04$ & 8  \\
HD~164073 &  B3 III/IV             & 8.03 & 0.20 & 961 & $344.171$ & $-12.56$ & 20  \\
HD~172140 &  B0.5 III              & 9.95 & 0.21 & 7318 & $  5.274$ & $-10.61$ & 27  \\
HD~193322 &  O9 V:((n))           & 5.83 & 0.38 & 727 & $ 78.110$ & $  2.78$ & 8  \\
HD~197512 &  B1 V                  & 8.56 & 0.32 & 1614 & $ 87.895$ & $  4.63$ & 16  \\
HD~197702 &  B1 III(n)           & 7.89 & 0.46 & 701 & $ 73.925$ & $ -6.83$ & 28  \\
HD~198781 &  B0.5 V                & 6.45 & 0.32 & 768 & $ 99.956$ & $ 12.62$ & 11  \\
HD~199216 &  B1 II                 & 7.03 & 0.74 & 1097 & $ 88.922$ & $  3.04$ & 29  \\
HD~204827 (Tr 37) &  B0 V                  & 7.94 & 1.11 &      835 & $ 99.168$ & $  5.56$ & 30  \\
HD~210072 &  B2 V                   & 7.66 & 0.49 & 659 & $100.844$ & $ -0.37$ & 26  \\
HD~210121 &  \nodata & 7.67 & 0.33 & \nodata & $ 56.861$ & $-44.47$ & \nodata \\
HD~217086 (Cep OB3) &  O7 Vn                & 7.64 & 0.92 &      724 & $110.215$ & $  2.73$ & 8  \\
HD~220057 &  B2 IV                 & 6.93 & 0.23 & 770 & $112.116$ & $  0.20$ & 31  \\
HD~228969 (Br 86) &  B2 II:                & 9.49 & 0.99 &     2429 & $ 76.615$ & $  1.29$ & 11  \\
HD~236960 &  B0.5 III             & 9.75 & 0.72 & 3474 & $134.601$ & $ -1.51$ & 26  \\
HD~239693 (Tr 37) &  B3 V                 & 9.51 & 0.44 &      835 & $ 98.824$ & $  4.72$ & 32  \\
HD~239722 (Tr 37) &  B2 IV                & 9.55 & 0.92 &      835 & $100.339$ & $  5.05$ & 32  \\
HD~239745 (Tr 37) &  B1 V                 & 8.91 & 0.53 &      835 & $ 99.432$ & $  2.99$ & 32  \\
HD~282485 &  B9 V                  & 9.88 & 0.47 & 412 & $172.220$ & $-10.29$ & 33  \\
HD~292167 &  O9 III:               & 9.25 & 0.69 & 2681 & $211.630$ & $ -1.17$ & 11  \\
HD~294264 (NGC 1977) &  B3 Vn                & 9.53 & 0.49 &      476 & $208.492$ & $-19.16$ & 34  \\
HD~303068 (NGC 3293) &  B1 V                 & 9.78 & 0.25 &     2910 & $285.688$ & $  0.07$ & 7  \\
NGC 2244 11 (NGC 2244) &  B2 V                  & 9.73 & 0.44 &     1670 & $206.422$ & $ -2.02$ & 35  \\
NGC 2244 23 (NGC 2244) &  B2.5 V               & 11.21 & 0.47 &     1670 & $206.311$ & $ -2.08$ & 36  \\
Trumpler 14 6 (Tr 14) &  B1 V                 & 11.23 & 0.49 &     3200 & $287.419$ & $ -0.59$ & 37  \\
Trumpler 14 27 (Tr 14) &  \nodata & 11.30 & 0.57 &     3200 & $287.399$ & $ -0.60$ & \nodata \\
VSS VIII-10 &  B8 V                 & 10.07 & 0.73 & 235 & $359.518$ & $-18.29$ & 38  \\
\enddata
\tablenotetext{a}{The stars are listed in alphabetical order using the most commonly adopted forms of 
their names. The first preference was ``HD~nnn'', followed by ``BDnnn'', etc.  There are 31 program 
stars that are members of open clusters or associations.  The identity of the cluster or association 
is given in parentheses after the star's name.}
\tablenotetext{b}{Spectral types were selected from those given in the SIMBAD database, and the 
source of the adopted types is shown in the ``Reference'' column.  When multiple types were available 
for a particular star, we selected one based on our own preferred ranking of the sources.}
\tablenotetext{c}{Values of \ebv\/ were inferred from the values of \efourfive\/ determined from the 
extinction curves, whose construction is described in \S\ref{secCURVES}. }
\tablenotetext{d}{The stellar distances are the same as reported by \citet{Fitzpatrick07}. For HD~142096 and HD142165, 
neither of which was in the \citet{Fitzpatrick07} survey, the distances were computed from $Hipparcos$ parallaxes 
\citep{vanLeeuwen07}.  For the star GSC03712-01870, the distance was taken from \citep{Smartt96}.}
\tablerefs{(1) \citet{Bouigue59}; (2) \citet{Schild65}; (3) \citep{Johnson55}; (4) \citet{Slettebak68}; 
(5) \citet{Racine68a}; (6) \citet{Schild71}; (7) \citet{Turner80}; (8) \citet{MaizApellaniz04}; 
(9) \citet{Massey93}; (10) \citet{Muzzio74}; (11) \citet{Morgan55}; (12) \citet{Racine68b}; 
(13) \citet{Osawa59}; (14) \citet{Cowley69}; (15) \citet{Metreveli68}; (16) \citet{Guetter68}; 
(17) \citet{Schild71b}; (18) \citet{Johnson53}; (19) \citet{Feast55}; (20) \citet{Houk78}; 
(21) \citet{Hoffleit56}; (22) \citet{Houk75}; (23) \citet{Feast57}; (24) \citet{Garrison67}; 
(25) \citet{Borgman60}; (26) \citet{Hiltner56}; (27) \citet{Hill70}; 
(28) \citet{Walborn71};
(29) \citet{Divan54}; 
(30) \citet{Morgan53}; (31) \citet{Boulon58}; (32) \citet{Garrison76}; (33) \citet{Roman55}; 
(34) \citet{Smith72}; (35) \citet{Meadows61}; (36) \citet{Johnson62}; (37) \citet{Morrell88}; 
(38) \citet{Vrba84}}
\label{tabSTARS}
\end{deluxetable*}

\subsection{Optical Data - \hst}
\label{HSTdata}

The optical data were all obtained as part of the \hst\/ SNAP Program 13760 and
consist of STIS spectrophotometry with the G430L ($2900 \leq \lambda \leq 5700$
\AA) and G750L ($5240 \leq \lambda \leq 10270$ \AA) first order gratings.  The
total wavelength coverage is 2900--10270 \AA\ with a resolution $\lambda/\Delta\lambda$ ranging from 530
to 1040 \citep[e.g.,][]{Bostroem11}. The G430L observations are essential
for determining the stellar parameters and had a minimum signal-to-noise (S/N)
goal of 20:1 per pixel. G750L observations are always more problematic because
of the sensitivity drop in the instrument of $\sim$5.5$\times$ from a peak at 7100~\AA\ to 10000~\AA\ and
because moderately reddened OB stars are much fainter at longer wavelengths.
For example, HD~27778 in Figure~\ref{figSED}, with \ebv\/ = 0.36, drops by a factor of 
$\sim$2.5$\times$ over the same wavelength interval.
Consequently, our G750L exposure times were limited to avoid saturation at the shortest 
wavelengths (assuming a mean extinction curve shape) and also to keep the
total on-target time under 30 minutes to increase the potential for execution of
our SNAP observations. As a result of these considerations, the S/N of the G750L
data vary considerably with wavelength and from star-to-star.

The G430L and G750L spectra are processed with the Instrument Definition Team
(IDT) pipeline software written in IDL by D. J. Lindler in 1996--1997. This IDL
data reduction is more flexible and offers the following advantages over the
STScI pipeline results that are available from the Mikulski Archive for Space 
Telescopes (\mast\footnote{http://archive.stsci.edu/hst/}): 
(1) better compensation for small shifts in pointing or instrumental flexure that 
may occur between the images of a cosmic ray split (CR-split) observation; (2) a 
larger extraction window height for the G750L data, to compensate for an increase 
in the spectral width in recent years; (3) elimination of ``hot'' pixels from the 
extracted spectrum; and (4) the use of tungsten lamp spectra to remove CCD fringing 
effects in the G750L spectra.  The processing technique is discussed more thoroughly 
in Appendix~\ref{appendix_stis}.

In addition to the above data extraction advantages, our post-processing
includes the charge transfer efficiency (CTE) corrections of \citet[]{Goudfrooij06}, 
and an up-to-date
correction for the changes in STIS sensitivity with time and temperature.
The observations of the three primary flux standards G191B2B, GD153, and GD71,
as corrected for CTE and the latest measures of sensitivity change with time,
produce the current absolute flux calibration \citep[{\it CALSPEC}; see][]{Bohlin14}. 

After completion of all the processing steps, the data were concatenated, with
G430L used below 5450~\AA\ and G750L used above 5450~\AA.

\subsection{Ultraviolet Data - \iue}

With the exception of GSC03712-01870, the stars included in this paper have all been 
incorporated in previous UV extinction studies, notably by \citet[][hereafter F05]{Fitzpatrick05a} 
and \citetalias{Fitzpatrick07}, and have \iue\/ spectrophotometric observations available 
from both the short-wavelength region (``SWP'' camera; 1150--1980~\AA) and long-wavelength 
region (``LWR'' or ``LWP'' cameras; 1980--3200~\AA).  Only SWP observations exist for 
GSC03712-01870.  The \iue\/ data were obtained by us from the \mast\ archive and were 
processed as described earlier (e.g., see \citetalias{Fitzpatrick07}).  The critical step in the 
processing of the 
archival data was the correction for deficiencies in the \iue\/ NEWSIPS calibration, using 
the results of \citet[][hereafter M00]{Massa00}.  These deficiencies include
systematic thermal and temporal effects and an 
incorrect absolute calibration.   
This is particularly important for our current study because we derive extinction curves 
using model atmosphere calculations, rather than unreddened comparison stars, and errors 
in the calibration of the data would be reflected directly in the resultant 
curves. The \citetalias{Massa00} corrections place the \iue\/ data on the absolute flux calibration 
system as it was in the year 2000.  The calibration has been modified in the intervening years and, to take
advantage of this, we additionally corrected the data to the current {\it CALSPEC} standard.  We did this by 
comparing \citetalias{Massa00}-corrected \iue\/ spectra of the flux standards G191B2B, GD71 and GD153 with
the current SEDs of those stars as found in the {\it CALSPEC} database.\footnote{The {\it CALSPEC} Calibration Database was accessed at 
http://www.stsci.edu/hst/instrumentation/reference-data-for-calibration-and-tools/astronomical-catalogs/calspec and data retreived from ftp://ftp.stsci.edu/cdbs/calspec/}  
This comparison indicated that the
flux levels of \citetalias{Massa00}-corrected \iue\/ spectra need to be reduced by an average of $\sim$0.9\% in 
the short-wavelength region and $\sim$1.8\% in the long-wavelength region (with a roughly linear wavelength dependence across the whole \iue\/ UV).  This final correction 
makes the calibration of the \iue\/ data fully consistent and compatible with the calibration of the optical 
\hst\/ spectrophotometry described above and the NIR photometry described below.  When the processing was completed, the UV 
spectra for each star were trimmed to the wavelength range 1150--3000~\AA, with the 
short- and long-wavelength camera data joined at a wavelength of 1978~\AA.\\

\subsection{NIR Data - \tmass}\label{sec2MASS}

The third dataset consists of NIR \jhk\/ photometry from \tmass.  These data were obtained from the 
2MASS All-Sky Point Source Catalog, accessed via the \tmass\/ website hosted by the Infrared Processing and 
Analysis Center (IPAC).\footnote{The \tmass\/ website can be found at https://www.ipac.caltech.edu/2mass/.} 
\jhk\/ measurements are available for all program stars although, due to lack of uncertainty estimates, we 
excluded the {\it J} and {\it H} data for the star CPD--57 3523.  The \tmass\/ data required no additional 
processing.

\begin{figure}
\epsscale{1.0}
\plotone{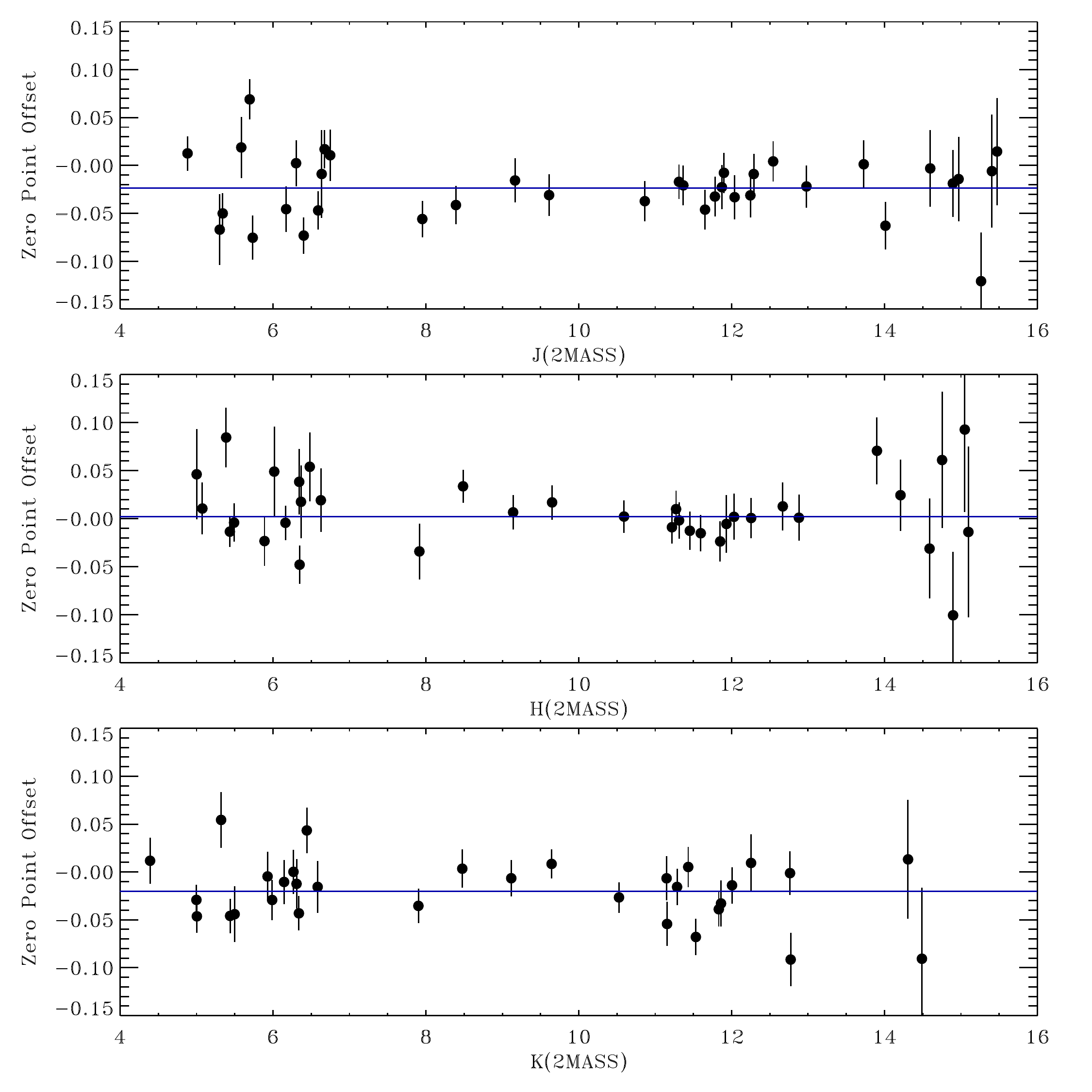}
\caption{Calibration of synthetic \tmass\/ \jhk\/ photometry.  Synthetic {\it JHK} magnitudes were computed for the 
model SEDs of 30-40 stars in the {\it CALSPEC} database, as described in \S \ref{sec2MASS}, and ``zero point offsets'' 
determined to match the synthetic values to the observed \tmass\/ values.  The top, middle, and bottom panels show these 
offsets for the $J$, $H$, and $K$ filters, respectively, plotted against the observed \tmass\/ magnitudes. Thick blue lines indicate 
the weighted mean offsets, with values of --0.023, 0.002, and --0.020 mag for $J$, $H$, and $K$, respectively.  The error bars 
are those associated with the observed magnitudes.
\label{fig2MASS}}
\end{figure}

The flux calibration of the \tmass\/ bands is important to our study because we derive NIR extinction 
by comparing the \tmass\/ magnitudes for reddened stars with synthetic values derived from the stars' 
assumed intrinsic SEDs.  Thus, the synthetic photometry must be calibrated in a manner 
consistent with the latest {\it CALSPEC} system. The standard \tmass\/ calibration was performed by 
\citet{Cohen03}, based on a synthetic spectrum of Vega \citep[see][]{Cohen92} and the assumption 
that Vega defines zero magnitude in the \tmass\/ NIR. To assure consistency with our new study, 
we rederived this calibration, based on the latest {\it CALSPEC} database. We did this by computing
synthetic \tmass\/ magnitudes for a subset of the {\it CALSPEC} calibration sample and then determining 
the transformation that maps these values to the observed \tmass\/ 
\jhk\/ magnitudes.  The procedure is essentially identical to that used by \citet{Fitzpatrick05b} in 
a general calibration of synthetic photometry for B and early A-type stars.  In this case, we began with the 40 flux 
standard stars from the {\it CALSPEC} Calibration Database for which NIR model fluxes were computed.  
This subset of the full calibration sample includes all the stars in Table 1 of the Database for which entries exist 
in Column (6), indicating the existence of a model SED, with the exception of SF1615+001A (no \tmass\/ data) 
and HZ43 (contamination of \tmass\/ by a red companion star).  The sample was further restricted to those stars with \tmass\/
magnitude uncertainties less than 0.1 mag, leaving 37, 36, and 32 stars for calibration of the $J$, $H$, and $K$ filters, respectively. 
We performed synthetic photometry on the model SEDs 
(in $f_{\lambda}$ units) using the \tmass\/ \jhk\/ filter response functions and $F_\lambda$(0 mag) values from 
\citet{Cohen03}.  The transformation to the observed \tmass\/ system is expressed in terms of 
``zero point offsets'' as defined in \citet{Cohen03} or, equivalently, in Equation 1 of \citet{MaizApellaniz07}.
The results for each filter are shown in Figure \ref{fig2MASS}, where we plot zero point offsets against the \tmass\/ 
magnitudes. The final adopted offsets are weighted means with values of --0.023, +0.002, and --0.020 mag for the $J$, $H$, and $K$ filters, respectively.  These offsets are 
almost identical (i.e., within $\pm$0.005 mag) to those found by \citet{MaizApellaniz18}, who also used the {\it CALSPEC} 
database to examine the \tmass\/ calibration.  Like \citet{MaizApellaniz18}, we find no evidence for color dependence in these
offsets, nor for a systematic brightness dependence (as can be seen in Figure \ref{fig2MASS}). Use of these zero point offsets assures that 
the \tmass\/ photometric data are calibrated consistently with our UV and optical spectrophotometric datasets.

\section{Creating the NIR-UV Extinction Curves}

We derive NIR through UV extinction curves for our 72 program stars 
using the ``Extinction-Without-Standards'' approach, first described by \citetalias{Fitzpatrick05a}.  This technique 
utilizes SEDs from stellar atmosphere models to represent the intrinsic SEDs of the reddened 
program stars, rather than observations of unreddened stars of similar spectral type.  The 
advantages of such an approach are described in detail in \citetalias{Fitzpatrick05a}.  In our previous applications 
of Extinction-Without-Standards (e.g., \citetalias{Fitzpatrick05a} and \citetalias{Fitzpatrick07}), we determined the stellar properties and 
the extinction properties simultaneously by fitting \iue\/ UV spectrophotometry and 
optical/near-IR photometry with model SEDs and a parameterized version of the interstellar 
extinction curve.  In this study, we adopt a slightly different approach, explicitly 
separating the process into two distinct steps: (1) first, we determine the appropriate 
stellar properties needed to characterize the model SED, and then (2) form a normalized 
ratio between the observed and model SEDs (i.e., the extinction curve).  This approach is 
made possible by the new \hst\/ spectrophotometry, which allows a detailed modeling of 
critical stellar diagnostics in the optical spectral region.  The advantage is that 
the resultant extinction curves are derived completely empirically, without any 
assumptions as to their underlying functional form.  The two steps in this process are 
described separately in the following subsections.

\subsection{Step 1: Determining the Stellar Properties}
\label{secINTRINSIC}

We estimate the intrinsic properties of our program stars by fitting their G430L spectra with 
stellar atmosphere models.  The \hst\/'s G430L data consist of spectrophotometric observations 
over the wavelength range 2900--5700~\AA\/ at a resolution better than 10~\AA.  As such, they 
provide a clear view of the wide range of spectroscopic features present in the classical 
blue spectral region, which formed the basis of the spectral classification system.  Thus, 
ample information is present in these data to determine stellar effective temperature (\teff), 
surface gravity (\logg) and, more coarsely, metallicity (\abund).

For most of our program stars, we fit the G430L data using the TLUSTY NLTE SED grids published by 
\citet[][``OSTAR2002'']{Lanz03} and  \citet[][``BSTAR2006'']{Lanz07}. The grids were computed at full 
spectral resolution (at the thermal broadening level) and, together, cover a wide range of metallicities 
(\abund\/ = +0.3 to --1.0), effective temperatures (\teff\/ = 15,000~K to 55,000~K), and surface gravities 
(\logg\/ = 4.75 to the modified Eddington limit).  The OSTAR2002 grid assumes a microturbulence velocity of 
\vturb\/ = 10~\kms, while the BSTAR2006 grid assumes \vturb\/ = 2~\kms, with a small number of low gravity 
models computed with \vturb\/ = 10~\kms. Both grids assume the solar abundance scale of \citet{Grevesse98}.  
The modeling process involves four steps: (1) begin with initial 
estimates of \teff, \logg, and \abund; (2) interpolate within the 
appropriate TLUSTY grid to produce a full-resolution SED, F$_\lambda$, with these properties in the G430L 
spectral range; (3) broaden the SED to account for the effects of stellar rotation and the instrumental 
profile of the G430L and (4) compare the model SED with the G430L spectrum, using $\chi^2$ as the measure 
of the goodness-of-fit. The initial estimates of the stellar properties were then adjusted and the process 
repeated iteratively until the minimum value of $\chi^2$ was achieved. The broadband  shapes and levels of 
the G430L SEDs are distorted relative to the models due to the effects of distance and interstellar 
extinction.  To remove these effects from the analysis, we mapped the overall shapes of the model continua 
onto the observed continua using a cubic spline, whose anchor points were iteratively adjusted along with 
the stellar properties.  Thus, the final stellar properties determined from the fitting procedure depend 
only on the absorption line spectrum (plus the Balmer jump) and not at all on the general shape of the 
spectrum.  The continuum mapping is not used in the subsequent derivation of the extinction curves.

The broadening applied to the model SEDs requires some additional discussion.  In early tests, we applied 
a Gaussian PSF with FWHM = 5.5~\AA, as taken from the STIS Instrument Handbook \citep{Riley18}\footnote{The STIS 
Instrument Handbook was accessed at 
http://www.stsci.edu/hst/stis/faqs/documents/handbooks/currentIHB/cover.html.}, and rotational broadening based 
on \vrot\/ values from the literature.  In a number of cases, however, the observed spectra appeared 
significantly more broadened, consistent (in the most extreme cases) with a PSF FWHM of 7.8~\AA.  The 
initially adopted value 5.5~\AA\/ appeared, in these tests, to be a minimum.  We attribute these 
discrepancies to a variable PSF width in our slitless spectra, arising from focus changes due to thermal 
cycling of the instrument (aka ``breathing'').  Ideally, we would deal with this by adopting measured 
\vrot\/ values for all the stars and allowing the FWHM of the PSF to be a free parameter in the fitting 
routine, adjusted to achieve a best-fit to the line profiles.  Unfortunately, only about one-third of our 
program stars have measured \vrot\/ values and so this is not feasible.  Instead, we adopted a modified 
approach and fixed the PSF FWHM at the minimum observed value of 5.5~\AA\/ and varied the width of the 
rotational broadening profile to achieve a best-fit to the spectra. This yields a value of ``$v_{broad}$'', 
which contains both the rotational broadening of the star and the effects of breathing.  We found that, 
in cases where \vrot\/ is known, these two different approaches yield essentially identical results, in 
both the values of $\chi^2$ and the model parameters.  Evidently, the resolution and S/N of our data are 
such that the differences between a Gaussian and a rotational profile are not significant.

Eleven of the program stars have \teff\/ values below the 15,000~K lower limit of the TLUSTY BSTAR2006 grid.  
The general procedure for determining their properties is similar to that described above, except that in 
step (2) we used a combination of the ATLAS9 LTE models from \citet{Kurucz91} to produce an interpolated 
atmospheric structure for the desired stellar properties, and then the spectral synthesis program SPECTRUM 
(Version 2.76e) developed by Richard Gray \citep[see][]{Gray94} to produce a fully resolved stellar 
spectrum over the G430L spectral window.  For consistency with the TLUSTY grids, these calculations also assumed 
the \citet{Grevesse98} solar abundance scale.

\startlongtable
\begin{deluxetable*}{lcccccc} 
\tablenum{2} 
\tabletypesize{\scriptsize}
\tablewidth{0pc} 
\tablecaption{Stellar Properties from G430L Analysis}
\tablehead{ 
\colhead{Star}              &
\colhead{$T_{eff}$}                          & 
\colhead{$\log g$}                           &
\colhead{[m/H]\tablenotemark{a}}                              & 
\colhead{$v_{turb}$\tablenotemark{b}}        & 
\colhead{$v_{broad}$\tablenotemark{c}}       &
\colhead{Atmosphere}                         \\
\colhead{}                                   &  
\colhead{(K)}                                & 
\colhead{}                                   & 
\colhead{}                                   & 
\colhead{$\rm (km/s)$}                       & 
\colhead{$\rm (km/s)$}                       &
\colhead{Model\tablenotemark{d}}             }
\startdata
BD+44 1080 & $ 12879$ & $ 3.42$ & $-0.25$ & $    2$ & $  303$ & ATLAS9/SPECTRUM \\
BD+56 517 & $ 23752$ & $ 3.76$ & $-0.38$ & $    2$ & $  192$ & TLUSTY B02 \\
BD+56 518 & $ 23415$ & $ 3.80$ & $-0.36$ & $    2$ & $  165$ & TLUSTY B02 \\
BD+56 576 & $ 23275$ & $ 3.48$ & $-0.16$ & $    2$ & $  169$ & TLUSTY B02 \\
BD+69 1231 & $ 11292$ & $ 4.23$ & $-0.16$ & $    2$ & $  413$ & ATLAS9/SPECTRUM \\
BD+71 92 & $ 11634$ & $ 4.04$ & $-0.21$ & $    2$ & $  325$ & ATLAS9/SPECTRUM \\
CPD-41 7715 & $ 23617$ & $ 3.98$ & $-0.23$ & $    2$ & $  322$ & TLUSTY B02 \\
CPD-57 3507 & $ 25921$ & $ 3.71$ & $-0.19$ & $    2$ & $  192$ & TLUSTY B02 \\
CPD-57 3523 & $ 25037$ & $ 3.10$ & $-0.34$ & $   10$ & $  255$ & TLUSTY B10 \\
CPD-59 2591 & $ 34169$ & $ 4.18$ & $-0.25$ & $   10$ & $  332$ & TLUSTY O10 \\
CPD-59 2600 & $ 38901$ & $ 3.68$ & $-0.22$ & $   10$ & $  338$ & TLUSTY O10 \\
CPD-59 2625 & $ 26310$ & $ 4.42$ & $-1.03$ & $    2$ & $  137$ & TLUSTY B02 \\
GSC03712-01870 & $ 33233$ & $ 3.98$ & $-0.50$ & $   10$ & $  229$ & TLUSTY O10 \\
HD~13338 & $ 22748$ & $ 3.61$ & $-0.38$ & $    2$ & $  204$ & TLUSTY B02 \\
HD~14250 & $ 24701$ & $ 3.53$ & $-0.16$ & $    2$ & $  350$ & TLUSTY B02 \\
HD~14321 & $ 23294$ & $ 3.57$ & $-0.33$ & $    2$ & $  224$ & TLUSTY B02 \\
HD~17443 & $ 11241$ & $ 4.15$ & $-0.16$ & $    2$ & $  397$ & ATLAS9/SPECTRUM \\
HD~27778 & $ 17362$ & $ 3.91$ & $-0.47$ & $    2$ & $  295$ & TLUSTY B02 \\
HD~28475 & $ 15570$ & $ 4.01$ & $-0.44$ & $    2$ & $  368$ & TLUSTY B02 \\
HD~29647 & $ 12356$ & $ 3.58$ & $-0.25$ & $    2$ & $  286$ & ATLAS9/SPECTRUM \\
HD~30122 & $ 16432$ & $ 3.65$ & $-1.05$ & $    2$ & $  341$ & TLUSTY B02 \\
HD~30675 & $ 16428$ & $ 3.97$ & $-0.37$ & $    2$ & $  318$ & TLUSTY B02 \\
HD~37061 & $ 28366$ & $ 4.12$ & $-0.51$ & $   10$ & $  332$ & TLUSTY O10 \\
HD~38087 & $ 17408$ & $ 4.20$ & $-0.34$ & $    2$ & $  294$ & TLUSTY B02 \\
HD~40893 & $ 31837$ & $ 3.60$ & $-0.19$ & $   10$ & $  335$ & TLUSTY O10 \\
HD~46106 & $ 29265$ & $ 4.10$ & $-0.58$ & $   10$ & $  339$ & TLUSTY O10 \\
HD~46660 & $ 31067$ & $ 3.98$ & $-0.53$ & $   10$ & $  303$ & TLUSTY O10 \\
HD~54439 & $ 25300$ & $ 3.71$ & $-0.27$ & $    2$ & $  315$ & TLUSTY B02 \\
HD~62542 & $ 17083$ & $ 4.07$ & $-0.55$ & $    2$ & $  302$ & TLUSTY B02 \\
HD~68633 & $ 18643$ & $ 3.91$ & $-0.52$ & $    2$ & $  277$ & TLUSTY B02 \\
HD~70614 & $ 18105$ & $ 3.25$ & $-0.39$ & $    2$ & $  333$ & TLUSTY B02 \\
HD~91983 & $ 24490$ & $ 3.36$ & $-0.12$ & $    2$ & $  232$ & TLUSTY B02 \\
HD~92044 & $ 24529$ & $ 3.06$ & $-0.28$ & $   10$ & $  247$ & TLUSTY B10 \\
HD~93028 & $ 33564$ & $ 4.04$ & $-0.21$ & $   10$ & $  252$ & TLUSTY O10 \\
HD~93222 & $ 36349$ & $ 3.66$ & $-0.24$ & $   10$ & $  296$ & TLUSTY O10 \\
HD~104705 & $ 29222$ & $ 3.44$ & $-0.02$ & $   10$ & $  307$ & TLUSTY O10 \\
HD~110336 & $ 11941$ & $ 4.15$ & $-0.17$ & $    2$ & $  300$ & ATLAS9/SPECTRUM \\
HD~110946 & $ 20582$ & $ 3.24$ & $-0.53$ & $   10$ & $  249$ & TLUSTY B10 \\
HD~112607 & $ 12980$ & $ 3.43$ & $-0.22$ & $    2$ & $  347$ & ATLAS9/SPECTRUM \\
HD~142096 & $ 18205$ & $ 4.34$ & $-0.29$ & $    2$ & $  330$ & TLUSTY B02 \\
HD~142165 & $ 14230$ & $ 4.20$ & $-0.31$ & $    2$ & $  398$ & ATLAS9/SPECTRUM \\
HD~146285 & $ 12328$ & $ 4.36$ & $-0.18$ & $    2$ & $  353$ & ATLAS9/SPECTRUM \\
HD~147196 & $ 12225$ & $ 3.94$ & $-0.45$ & $    2$ & $  544$ & ATLAS9/SPECTRUM \\
HD~147889 & $ 22061$ & $ 4.13$ & $-0.54$ & $    2$ & $  229$ & TLUSTY B02 \\
HD~149452 & $ 31816$ & $ 3.55$ & $-0.00$ & $   10$ & $  405$ & TLUSTY O10 \\
HD~164073 & $ 17573$ & $ 3.99$ & $-0.43$ & $    2$ & $  333$ & TLUSTY B02 \\
HD~172140 & $ 27347$ & $ 3.77$ & $ 0.06$ & $   10$ & $  232$ & TLUSTY O10 \\
HD~193322 & $ 33358$ & $ 3.90$ & $-0.19$ & $   10$ & $  387$ & TLUSTY O10 \\
HD~197512 & $ 23773$ & $ 3.83$ & $-0.41$ & $    2$ & $  165$ & TLUSTY B02 \\
HD~197702 & $ 23854$ & $ 2.99$ & $-0.38$ & $   10$ & $  430$ & TLUSTY B10 \\
HD~198781 & $ 25529$ & $ 3.13$ & $-0.49$ & $   10$ & $  284$ & TLUSTY B10 \\
HD~199216 & $ 24140$ & $ 3.04$ & $-0.34$ & $   10$ & $  189$ & TLUSTY B10 \\
HD~204827 & $ 32037$ & $ 3.96$ & $-0.28$ & $   10$ & $  345$ & TLUSTY O10 \\
HD~210072 & $ 19186$ & $ 3.08$ & $-0.55$ & $   10$ & $  312$ & TLUSTY B10 \\
HD~210121 & $ 16498$ & $ 4.06$ & $-0.29$ & $    2$ & $  290$ & TLUSTY B02 \\
HD~217086 & $ 35722$ & $ 3.65$ & $-0.34$ & $   10$ & $  443$ & TLUSTY O10 \\
HD~220057 & $ 19736$ & $ 4.21$ & $-0.38$ & $    2$ & $  396$ & TLUSTY B02 \\
HD~228969 & $ 28818$ & $ 3.84$ & $-0.28$ & $   10$ & $  378$ & TLUSTY O10 \\
HD~236960 & $ 26572$ & $ 3.47$ & $ 0.11$ & $    2$ & $  229$ & TLUSTY B02 \\
HD~239693 & $ 18885$ & $ 4.11$ & $-0.33$ & $    2$ & $  357$ & TLUSTY B02 \\
HD~239722 & $ 20146$ & $ 3.79$ & $-0.62$ & $    2$ & $  210$ & TLUSTY B02 \\
HD~239745 & $ 27059$ & $ 4.12$ & $-0.37$ & $    2$ & $  215$ & TLUSTY B02 \\
HD~282485 & $ 12927$ & $ 3.88$ & $-0.26$ & $    2$ & $  309$ & ATLAS9/SPECTRUM \\
HD~292167 & $ 28817$ & $ 3.01$ & $-0.21$ & $   10$ & $  225$ & TLUSTY O10 \\
HD~294264 & $ 19095$ & $ 4.23$ & $-0.37$ & $    2$ & $  300$ & TLUSTY B02 \\
HD~303068 & $ 24810$ & $ 3.80$ & $-0.29$ & $    2$ & $  203$ & TLUSTY B02 \\
NGC 2244 11 & $ 27166$ & $ 4.13$ & $-0.37$ & $    2$ & $  296$ & TLUSTY B02 \\
NGC 2244 23 & $ 19091$ & $ 4.06$ & $-0.29$ & $    2$ & $  440$ & TLUSTY B02 \\
Trumpler 14 6 & $ 28047$ & $ 4.09$ & $-0.22$ & $    2$ & $  243$ & TLUSTY B02 \\
Trumpler 14 27 & $ 25998$ & $ 4.17$ & $-0.45$ & $    2$ & $  178$ & TLUSTY B02 \\
VSS VIII-10 & $ 11525$ & $ 3.76$ & $-0.30$ & $    2$ & $  393$ & ATLAS9/SPECTRUM \\
\enddata
\tablenotetext{a}{Values of \abund\/ are measured with respect to the solar abundance scale of \citet{Grevesse98}.}
\tablenotetext{b}{Values of the microturbulence velocity \vturb, either 2~\kms\/ or 
10~\kms, were determined by the grid of model atmospheres used for modeling each star.}
\tablenotetext{c}{As discussed in \S \ref{secINTRINSIC}, $v_{broad}$ incorporates the rotational 
broadening of the star and any additional broadening of the spectra above the nominal PSF FHM of 
5.5~\AA, due to ``breathing'' of the instrument.}
\tablenotetext{d}{This column indicates the model atmosphere grid used to compute the intrinsic 
SED for each star. ``TLUSTY O10'' refers to the NLTE OSTAR2002 grid of \citet{Lanz03}, 
computed with a value of \vturb\/ = 10~\kms. ``TLUSTY B2'' and ``TLUSTY B10'' refer to the 
BSTAR2006 grid from \citet{Lanz07}, computed with \vturb = 2~\kms\/ and 10~\kms, 
respectively. In the \teff\/ and \logg\/ region where these models overlap, the model which 
provided the best fit to the data was adopted.  ``ATLAS/SPECTRUM'' refers to atmosphere 
structures taken from Kurucz's 1991 ATLAS9 models (\vturb\/ = 2~\kms) and SEDs computed 
using the spectral synthesis program SPECTRUM from \citet{Gray94}.  These models 
were used for the coolest stars in the sample, with \teff\/ $\le$ 15000~K.}
\label{tabPARS}
\end{deluxetable*}

\begin{figure*}
\figurenum{3a}
\epsscale{0.9}
\plotone{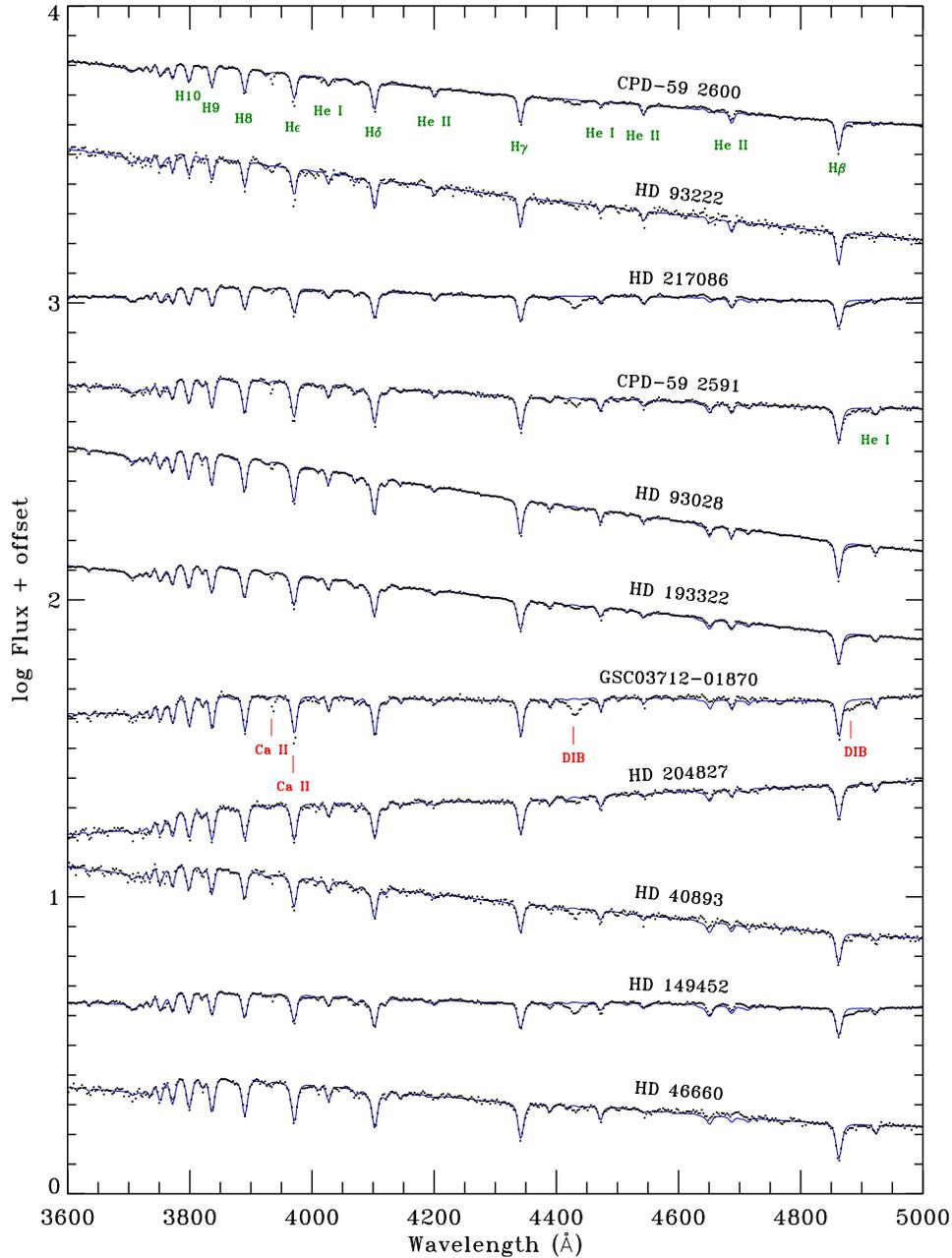}
\caption{Model atmosphere fits to the G430L spectra of the program stars.  Small black circles show the 
observed G430L spectrum over the wavelength range 3600--5000~\AA.  Solid blue curves show the best-fitting 
stellar atmosphere models.  The fitting procedure is described in \S \ref{secINTRINSIC} and included the full 
range of the G430L spectra (2900--5700~\AA).  Most of the spectral features are contained in the narrower 
range shown in the figure.  Prominent stellar features are identified in green and interstellar features, including 
gas-phase absorption lines and diffuse interstellar bands (``DIB''), are 
identified in red.  The stars are ordered in decreasing \teff, throughout the figures.  Stellar properties 
derived from the fits are listed in Table \ref{tabPARS}.
\label{figFITS}}
\end{figure*}

\begin{figure*}
\figurenum{3b}
\plotone{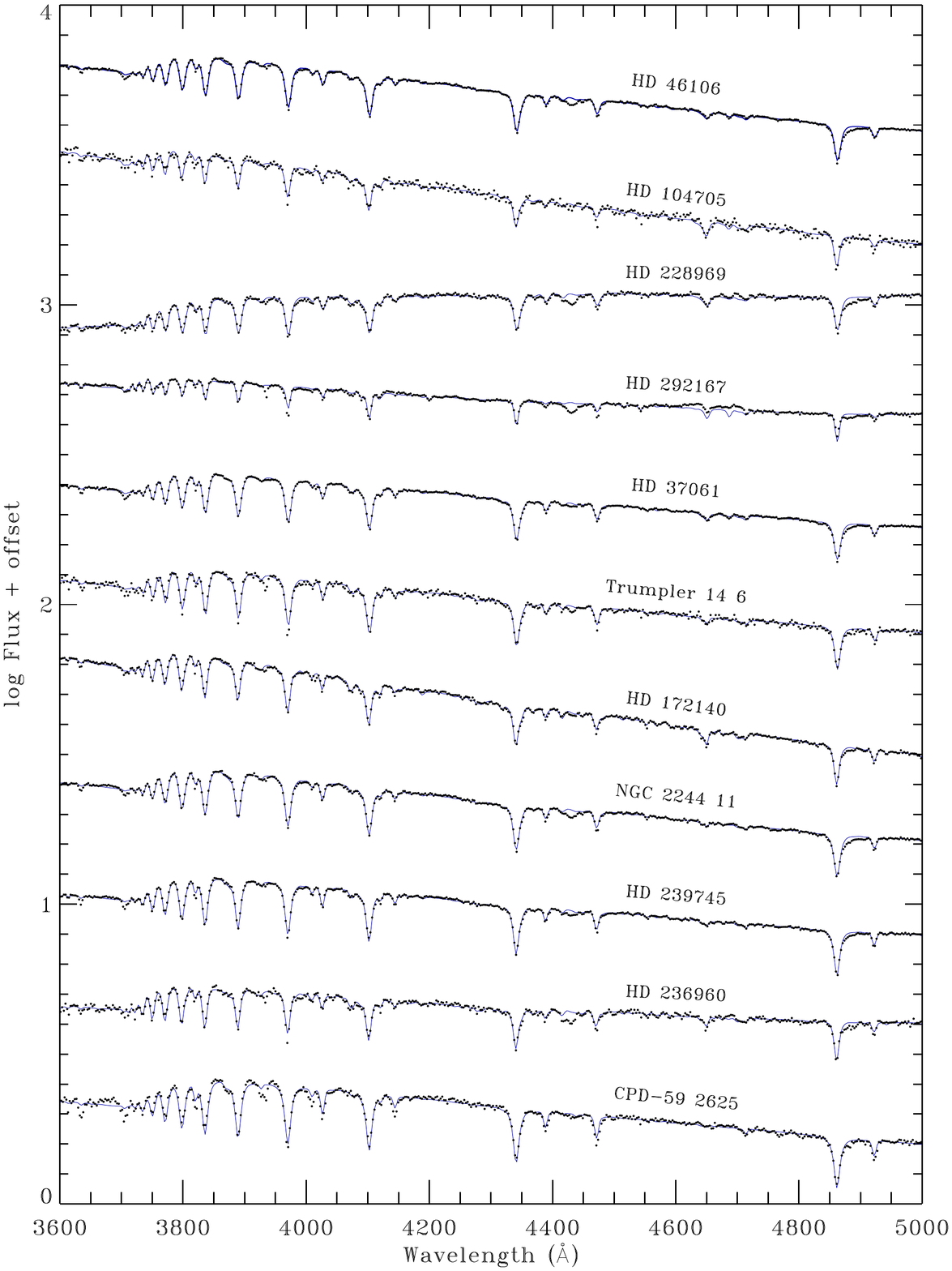}
\caption{Same as Figure 3a.}
\end{figure*}

\begin{figure*}
\figurenum{3c}
\plotone{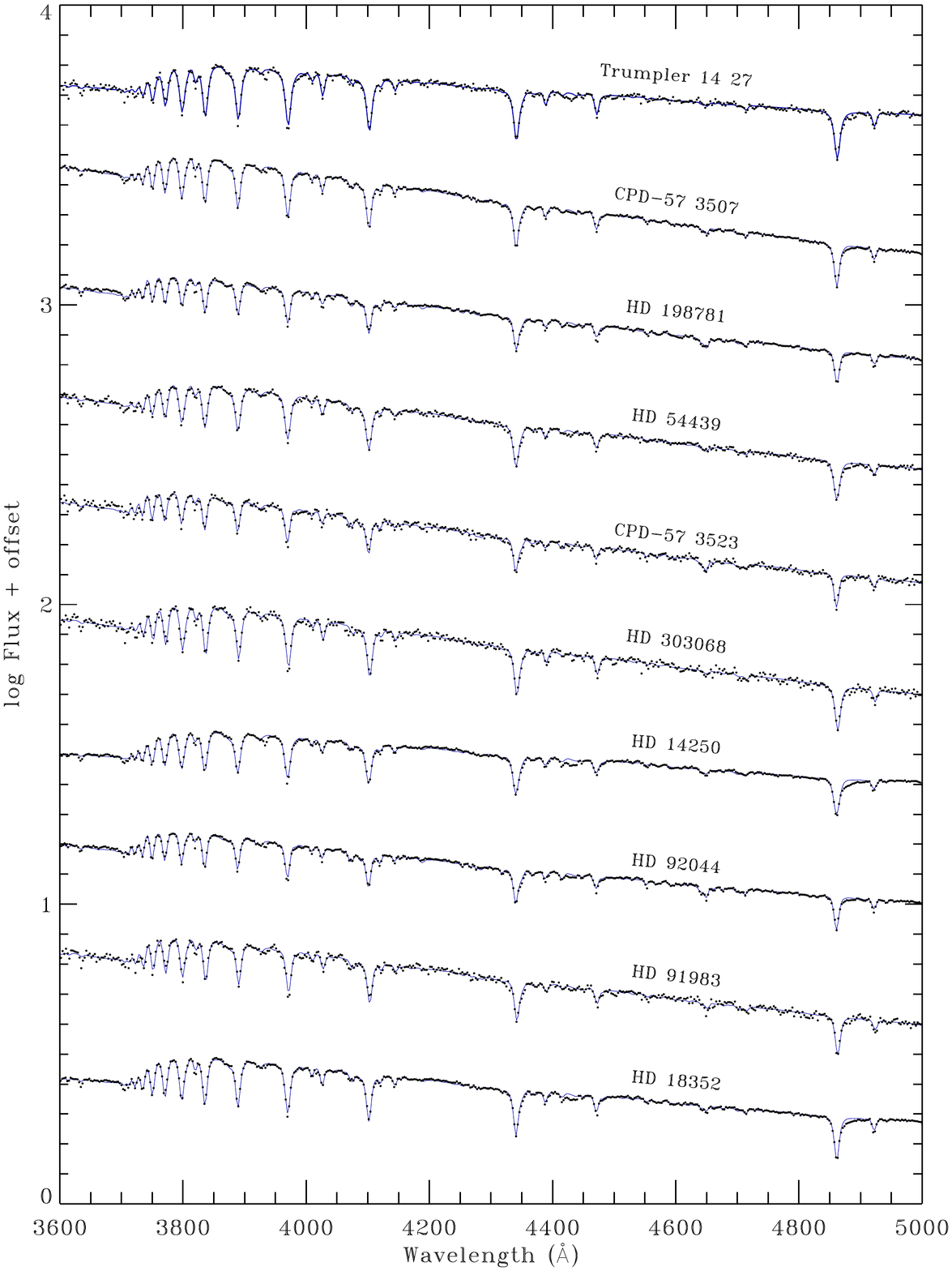}
\caption{Same as Figure 3a.}
\end{figure*}

\begin{figure*}
\figurenum{3d}
\plotone{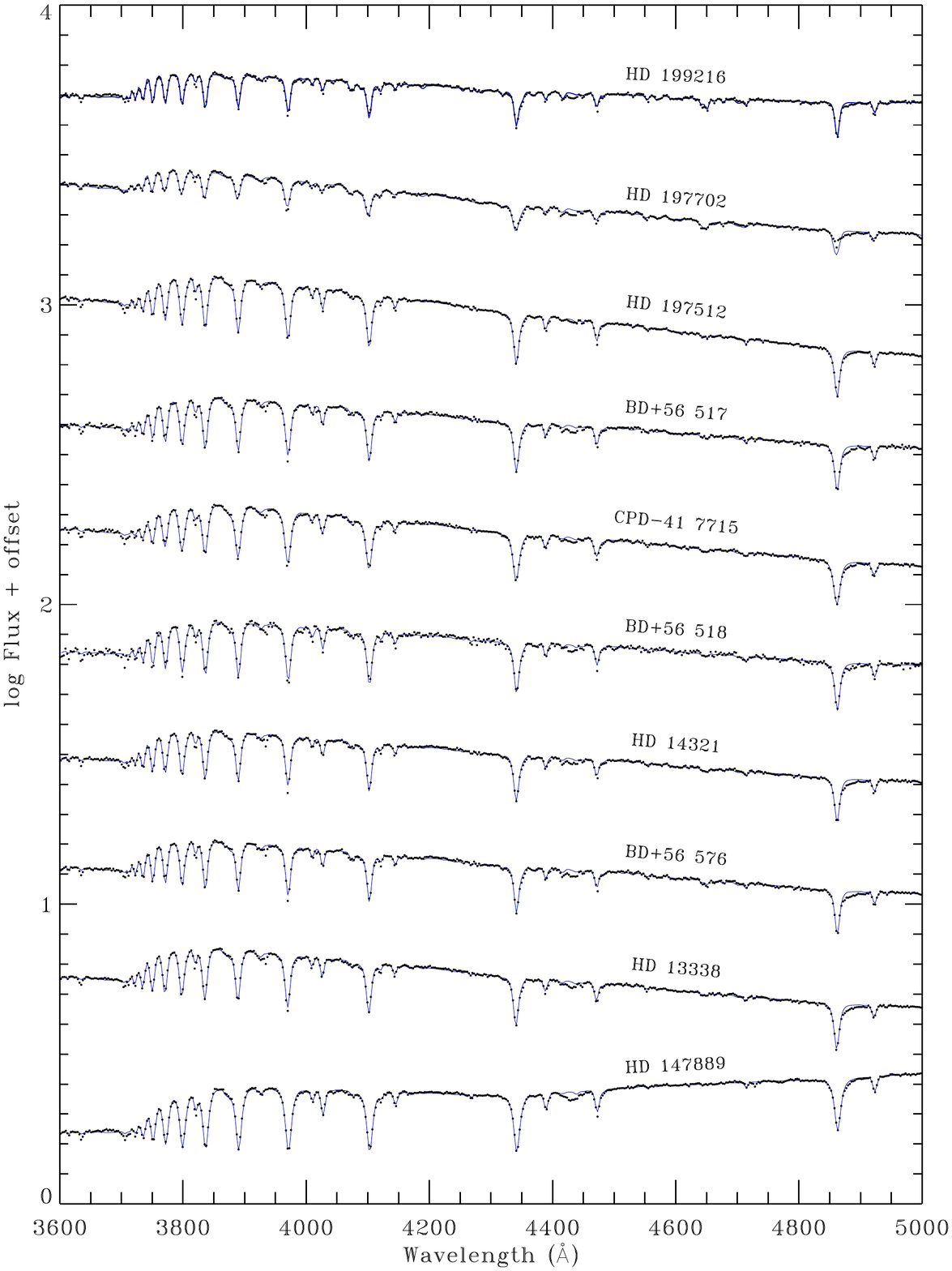}
\caption{Same as Figure 3a.}
\end{figure*}

\begin{figure*}
\figurenum{3e}
\plotone{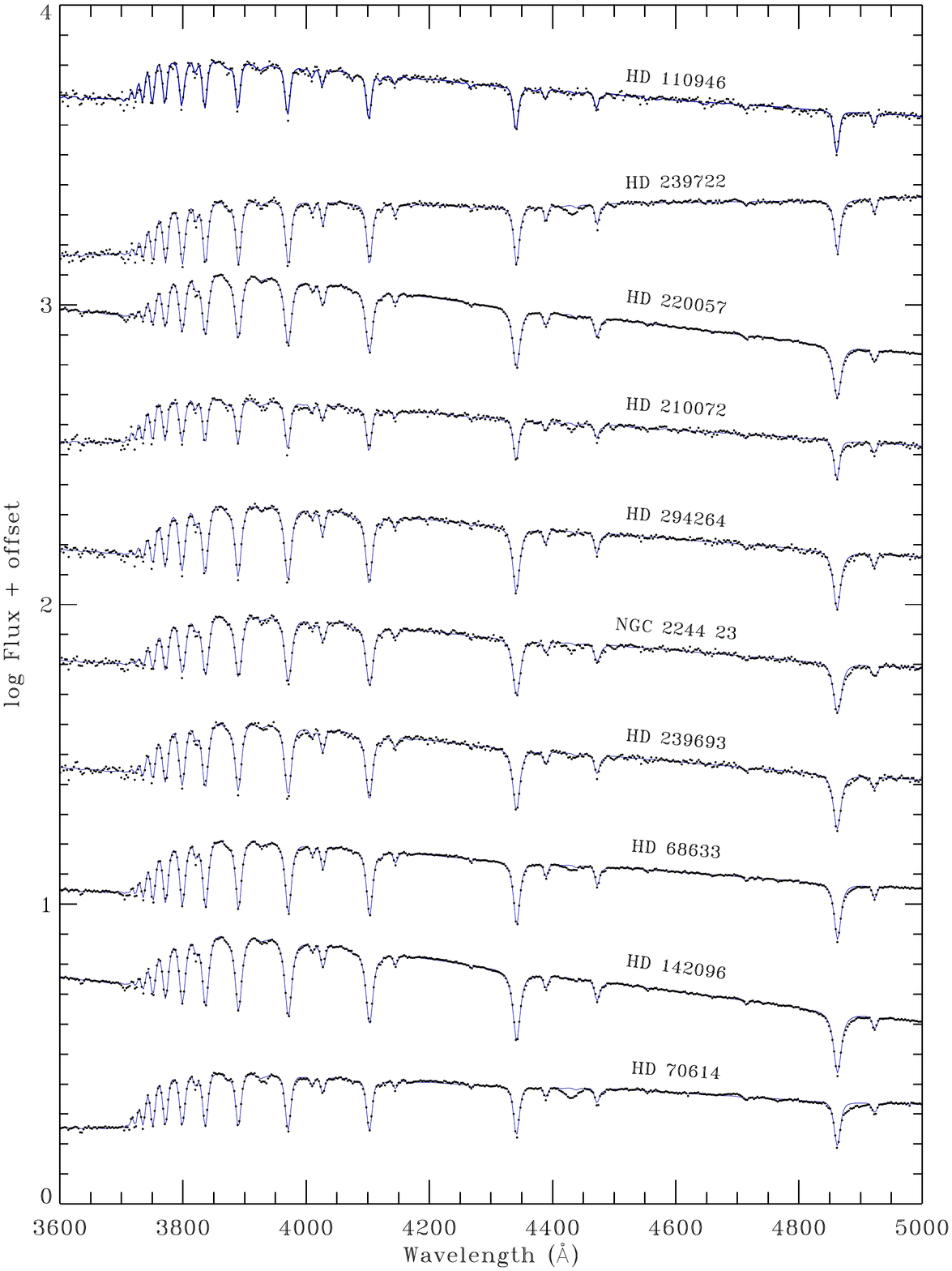}
\caption{Same as Figure 3a.}
\end{figure*}

\begin{figure*}
\figurenum{3f}
\plotone{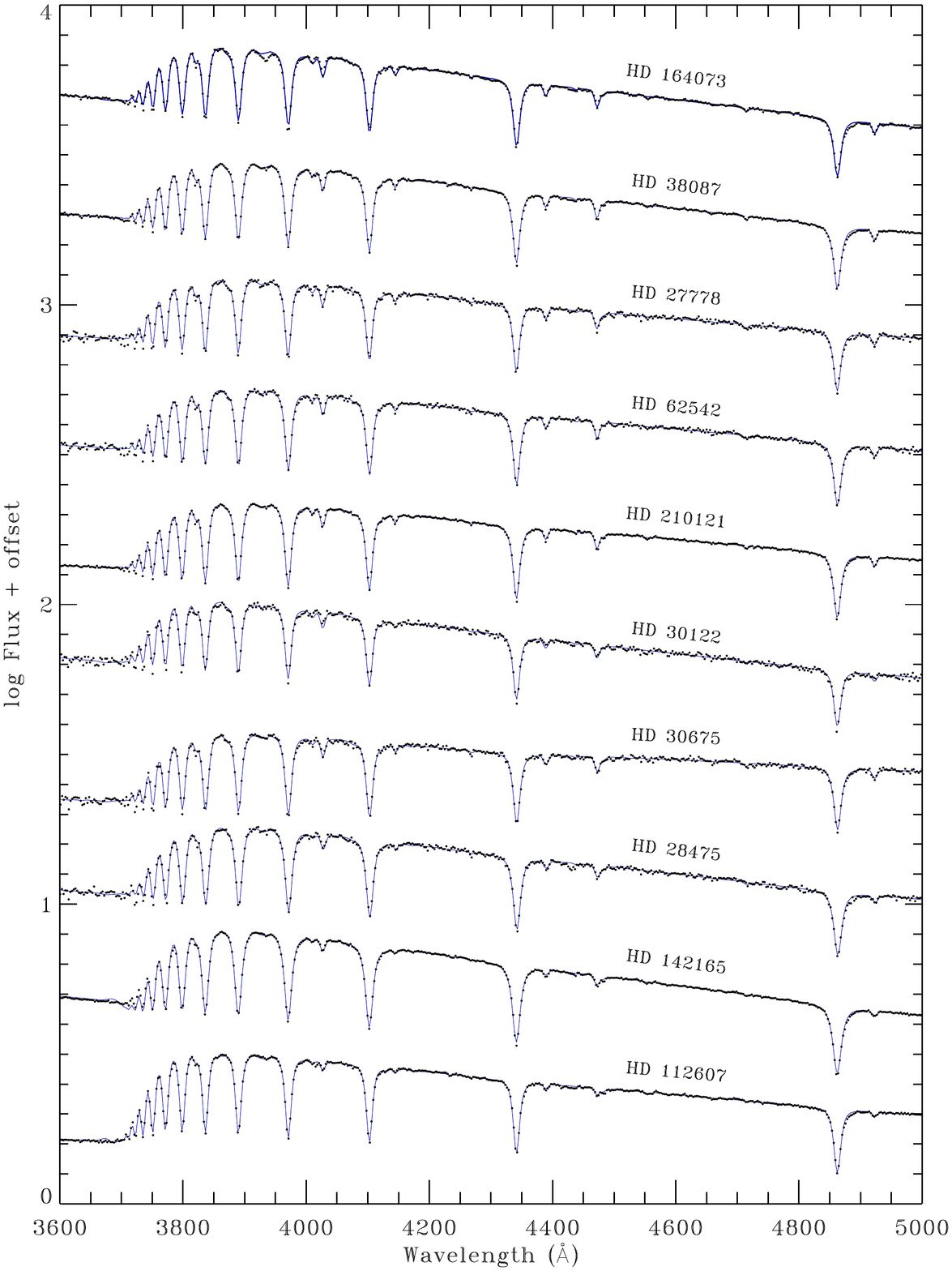}
\caption{Same as Figure 3a.}
\end{figure*}

\begin{figure*}
\figurenum{3g}
\plotone{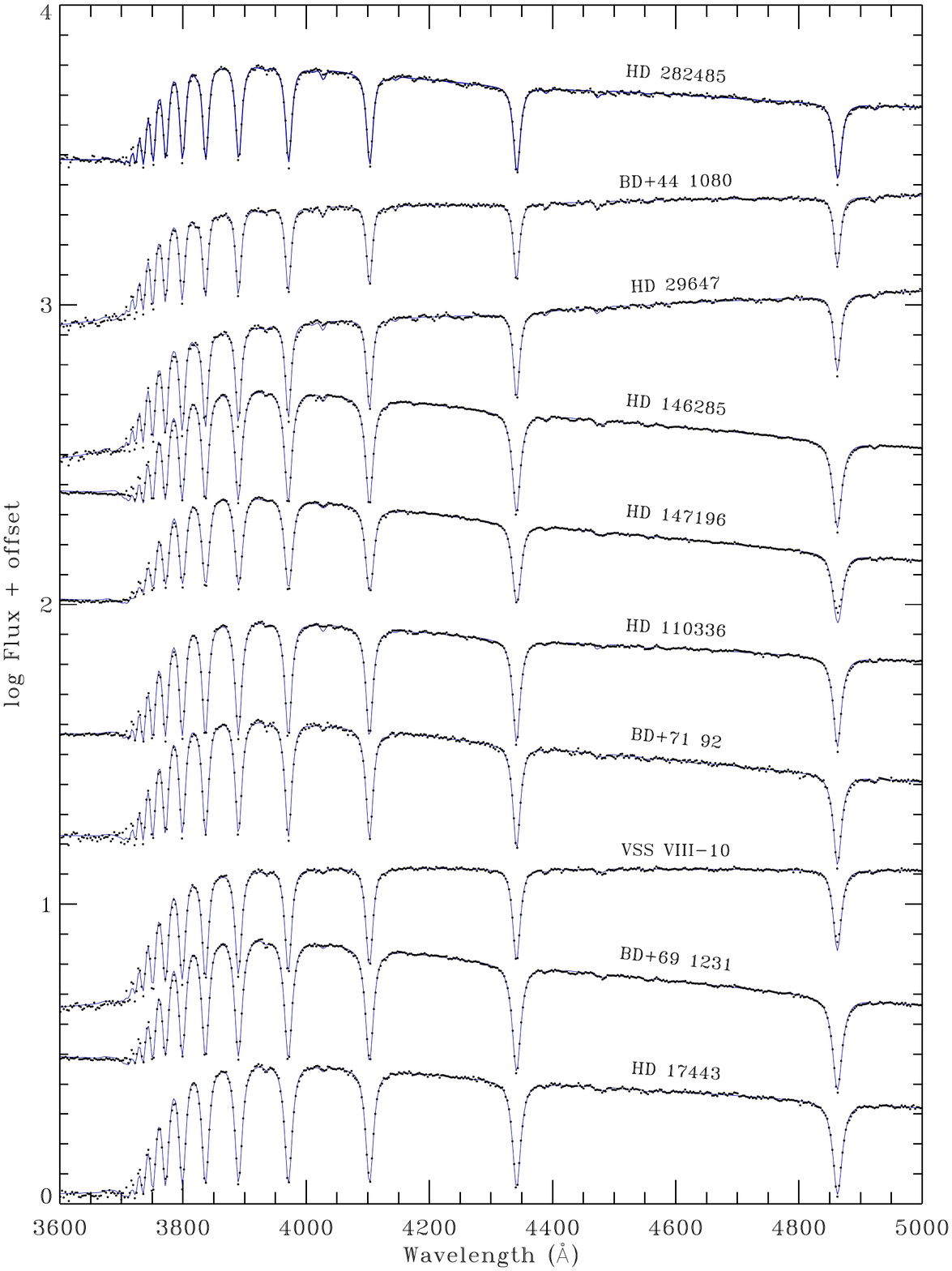}
\caption{Same as Figure 3a.}
\end{figure*}

The ability of the stellar models to reproduce the G430L spectra is demonstrated in Figure~3. 
The various panels show the observed G430L data (small black filled circles) overlaid with the best fit 
models (solid blue lines). For clarity, the stars are ordered by decreasing \teff, from the top of Figure 
3a (hottest star) to the bottom of Figure 3g (coolest star).  Prominent stellar features are identified 
and labeled in green, while interstellar features are labeled in red.  As seen in the figures, the models reproduce the 
observed stellar features very well.  The only significant and systematic discrepancies noticeable 
in Figure 3 are due to interstellar features, notably the broad diffuse interstellar bands (``DIBs'') at 
4430~\AA\/ and 4870~\AA\/ and the Ca II H and K absorption lines at 3968~\AA\/ and 3933~\AA, 
respectively, which are identified in the first panel of Figure 3.  Four stars in the sample (HD~147196, HD~197702, HD~292167, 
and VSS~VII-10) show evidence for mild H$\alpha$ emission in
their G750L spectra, which is not reproduced by the atmosphere models.  Two of these stars (HD~197702 and HD~292167) 
are early-type giants at the low gravity edge of our adopted model grids (\logg\/ $\sim$3.0). The other two are late-B main 
sequence stars and may be rapid rotators.  Indeed, HD~147196 has the largest value of $v_{broad}$ among our sample.  In all 
four cases, however, the blue region of the spectra appear normal (as seen in Figure 3), as does the UV region, and we 
believe that the stellar parameters and SEDs are well 
reproduced by the models used here.  (Note that the rapidly rotating HD~147196 will ultimately be excluded from the final extinction
sample analyzed below due to its low value of \ebv.) 

The final set of parameters describing 
the best-fit models are listed in Table \ref{tabPARS}.  In addition to the values of \teff, \logg\, 
\abund, \vturb, and $v_{broad}$, we also indicate -- in the last column of the table -- the particular stellar 
atmosphere grid used for each star.  In cases where the stellar properties fell into regions in which the 
different TLUSTY grids overlapped, we choose the model which yielded the best fit to the data. 

The typical statistical uncertainties in the stellar parameters derived from the fitting procedure are $\pm$200~K in \teff, $\pm$0.02 in 
\logg, $\pm$0.1 in \abund, and $\pm$10~\kms\/ in $v_{broad}$.  These, however do not take into account 
systematic effects that could arise from the fundamental assumptions in the modeling process.  For example, the 
solar abundances of \citet{Grevesse98} have been superceded since the TLUSTY grids were computed \citep{Grevesse13}.  
In addition,
the overlap region between the cool TLUSTY BSTAR2006 models and the 
ATLAS9/SPECTRUM models is illustrative.  For any star for which the BSTAR2006 models indicated 
\teff\/ $<$ 15,000~K, we adopted the ATLAS9/SPECTRUM combination described above.  However, when 
experimenting with the ATLAS9/SPECTRUM combination for stars warmer the 15,000~K, we found that they 
yielded \teff\/ values several hundred K hotter than the best-fit TLUSTY models.  This is undoubtedly 
due to the fundamental difference in the LTE vs. NLTE approaches and, possibly, to differences in the 
base opacities and atomic parameters assumed in the models.  It would be useful to repeat this comparison with
other LTE model grids, e.g., the BOSZ grid of \citet{Bohlin17}. However, while this is an interesting stellar atmospheres 
issue, fortunately for us it has little impact on the derived extinction curves.  As was shown by 
\citet{Massa83}, the effects of temperature mismatch in extinction curves formed from early-type 
stars tend to be self-canceling due to the optical normalization (which, with some modifications, we 
will adopt here).  Moreover, the wide range in intrinsic extinction curve properties also serves to 
lessen the impact of the relatively small uncertainties in the intrinsic stellar properties.

\subsection{Step 2: Constructing the Extinction Curves}
\label{secCURVES}

Once the properties of the reddened stars were determined, the production of the normalized 
extinction curves was straightforward.  $A(\lambda)$, the total extinction at a wavelength 
$\lambda$, is given by: 
\begin{equation}
A(\lambda) = -2.5\times\log\frac{f_\lambda}{\rm F_\lambda} + 5\times\log\frac{R_*}{d}
\label{eq:Alam}
\end{equation}
where $f_\lambda$ and $\rm F_\lambda$ are the observed and intrinsic SEDs, respectively, and 
the quantity $R_*/d$ is the angular radius of the star, i.e., its physical radius $R_*$ divided 
by its distance $d$.  Since these latter two quantities are rarely known accurately, 
extinction is usually presented in a normalized form, in which the angular radius cancels 
out. In this paper, we adopt the normalization:
\begin{equation}
k(\lambda-55) \equiv \frac{E(\lambda-55)}{E(44-55)} = \frac{A(\lambda) - A(55)}
{A(44) - A(55)}
\label{eq:klam}
\end{equation}
and:
\begin{equation}
R(55) \equiv \frac{A(55)}{E(44-55)} 
\label{eq:r55}
\end{equation}
These are analogous to the most commonly used normalizations of extinction, i.e., $k(\lambda-V) 
\equiv E(\lambda-V)/E(B-V)$ and $R(V)$, but with monochromatic measures of the extinction at 4400~\AA\/ 
and 5500~\AA\/ substituting for measurements with the Johnson $B$ and $V$ filters, 
respectively.  The monochromatic values were determined by interpolation of a quadratic fit to the extinction curves 
over a range $\pm$0.1\invmic\/ from the desired wavelength. The benefit of the 
monochromatic normalization is that it eliminates bandpass effects in the 
extinction  measurements.  Such effects will be illustrated further in \S \ref{secPHOTOM}. Because 
the effective wavelengths of the broad Johnson filters depend on the 
shape of a stellar SED (and thus on \ebv\/ and \teff\/), there is no unique transformation 
between \klam\/ and \kfive. For our whole sample, which consists -- on average -- of middle 
B stars with \ebv\/ $\simeq 0.5$, the mean linear transformation is:
\begin{equation}
k(\lambda-55) = \alpha k(\lambda-V) + \beta   
\label{eq:klamtransform}
\end{equation}
where $\alpha = 0.990$ and $\beta = 0.049$. 
This shows that the two normalizations are actually quite similar.  This relationship implies that:
\begin{equation}
R(55) = \alpha R(V) - \beta 
\label{eq:rvtransform}
\end{equation}

To determine \kfive\/ for a given reddened star, we first produced a fully resolved 
intrinsic SED $F_\lambda$ using the models and stellar properties described above in 
\S\ref{secINTRINSIC} and covering the complete range of the available data (i.e., 
$\simeq$1150~\AA\/ to 2.5~\mic).  We then broadened it and resampled the $F_\lambda$ to 
match the various datasets shown in Figure \ref{figSED}.  
We adopted a 9.8~\AA\/ 
Gaussian (FWHM) for the \hst\/ G750L data \citep{Riley18} and a 5.5~\AA\/ Gaussian 
(FWHM) for the \iue\/ data (both SW and LW; as based on our own experience with \iue).  
In the NIR region, synthetic photometry was performed on the intrinsic SED, as 
described in \S \ref{sec2MASS}, to arrive at intrinsic values for the \tmass\/ $J$, $H$, 
and $K$ magnitudes.  The normalized extinction curve was then computed as in Equations (\ref{eq:Alam}) 
and (\ref{eq:klam}).


\begin{figure*}
\figurenum{4a}
\plotone{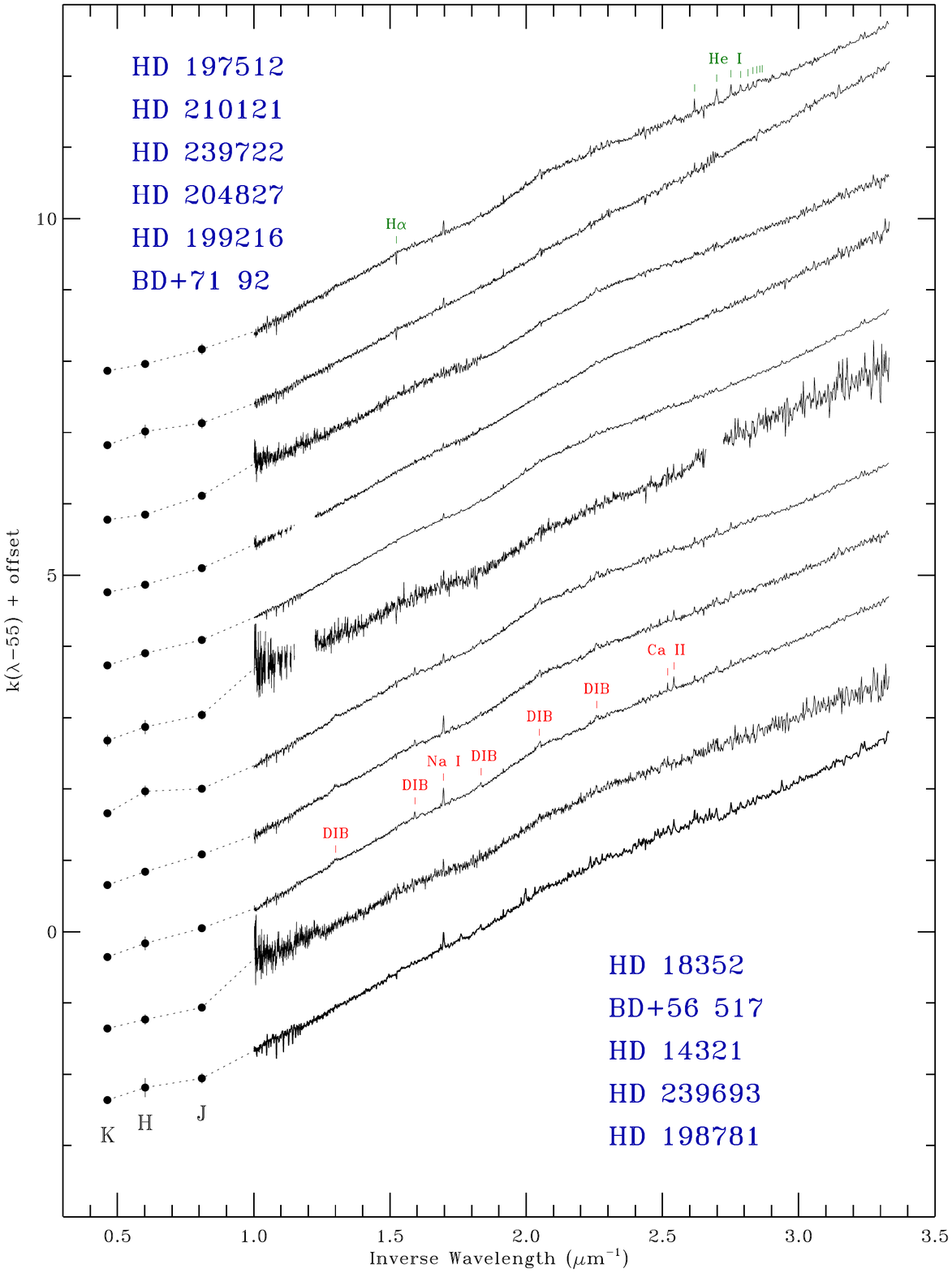}
\caption{Normalized NIR-through-optical extinction curves for the program stars, in the form 
$\kfive\/\equiv\elamfive/\efourfive$. The large filled circles indicate the extinction in the 
\tmass\/ \jhk\/ bands.  The solid curves show the spectrophotometric extinction curves derived from 
the \hst\/ G430L and G750L data, which are joined at 5450~\AA.  Narrow features in the curves 
resulting from stellar spectral mismatch are labeled in green.  Interstellar features, including 
gas-phase absorption lines and diffuse interstellar bands (``DIB''), are labeled in red.  Gaps 
appear in some of the curves near the head of the Paschen series ($\invlam \simeq 1.2\,\invmic$) 
and the head of the Balmer series ($\invlam \simeq\/2.7\,\invmic$) where stellar mismatch features 
were removed. See the discussion in \S \ref{secCURVES}.  The curves are organized in order of 
increasing $|k(K\!-\!55)|$, throughout the figures.  Arbitrary offsets are added for clarity.
\label{figOPIR}}
\end{figure*}

\begin{figure*}
\figurenum{4b}
\plotone{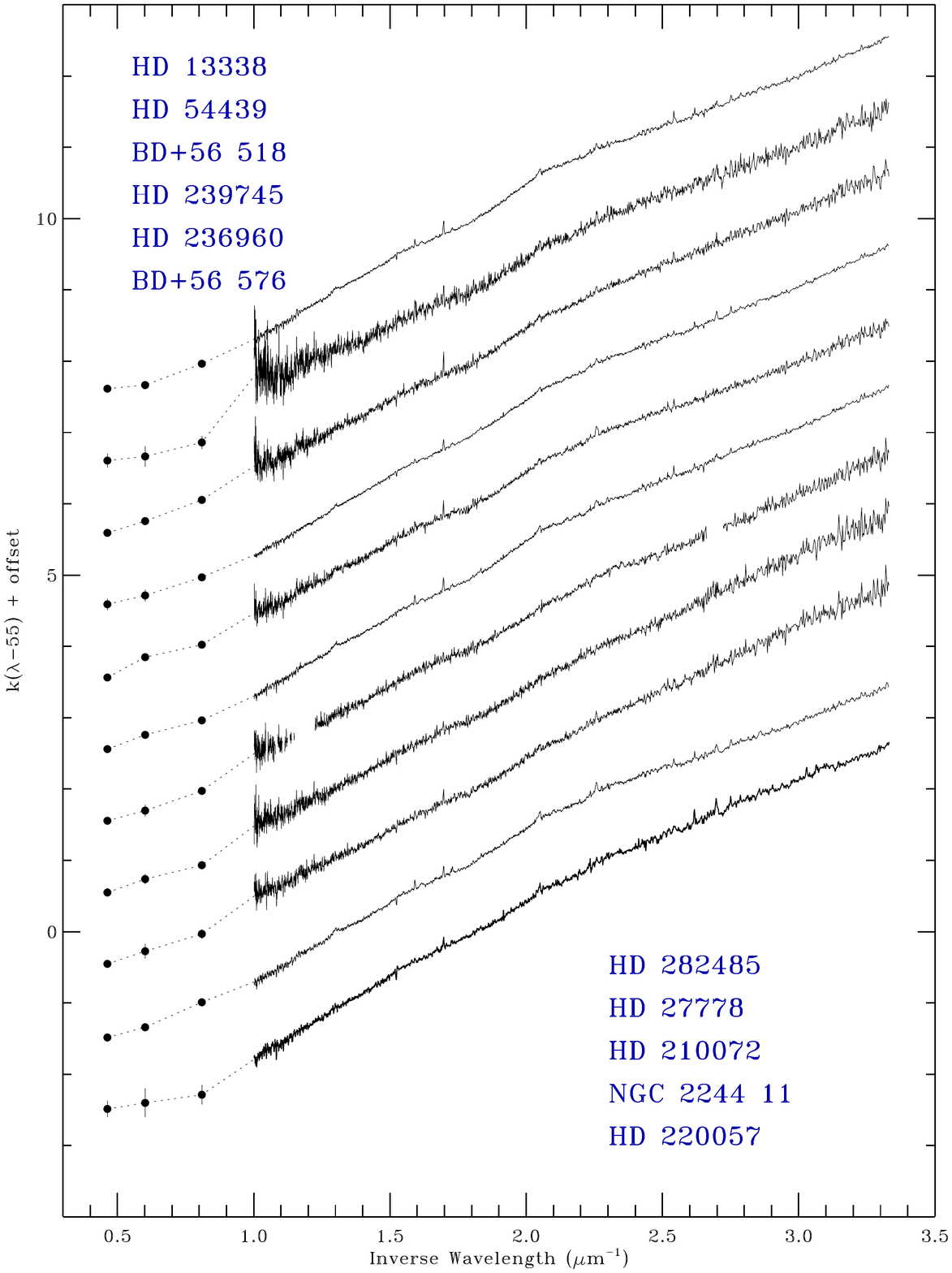}
\caption{Same as Figure 4a.}
\end{figure*}

\begin{figure*}
\figurenum{4c}
\plotone{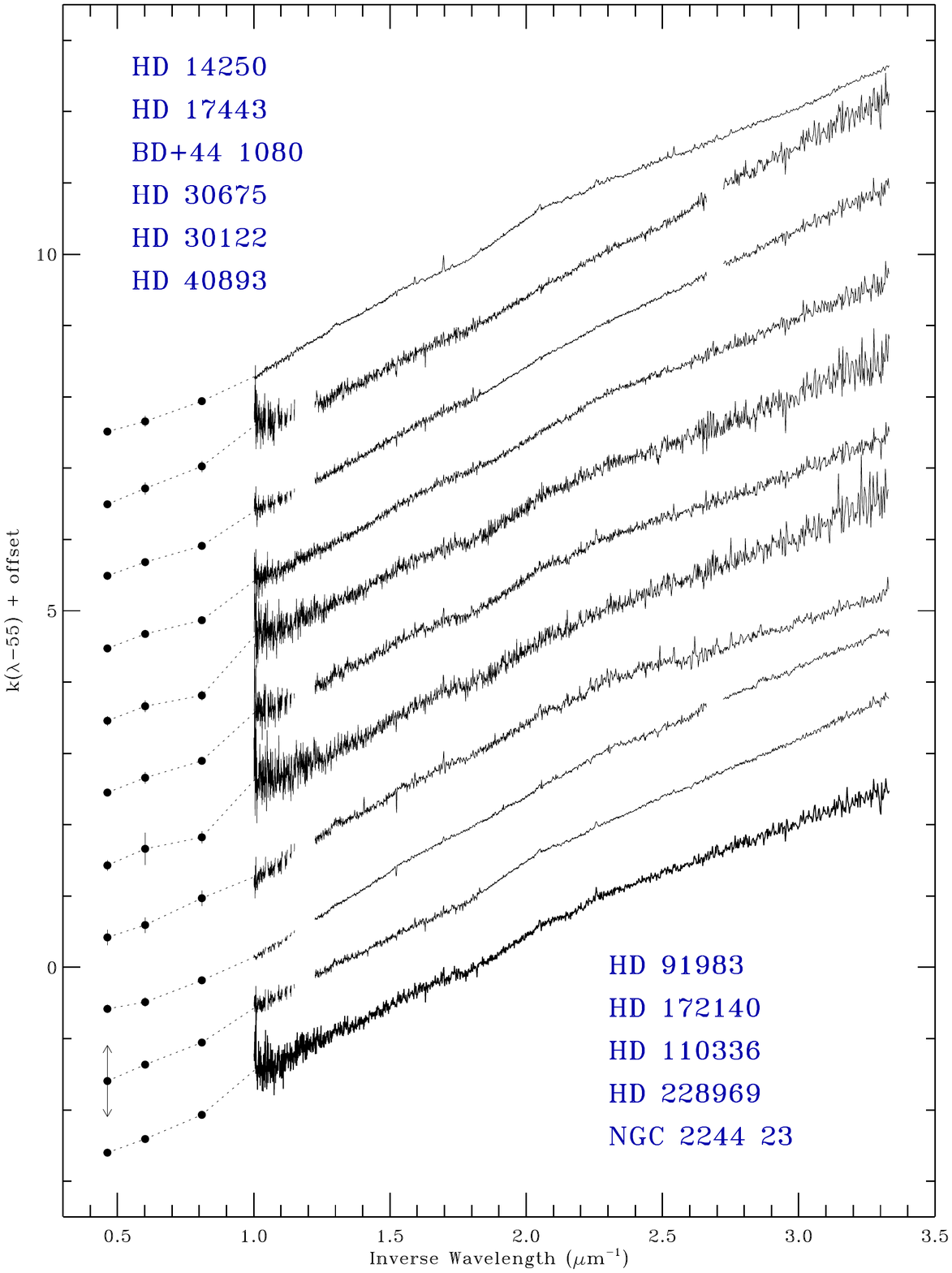}
\caption{Same as Figure 4a.}
\end{figure*}

\begin{figure*}
\figurenum{4d}
\plotone{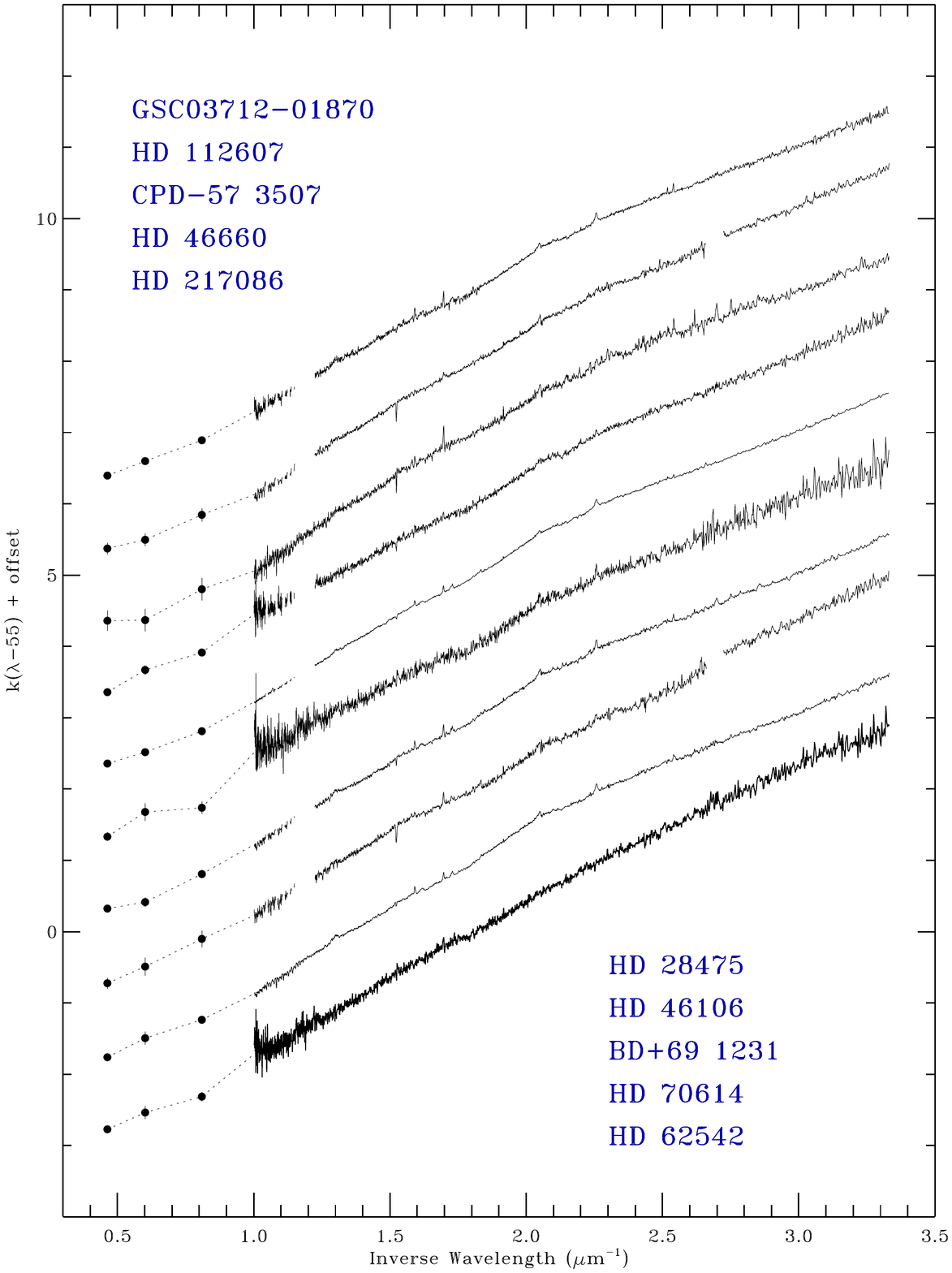}
\caption{Same as Figure 4a.}
\end{figure*}

\begin{figure*}
\figurenum{4e}
\plotone{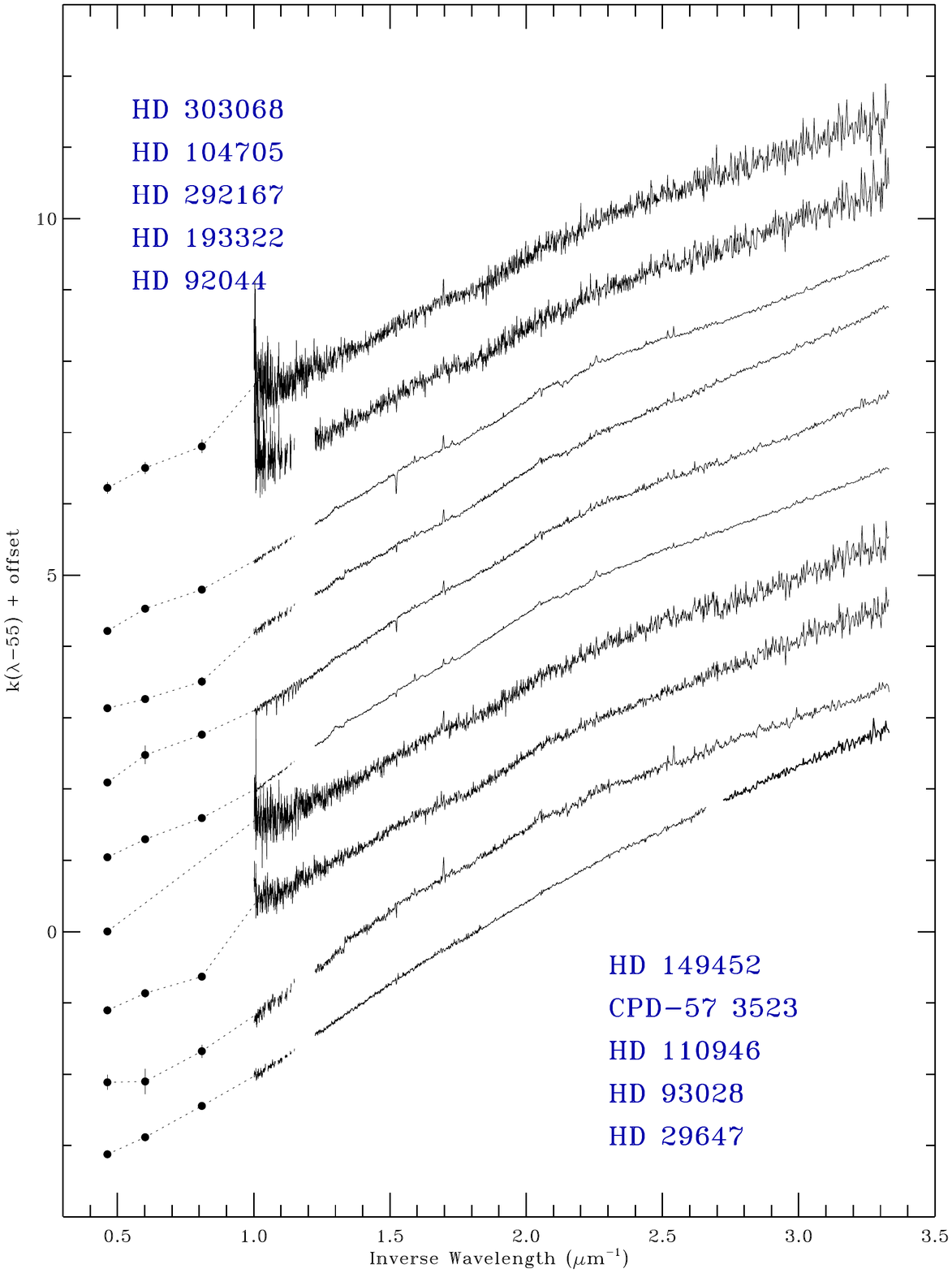}
\caption{Same as Figure 4a.}
\end{figure*}

\begin{figure*}
\figurenum{4f}
\plotone{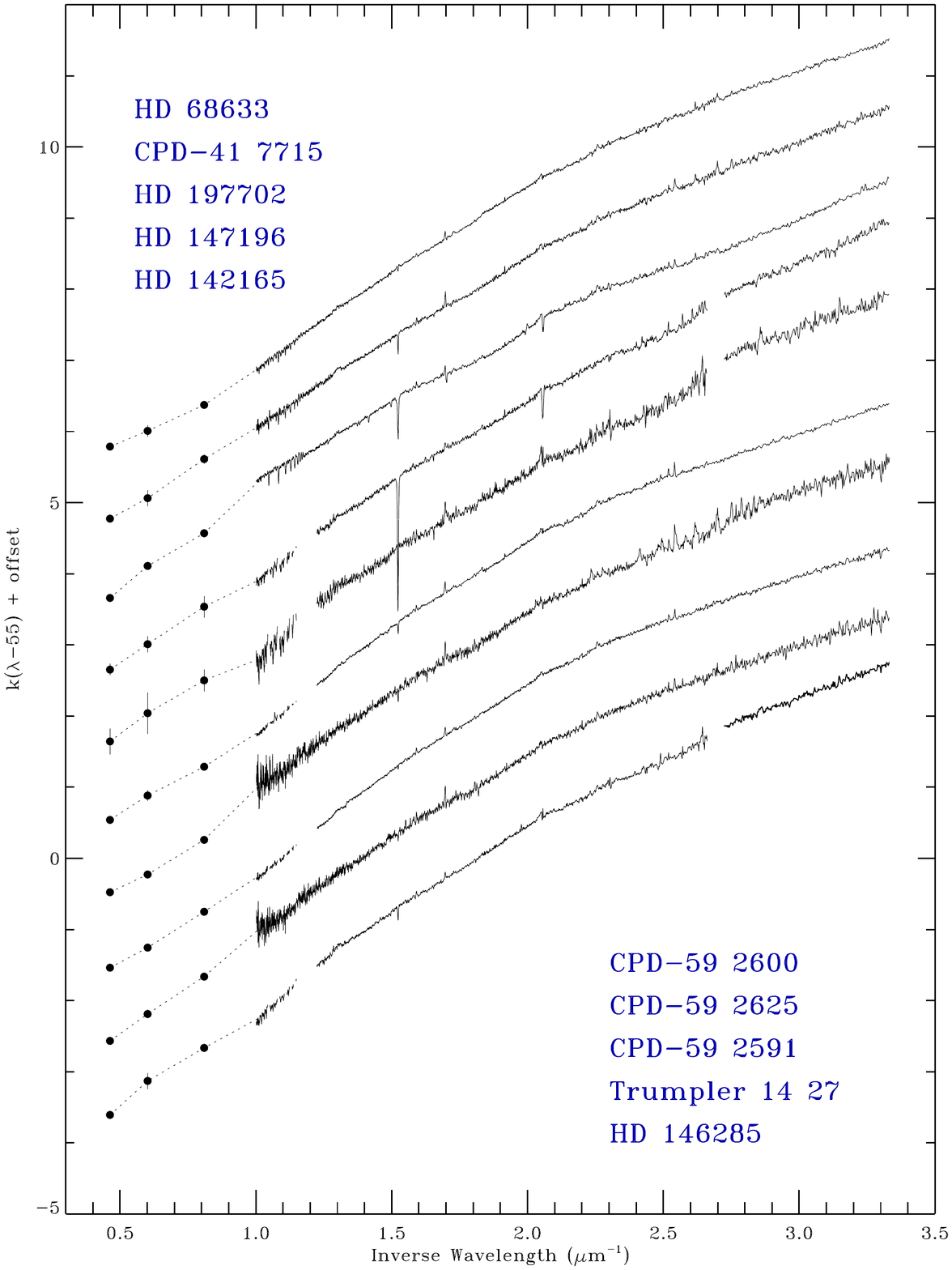}
\caption{Same as Figure 4a.}
\end{figure*}

\begin{figure*}
\figurenum{4g}
\plotone{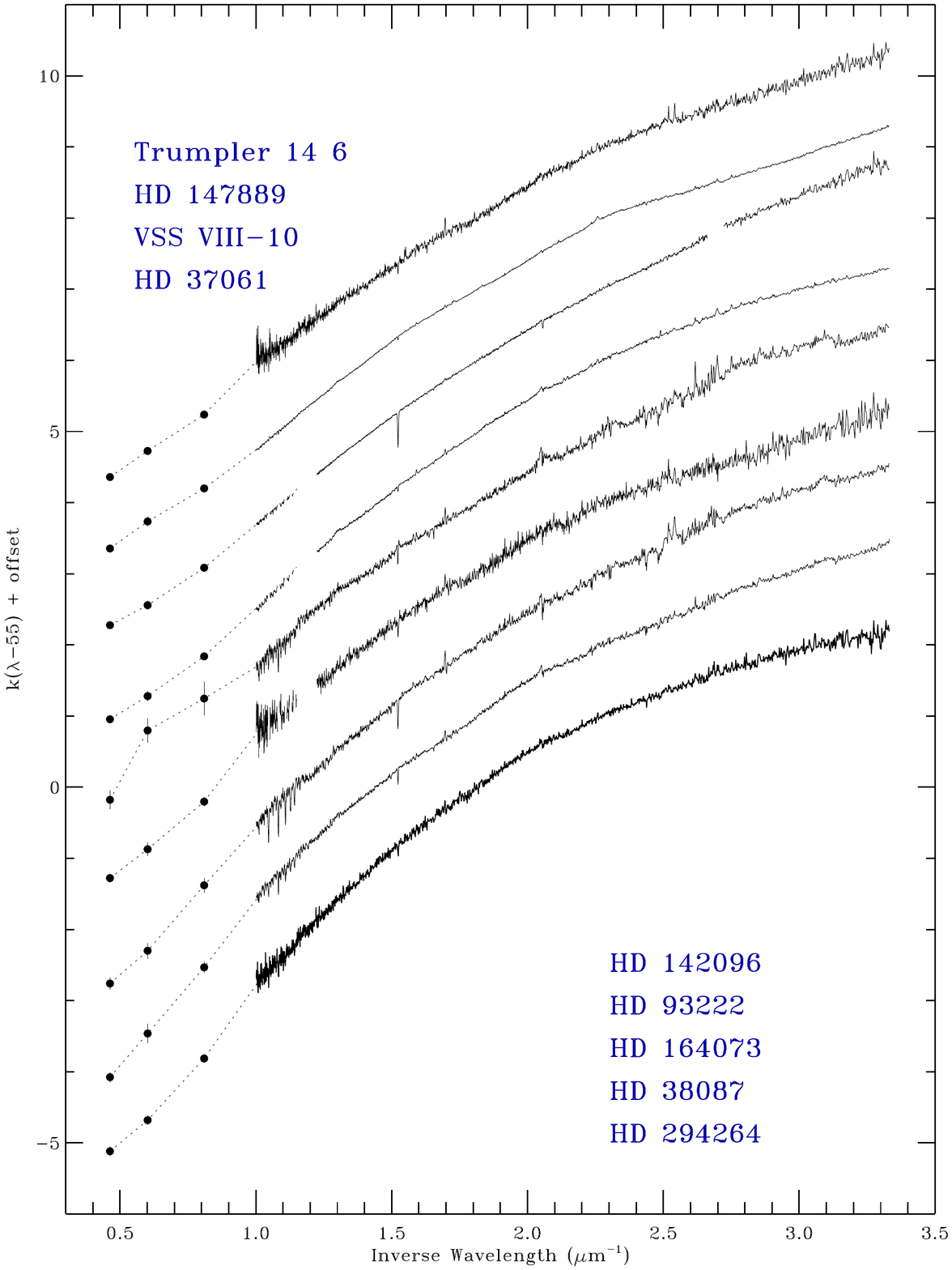}
\caption{Same as Figure 4a.}
\end{figure*}

The normalized optical and NIR extinction curves for all 72 of our program stars are shown in 
Figure~4.  The portions of the curves derived from \tmass\/ \jhk\/ data are shown as discrete 
points at 0.46, 0.62, and 0.81~\invmic, while the portions derived from the combined \hst\/ G750L and G430L 
data are shown from 1.0~\invmic\/ to 3.33~\invmic.  For clarity, the stars are sorted by their $K$-band 
extinction, with the smallest values of $|k(K-55)|$ at the top of panel 4a and the largest at the bottom of 
panel 4g.  The UV portion of the curves for 71 of the 72 program stars have already been presented in \citetalias{Fitzpatrick05a}
and \citetalias{Fitzpatrick07} and, apart from the normalization, there are no substantial differences between the results here and 
those of the earlier work.  There is no previously published UV extinction curve for the heavily reddened O 
star GSC03712-01870 [$\ebv \simeq 1$], located in the direction of -- but beyond -- the Perseus arm 
\citep{Muzzio74}.  Its far-UV curve (i.e., for $\lambda < 2000$~\AA, since no long wavelength \iue\/ spectra 
exist) is unremarkable, closely resembling the Milky Way average curve.

A number of the curves in Figure 4 show gaps near 1.2~\invmic\/ and/or 2.7~\invmic.  The SPECTRUM models fail 
to precisely reproduce the hydrogen lines near the head of the Balmer series (e.g., see Figure 3g) -- which is 
not surprising given the complexity in modeling the overlapping lines -- and this produces ``mismatch features'' 
in the curves of the coolest stars in our sample. To avoid these distracting features, we eliminated the region 
between 3670~\AA\/ and 3760~\AA\/ ($\sim$2.7 \invmic) for these stars from the plots.  The cool star models also 
failed to precisely reproduce the higher Paschen lines of hydrogen and so, likewise, we eliminated the region 
near 8200~\AA\/ ($\sim$1.2~\invmic).  The TLUSTY OSTAR2002 models used for the hottest of our stars did not 
include the individual Paschen features and so the regions near these lines were also eliminated.  All other 
features of the curves are on display in the panels of Figure 4.  Some discrete stellar mismatch features still 
exist in some of the curves, but most of the observed structure -- from the narrowest to the broadest scales 
visible on the plots -- is a product of the interstellar medium.  The most common and prominent stellar mismatch 
feature arises from non-photospheric H$\alpha$ emission, which produces a narrow dip in the curves of some of 
the stars at 6563 \AA\/ (1.52 \invmic).  This feature is labeled (in green) in the top curve of Figure 4a.  
Several mismatched stellar He I lines from the $2^3P^0 - n^3D$ series are also seen (and labeled in green) in 
this curve.  These absorption lines are stronger in the star (HD~197512) than in the best-fit model and produce 
narrow spikes in the extinction ++curve.  This is not common in our sample and suggests that HD~197512 is a member 
of the class of helium-strong B stars as seen in the Orion nebula \citep[e.g.,][]{Morgan78}. Narrow 
features in the curves arising from diffuse interstellar bands \citep[``DIBs,''][]{Herbig75} and interstellar 
Na I and Ca II absorption are labeled (in red) in panel 4a for the HD~14321 curve.

The collection of spectrophotometric extinction curves in Figure 4 exhibits sightline-to-sightline variability on 
all scales, ranging from the narrow DIB's with widths of a fraction of 1~\invmic\/ to the broadest view-able 
scales.  This largest scale structure can be seen most clearly in a comparison between, for example, the nearly 
linear curve for HD~210121 in Figure 4a and the nearly quadratic curves for HD~294264 or HD~37061 in Figure 4g. 
Intermediate scale structure can also be discerned in many of the curves, notably in HD~197512 or HD~199216 in 
Figure 4a.  Generally speaking, this structure resembles a broad dip in the extinction at $\sim$1.6~\invmic\/ 
and/or a rise in extinction at $\sim$2~\invmic and corresponds to the ``Very Broad Structure'' referred to in \S 1.
In a forthcoming paper, \cite{Massa20} 
analyze this intermediate scale structure and its variability, and search for relationships with other 
aspects of NIR through UV extinction. In the remainder of this paper, our focus will be to determine the 
mean properties of NIR through UV extinction and quantify the broadest scale of sightline-to-sightline 
variations.

\section{Results}
\subsection{The $R$-dependent Milky Way Extinction Curve}\label{secAVERAGE}

It was first shown by \citep[][hereafter CCM]{Cardelli89} that some large scale properties of UV extinction 
(particularly, the general level in the far-UV) appear correlated with $R(V)$.  This wavelength coherence allows 
a family of $R(V)$-dependent 
extinction curves to be derived, potentially reducing the uncertainties in applying extinction corrections to 
sightlines with measured values of $R(V)$ but otherwise unknown extinction properties.  Subsequent studies 
(e.g., \citetalias{Fitzpatrick07})
have shown that much of this correlated behavior is driven by a relatively small number of 
sightlines with extreme values of $R(V)$ and that extinction curves with similar values of $R(V)$ may have a wide 
range in UV properties \citep[e.g.,][]{Mathis92, Valencic04}.  Nevertheless, the $R(V)$-dependence is important since, 
with appropriate estimation of the uncertainties, it can (1) potentially reduce dereddening errors, (2) allow a 
meaningful definition of a Milky Way average extinction curve, i.e., as that which corresponds to the mean 
observed value of $R(V)$, and (3) provide insight into the link between interstellar environment and dust grain 
properties \citep[see the discussion in][]{Fitzpatrick99review}.

\begin{figure}
\figurenum{5}
\epsscale{1.15}
\plotone{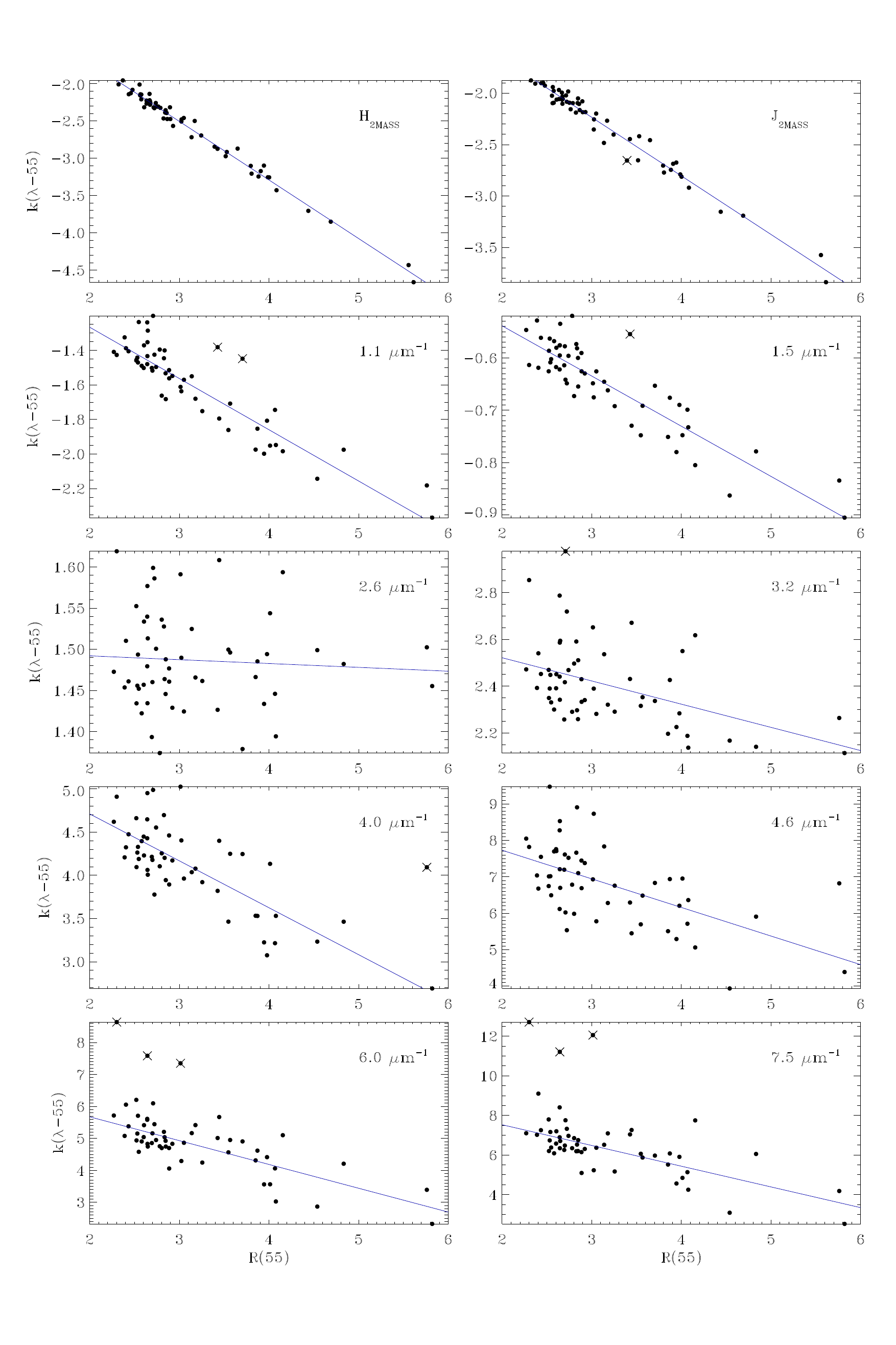}
\caption{Examples of the $R(55)$-dependence of the normalized extinction curves, where $R(55) \equiv 
A(55)/\efourfive$, for the 55 stars in our sample with $E(B-V) > 0.30$ (solid symbols).  The specific wavelengths 
shown include the NIR ($H$, $J$ and 1.1~\invmic), optical (1.5, 2.6 and 3.2~\invmic), and UV (4.0, 4.6, 6.0, and 7.5~\invmic) 
spectral regions.  The values of $R(55)$ for each star were determined from Equation 
\ref{eq:r55K55}, as discussed in \S\ref{secAVERAGE}. The blue lines show linear fits to \kfive\/ 
vs. $R(55)$ at each wavelength. Crossed out points were excluded from the fit using a ``sigma-clipping'' criterion 
set at 2.5$\sigma$. Fits such as these were performed for each of the 2700 data points defining the full 
extinction  curves.
\label{figSLOPES}}
\end{figure}

We examined the $R$-dependence of the curves in our sample in the simplest way, by plotting the values 
\kfive\/ versus $R$ at each wavelength point in the dataset.  We determined values of $R$ for by using the results
from \citetalias{Fitzpatrick07}, who found that -- when fitting the NIR data with a power law 
formula -- the values of $R(V)$ were related to the extinction at the $K$ band by the relation:
\begin{equation}
R(V) = -1.19 \times k(K-V) - 0.26   
\label{eq:rvKV}
\end{equation}
This can be converted to our normalization system by using the mean transformation in 
Equation (\ref{eq:rvtransform}), 
and yields:
\begin{equation}
R(55) = -1.19 \times k(K-55) - 0.25   
\label{eq:r55K55}
\end{equation}
Plots of \kfive\/ vs.\ $R(55)$ are shown in Figure \ref{figSLOPES} for 10 representative wavelengths (out of a 
total of 2700 points per curve), spanning the NIR through UV region.  The $R(55)$-dependence is noisy but 
reasonably well defined for most data points.  To quantify the relationship, we fit a simple linear function at 
each wavelength, recording the intercept, slope, and standard deviation about the fit.  A single iteration 
2.5-$\sigma$ $\sigma$-clip was performed to eliminate extreme points from the fit.  In Figure \ref{figSLOPES}, 
the blue lines show the fits and points that are crossed out indicate those rejected by the $\sigma$-clipping.  
This procedure was restricted to the 55 sightlines in our sample for which \ebv $> 0.30$, which minimizes the impact
of the larger random errors inherent in the low \ebv\/ curves.

\begin{figure*}
\figurenum{6}
\epsscale{0.9}
\plotone{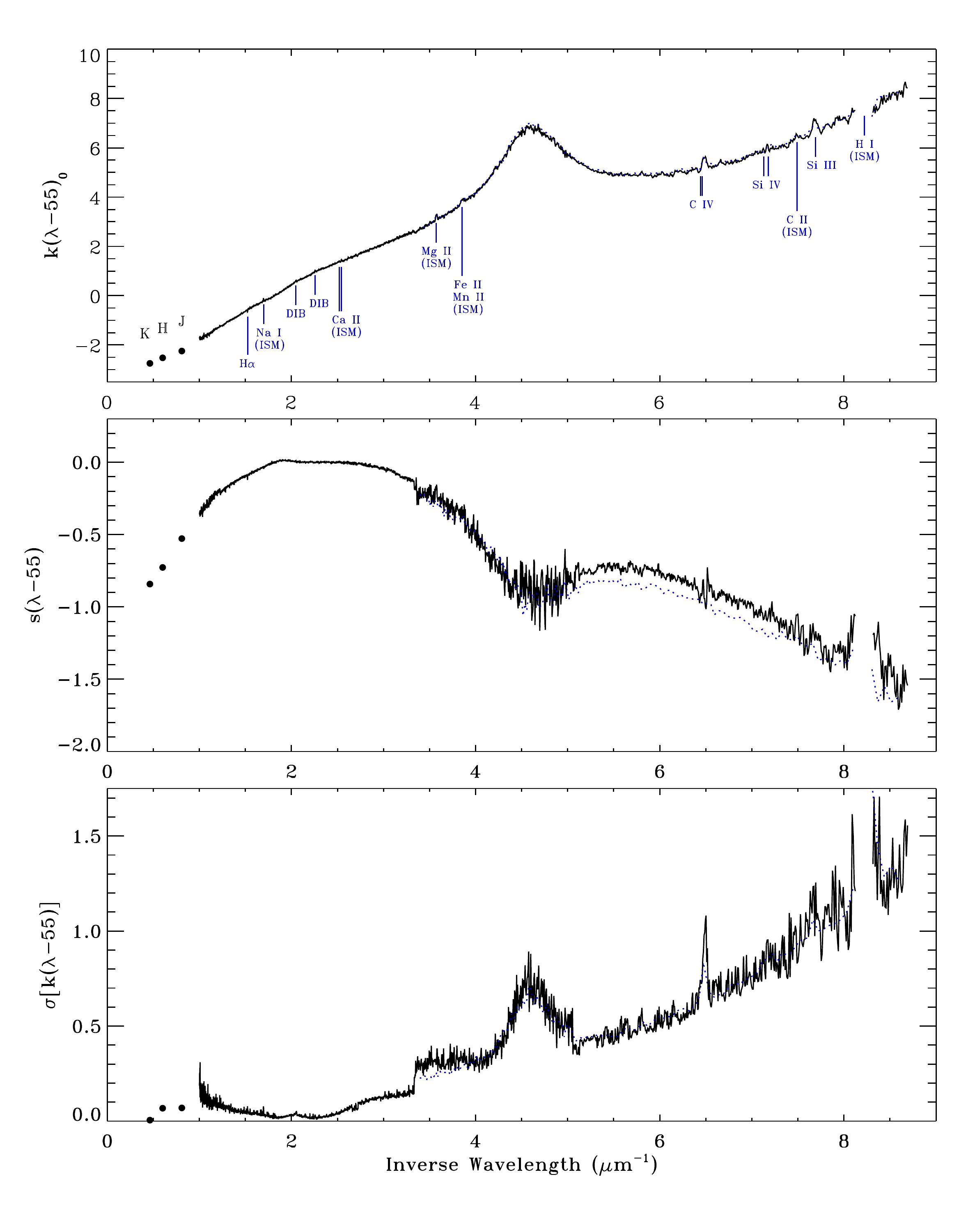}
\caption{The $R(55)$-Dependent Relationship. {\it Top panel:} The mean NIR thru UV extinction curve in the 
form $\kfive_0 \equiv \elamfive/\efourfive$ for the case of $R(55) = 3.02$, as derived from the linear fits to 
\kfive\/ vs. $R(55)$ described in \S \ref{secAVERAGE} and illustrated in Figure \ref{figSLOPES}. This value of 
$R(55)$ corresponds to $R(V) = 3.10$, which is generally considered to be the Milky Way mean. Features in the UV 
curve due to stellar mismatch, interstellar gas absorption lines, or geocoronal Ly$\alpha$ emission are 
indicated. {\it Middle panel:} The linear dependence of the mean curve on $R(55)$.
These values are the slopes of the linear fits described in \S \ref{secAVERAGE}. {\it Bottom 
panel:} Standard deviation of the individual curves about the mean $R(55)$-dependence.  The dotted blue curves in the
UV region of each panel show the results of a similar analysis performed by us on the much larger set of UV extinction curves presented 
by \citetalias{Fitzpatrick07}; see the discussion in \S \ref{secCOMPARE}.
\label{figMEANCURVE}}
\end{figure*}

Figure \ref{figMEANCURVE} graphically illustrates the overall results of fitting the linear $R$-dependence in 
our dataset.  The curve in the top panel shows $\kfive_0$, the values of the fits at $R(55)$ = 3.02, 
which corresponds to the Milky Way mean value of $R(V)$ = 3.10 (see, Eq. \ref{eq:rvtransform}).  
Thus, this curve shows our measurement of the extinction representative of the diffuse ISM over the 
wavelength range 1150~\AA\/ to 2.5~\mic.  Prominent stellar and interstellar mismatch features are 
labeled in blue.  The middle panel shows the slopes of the fits, $s(\lambda-55) \equiv \Delta$\kfive / 
$\Delta R(55)$, 
illustrating the $R(55)$-dependence.  The coherence and magnitude of the structure in the  $s(\lambda-55)$ plot, as 
compared to its point-to point scatter, indicate that the fitting procedure has identified a true $R$-dependence 
in the data, rather than just quantifying the noise.  Finally, the bottom panel shows $\sigma[\kfive]$, the 
standard deviation of the individual sightlines against the mean relationship, which quantifies the combination 
of measurement noise and real deviations from the $R$-dependent relation.

A comparison between the functional dependence of extinction on $R$ adopted here and that used 
by \citetalias{Cardelli89} is given in Appendix \ref{appendix_ccm}.

\subsection{A Tabular Form of the R-dependent Relationship}\label{secTABFORM}

\begin{figure*}
\figurenum{7}
\epsscale{1.1}
\plottwo{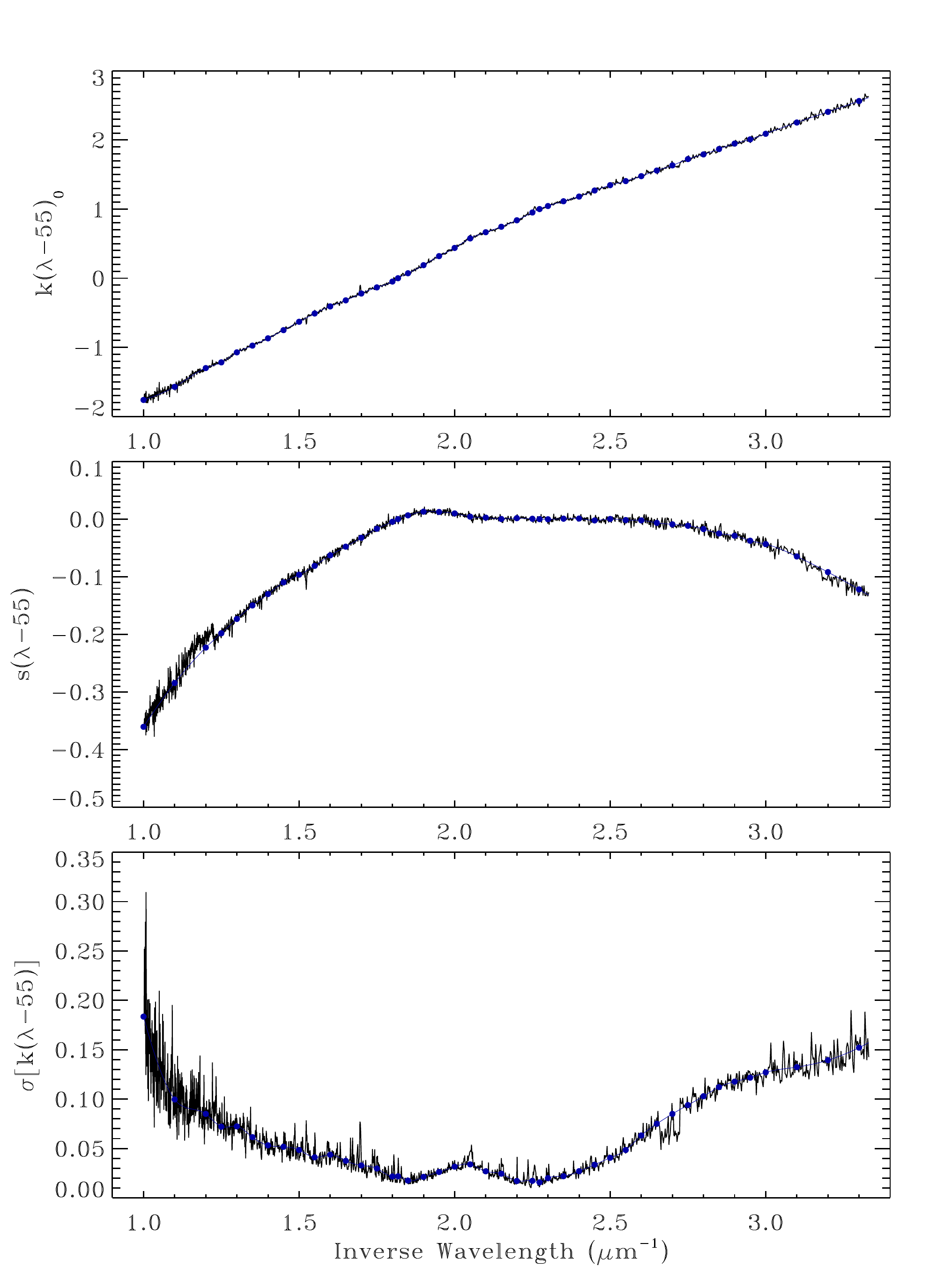}{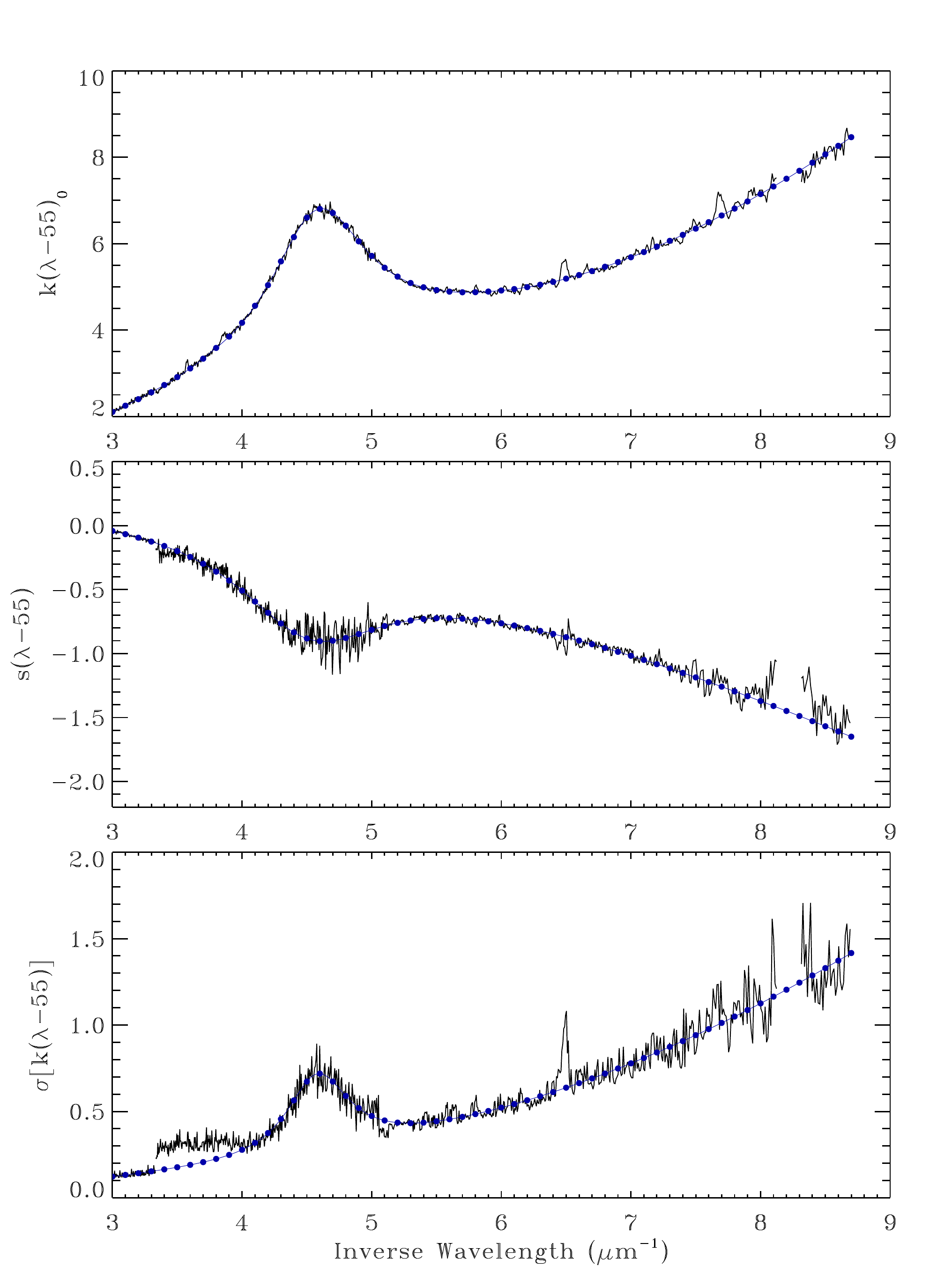}
\caption{Creating a tabular form of the $R$-dependence of the Milky Way Extinction Curve
in the optical/NIR (left) and UV (right) spectral regions.  In the optical, a cubic spline was fit to the \kfive$_0$\/ 
({\it top panel}), $s(\lambda-55)$ ({\it middle panel}), and $\sigma[\kfive]$ ({\it bottom panel}) curves.
The anchor points were spaced at 0.05 \invmic\/ intervals between 1.2~\invmic\/ and 3.0~\invmic\/ and at intervals 
of 0.1~\invmic\/ outside that range.
The values of the anchors were adjusted iteratively to achieve the best fit to the respective curves. 
Additional points were added at 1.818~\invmic\/ (5500~\AA) and 2.273~\invmic\/ (4400~\AA) to ensure 
the normalization.  The cubic spline fits are shown by the blue curves and the anchor points are 
indicated by the filled blue circles.  In the UV, the data were fit by variants of the UV fitting 
functions adopted by \citetalias{Fitzpatrick07}. The fits are shown by the blue curves and their values at 0.1~\invmic\/ 
intervals are shown by the filled blue circles.  To assure 
continuity with the optical results, the UV fits were extended into the optical region (between 3.0 and 
3.3~\invmic) and low weight was given to the longest wavelength UV data, due to instability in the 
calibration of \iue\;'s long wavelength camera at these wavelengths.  A tabular form of these curves is 
given in Table \ref{tabAVERAGE}.
See the discussion is \S\ref{secTABFORM}.
\label{figTABFORM}}
\end{figure*}

\begin{figure*}
\figurenum{8}
\epsscale{1.0}
\plotone{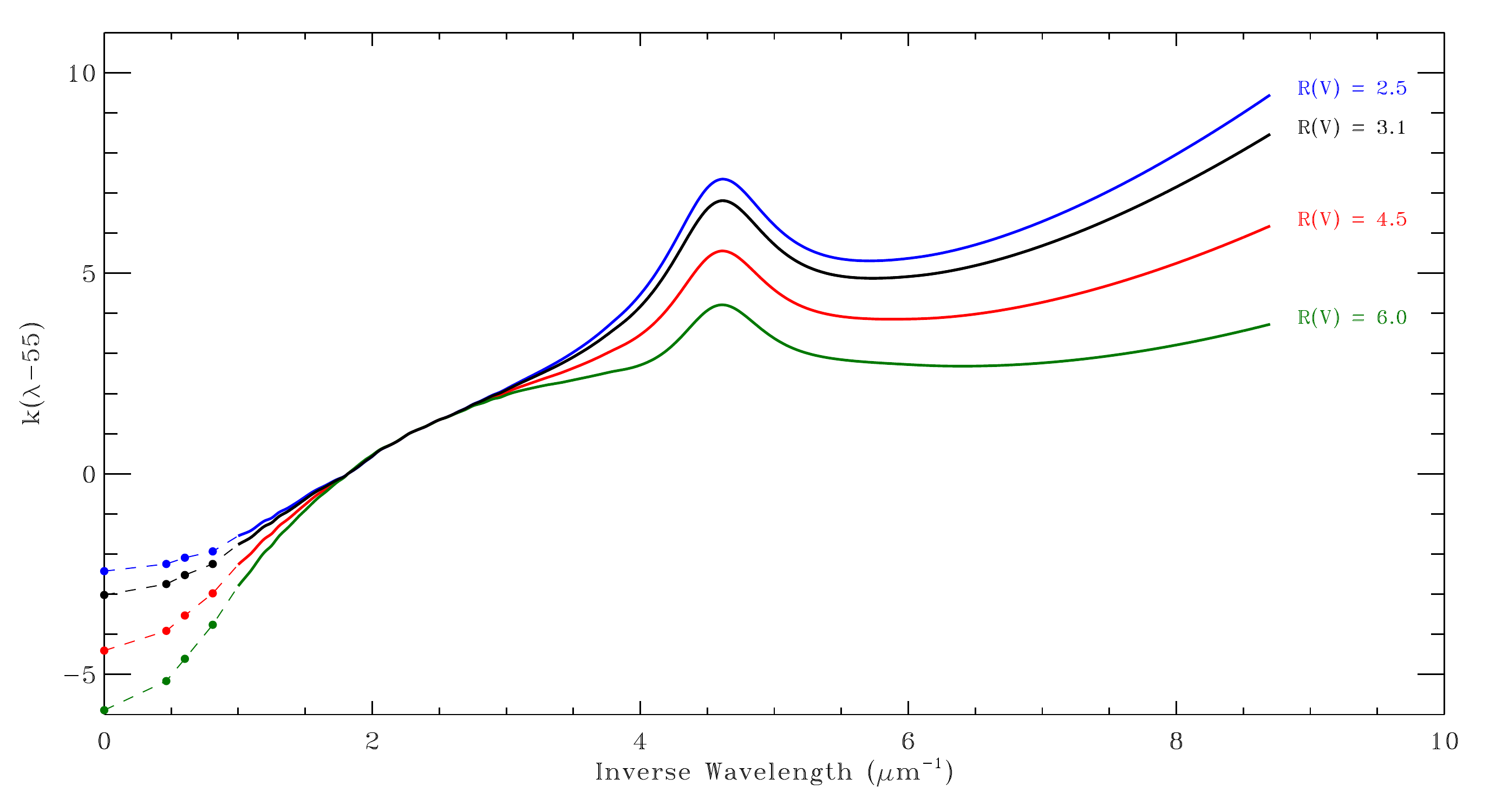}
\caption{$R(V)$-dependent NIR through UV extinction curves from this study, shown for several 
values of $R(V)$. The curves were produced from the data in Table \ref{tabAVERAGE} using Equation 
(\ref{eqTABVALUES2}).
\label{figRCURVES}} 
\end{figure*}

\begin{deluxetable*}{cccccccccc} 
\tabletypesize{\scriptsize}
\tablenum{3} 
\tablewidth{0pc} 
\tablecaption{Average Milky Way Extinction Curve}
\tablehead{ 
\colhead{x\tablenotemark{a}}                                            &
\colhead{$k(\lambda-55)_0$\tablenotemark{b}}                              &
\colhead{$s(\lambda-55)$\tablenotemark{c}}  &
\colhead{$\sigma [k(\lambda-55)]$\tablenotemark{d}}                       &
\colhead{}                                                              &
\colhead{}                                                              &  
\colhead{x\tablenotemark{a}}                                            &
\colhead{$k(\lambda-55)_0$\tablenotemark{b}}                              &
\colhead{$s(\lambda-55)$\tablenotemark{c}}  &
\colhead{$\sigma [k(\lambda-55)]$\tablenotemark{d}}                    }
\startdata
$ 0.000$ & $-3.020$ & $-1.000$ & $ 0.000$ & & & $ 3.700$ & $ 3.337$ & $-0.299$ & $ 0.207$ \\
K$_{2MASS}$ & $-2.747$ & $-0.842$ & $ 0.000$ & & & $ 3.800$ & $ 3.586$ & $-0.360$ & $ 0.225$ \\
H$_{2MASS}$ & $-2.523$ & $-0.727$ & $ 0.068$ & & & $ 3.900$ & $ 3.849$ & $-0.430$ & $ 0.248$ \\
J$_{2MASS}$ & $-2.247$ & $-0.528$ & $ 0.069$ & & & $ 4.000$ & $ 4.169$ & $-0.508$ & $ 0.278$ \\
$ 1.000$ & $-1.761$ & $-0.361$ & $ 0.184$ & & & $ 4.100$ & $ 4.562$ & $-0.594$ & $ 0.318$ \\
$ 1.100$ & $-1.572$ & $-0.284$ & $ 0.100$ & & & $ 4.200$ & $ 5.039$ & $-0.682$ & $ 0.375$ \\
$ 1.200$ & $-1.300$ & $-0.223$ & $ 0.085$ & & & $ 4.300$ & $ 5.589$ & $-0.765$ & $ 0.456$ \\
$ 1.250$ & $-1.217$ & $-0.198$ & $ 0.072$ & & & $ 4.400$ & $ 6.151$ & $-0.835$ & $ 0.564$ \\
$ 1.300$ & $-1.073$ & $-0.173$ & $ 0.072$ & & & $ 4.500$ & $ 6.603$ & $-0.882$ & $ 0.673$ \\
$ 1.350$ & $-0.976$ & $-0.150$ & $ 0.061$ & & & $ 4.600$ & $ 6.804$ & $-0.903$ & $ 0.719$ \\
$ 1.400$ & $-0.870$ & $-0.130$ & $ 0.053$ & & & $ 4.700$ & $ 6.712$ & $-0.900$ & $ 0.673$ \\
$ 1.450$ & $-0.751$ & $-0.110$ & $ 0.052$ & & & $ 4.800$ & $ 6.416$ & $-0.879$ & $ 0.590$ \\
$ 1.500$ & $-0.630$ & $-0.096$ & $ 0.048$ & & & $ 4.900$ & $ 6.053$ & $-0.848$ & $ 0.520$ \\
$ 1.550$ & $-0.511$ & $-0.081$ & $ 0.041$ & & & $ 5.000$ & $ 5.716$ & $-0.815$ & $ 0.474$ \\
$ 1.600$ & $-0.408$ & $-0.063$ & $ 0.044$ & & & $ 5.100$ & $ 5.443$ & $-0.785$ & $ 0.448$ \\
$ 1.650$ & $-0.320$ & $-0.048$ & $ 0.037$ & & & $ 5.200$ & $ 5.237$ & $-0.760$ & $ 0.435$ \\
$ 1.700$ & $-0.221$ & $-0.032$ & $ 0.033$ & & & $ 5.300$ & $ 5.089$ & $-0.742$ & $ 0.432$ \\
$ 1.750$ & $-0.133$ & $-0.017$ & $ 0.030$ & & & $ 5.400$ & $ 4.989$ & $-0.730$ & $ 0.435$ \\
$ 1.800$ & $-0.049$ & $-0.005$ & $ 0.022$ & & & $ 5.500$ & $ 4.925$ & $-0.724$ & $ 0.443$ \\
$ 1.818$ & $\phm{-} 0.000$ & $\phm{-} 0.000$ & $ 0.022$ & & & $ 5.600$ & $ 4.889$ & $-0.724$ & $ 0.455$ \\
$ 1.850$ & $\phm{-} 0.072$ & $\phm{-} 0.007$ & $ 0.017$ & & & $ 5.700$ & $ 4.875$ & $-0.728$ & $ 0.468$ \\
$ 1.900$ & $\phm{-} 0.188$ & $\phm{-} 0.013$ & $ 0.021$ & & & $ 5.800$ & $ 4.876$ & $-0.737$ & $ 0.485$ \\
$ 1.950$ & $\phm{-} 0.319$ & $\phm{-} 0.012$ & $ 0.026$ & & & $ 5.900$ & $ 4.891$ & $-0.749$ & $ 0.502$ \\
$ 2.000$ & $\phm{-} 0.438$ & $\phm{-} 0.010$ & $ 0.032$ & & & $ 6.000$ & $ 4.916$ & $-0.764$ & $ 0.522$ \\
$ 2.050$ & $\phm{-} 0.576$ & $\phm{-} 0.004$ & $ 0.034$ & & & $ 6.100$ & $ 4.949$ & $-0.781$ & $ 0.543$ \\
$ 2.100$ & $\phm{-} 0.665$ & $\phm{-} 0.003$ & $ 0.027$ & & & $ 6.200$ & $ 4.993$ & $-0.801$ & $ 0.565$ \\
$ 2.150$ & $\phm{-} 0.744$ & $\phm{-} 0.000$ & $ 0.024$ & & & $ 6.300$ & $ 5.049$ & $-0.823$ & $ 0.588$ \\
$ 2.200$ & $\phm{-} 0.838$ & $\phm{-} 0.002$ & $ 0.017$ & & & $ 6.400$ & $ 5.115$ & $-0.847$ & $ 0.612$ \\
$ 2.250$ & $\phm{-} 0.951$ & $\phm{-} 0.001$ & $ 0.017$ & & & $ 6.500$ & $ 5.190$ & $-0.872$ & $ 0.638$ \\
$ 2.273$ & $\phm{-} 1.000$ & $\phm{-} 0.000$ & $ 0.016$ & & & $ 6.600$ & $ 5.273$ & $-0.899$ & $ 0.664$ \\
$ 2.300$ & $\phm{-} 1.044$ & $-0.000$ & $ 0.020$ & & & $ 6.700$ & $ 5.365$ & $-0.927$ & $ 0.691$ \\
$ 2.350$ & $\phm{-} 1.113$ & $\phm{-} 0.001$ & $ 0.022$ & & & $ 6.800$ & $ 5.465$ & $-0.956$ & $ 0.719$ \\
$ 2.400$ & $\phm{-} 1.181$ & $\phm{-} 0.001$ & $ 0.027$ & & & $ 6.900$ & $ 5.571$ & $-0.987$ & $ 0.748$ \\
$ 2.450$ & $\phm{-} 1.269$ & $-0.002$ & $ 0.034$ & & & $ 7.000$ & $ 5.685$ & $-1.018$ & $ 0.778$ \\
$ 2.500$ & $\phm{-} 1.346$ & $\phm{-} 0.000$ & $ 0.041$ & & & $ 7.100$ & $ 5.805$ & $-1.050$ & $ 0.809$ \\
$ 2.550$ & $\phm{-} 1.405$ & $-0.002$ & $ 0.048$ & & & $ 7.200$ & $ 5.932$ & $-1.083$ & $ 0.841$ \\
$ 2.600$ & $\phm{-} 1.476$ & $-0.002$ & $ 0.063$ & & & $ 7.300$ & $ 6.065$ & $-1.117$ & $ 0.874$ \\
$ 2.650$ & $\phm{-} 1.558$ & $-0.006$ & $ 0.075$ & & & $ 7.400$ & $ 6.203$ & $-1.151$ & $ 0.907$ \\
$ 2.700$ & $\phm{-} 1.632$ & $-0.009$ & $ 0.085$ & & & $ 7.500$ & $ 6.347$ & $-1.186$ & $ 0.941$ \\
$ 2.750$ & $\phm{-} 1.724$ & $-0.012$ & $ 0.094$ & & & $ 7.600$ & $ 6.497$ & $-1.222$ & $ 0.976$ \\
$ 2.800$ & $\phm{-} 1.791$ & $-0.017$ & $ 0.103$ & & & $ 7.700$ & $ 6.652$ & $-1.259$ & $ 1.012$ \\
$ 2.850$ & $\phm{-} 1.868$ & $-0.025$ & $ 0.112$ & & & $ 7.800$ & $ 6.812$ & $-1.296$ & $ 1.049$ \\
$ 2.900$ & $\phm{-} 1.947$ & $-0.029$ & $ 0.118$ & & & $ 7.900$ & $ 6.978$ & $-1.333$ & $ 1.087$ \\
$ 2.950$ & $\phm{-} 2.008$ & $-0.037$ & $ 0.122$ & & & $ 8.000$ & $ 7.148$ & $-1.371$ & $ 1.125$ \\
$ 3.000$ & $\phm{-} 2.090$ & $-0.043$ & $ 0.127$ & & & $ 8.100$ & $ 7.323$ & $-1.410$ & $ 1.164$ \\
$ 3.100$ & $\phm{-} 2.252$ & $-0.065$ & $ 0.132$ & & & $ 8.200$ & $ 7.502$ & $-1.449$ & $ 1.204$ \\
$ 3.200$ & $\phm{-} 2.406$ & $-0.092$ & $ 0.139$ & & & $ 8.300$ & $ 7.686$ & $-1.488$ & $ 1.245$ \\
$ 3.300$ & $\phm{-} 2.563$ & $-0.122$ & $ 0.152$ & & & $ 8.400$ & $ 7.874$ & $-1.528$ & $ 1.287$ \\
$ 3.400$ & $\phm{-} 2.728$ & $-0.160$ & $ 0.165$ & & & $ 8.500$ & $ 8.067$ & $-1.568$ & $ 1.329$ \\
$ 3.500$ & $\phm{-} 2.911$ & $-0.200$ & $ 0.177$ & & & $ 8.600$ & $ 8.264$ & $-1.609$ & $ 1.373$ \\
$ 3.600$ & $\phm{-} 3.112$ & $-0.245$ & $ 0.191$ & & & $ 8.700$ & $ 8.464$ & $-1.650$ & $ 1.417$ \\
\enddata
\tablenotetext{a}{Inverse wavelength in units of $\rm \mu m^{-1}$.}
\tablenotetext{b}{Mean extinction curve in units of $\kfive\/ \equiv \elamfive/\efourfive$ for the case 
$R(55) = 3.02$, corresponding to the Milky Way mean value of $R(V) = 3.10$.}
\tablenotetext{c}{Linear dependence of \kfive\/ on $R(55)$ over the range $2.5 \leq R(55) \leq 6.0$.}
\tablenotetext{d}{Curve-to-curve standard deviation as a function of wavelength.}
\label{tabAVERAGE}
\end{deluxetable*}

To make the new results in Figure \ref{figMEANCURVE} useful for other extinction investigations, we 
created a tabular form of these data. This was done by fitting the three curves in Figure~\ref{figMEANCURVE} 
with smooth functions and then recording the values of the functions at selected wavelengths.  The details 
of this process 
are illustrated in Figure~\ref{figTABFORM} where the smooth blue curves show the 
fits to the data and the blue circles indicate the selected points.  

In the optical/NIR, i.e., for the \hst\/ spectrophotometric data, we fit cubic splines to the three curves, 
with anchor points spaced every 
0.1~\invmic in the interval 1.0~\invmic\/ (10000~\AA) to 3.3~\invmic\/ (3030~\AA), with intermediate points  
added between 1.2~\invmic\/ and 2.9~\invmic\/ and two special points -- at 1.818~\invmic (5500~\AA) 
and 2.273~\invmic (4400~\AA) -- added to enforce the normalization.  Regions 
near the Balmer and Paschen jumps that are affected by spectral mismatch were given zero 
weight so that the fits passed smoothly through these regions in the $s(\lambda-V)$ and 
$\sigma$[\kfive]\/ curves.  Likewise, data affected by interstellar absorption features (including 
DIB's) and stellar mismatch features were given zero weight. 
Notice that the values of $\sigma$[\kfive]\/ do not go to zero at the normalization points.  
This is because (as noted in \S\ref{secCURVES}) the normalization was performed using interpolation within a bin containing
the normalization wavelengths.  Thus, these small values of $\sigma$[\kfive]\/ 
$\simeq$ 0.02 reflect the point-to-point scatter in the normalization 
regions of the curves, rather than sightline-to-sightline variations.  The large values of 
$\sigma$[\kfive]\/ at the longest optical wavelengths are likely dominated by statistical 
noise in the \hst\/ G750L data rather than curve-to-curve variability.  The true 
curve-to-curve scatter is probably at the level of $\sigma$[\kfive]\/ $\simeq$ 0.1, 
similar to the shorter wavelength G750L data and the \tmass\/ $J$ and $H$ values.

In the UV, we used variations of the \citetalias{Fitzpatrick07} fitting function to achieve smooth representations 
of the curves.  The fits were performed over the wavelength range of the \iue\/ data, i.e., 3.3~\invmic 
\/(3030~\AA) to 8.7~\invmic\/ (1150~\AA), plus a small section of the \hst\/ data, from 3.0 to 3.3~\invmic, 
which was included to assure continuity with the optical region. 
Spectral mismatch features were given zero weight in the fits.  Figures \ref{figMEANCURVE} and \ref{figTABFORM} 
show that there is a significant
discontinuity between the longest UV wavelengths and the shortest optical wavelengths 
in the $\sigma$[\kfive]\/ curves.  We attribute this to instability 
in the calibration of \iue\/'s LWR and LWP cameras at their longest wavelengths, where the 
camera sensitivities are highly dependent on the placement of the spectra on the camera face.  
We gave the longest UV wavelengths zero weight so that the fits would join smoothly with the 
optical \hst\/ data.  There is also a small discontinuity noticeable in $\sigma$[\kfive]\/ 
at the joint between \iue\/'s short and long wavelength cameras at 5.05~\invmic\/ (1980~\AA) 
where, once again, the extreme sensitivity of the long wavelength cameras to spectrum 
placement (and the lower general sensitivity of the cameras) manifests as enhanced scatter.  Values of the
fits at 0.1~\invmic\/ intervals over the range 3.3~\invmic\/ to 8.7~\invmic, as indicated by 
the filled circles in Figure \ref{figTABFORM}, were recorded.

The final tabular form of our extinction results are given in Table \ref{tabAVERAGE}. The columns in the table 
list the wavelengths (in \invmic), followed by the values of \kfive$_0$, $s(\lambda-55)$ and $\sigma[\kfive]$.  
For the spectrophotometric data, these all correspond to the filled circles in Figure \ref{figTABFORM}. 
For the \tmass\/
$JHK$ data, the values in the table are simply those shown in Figure \ref{figMEANCURVE}. The result that 
$\sigma [k(K-55)] = 0$ arises because our values of $R(55)$ are tied directly to the $K$-band 
measurement via Equation (\ref{eq:r55K55}).

The data in Table \ref{tabAVERAGE} can be used to reconstruct an extinction curve in the form 
of \kfive\/ over the indicated wavelength range for any value of $R(55)$ or $R(V)$ ranging 
from 2.5 to 6.0.  The restriction on the range of valid $R$ values comes directly from the range of $R$ 
in our data (see Figure \ref{figSLOPES}).  The curve can be constructed using the formula:
\begin{equation}
\scriptsize
k(\lambda-55)_{R(55)} = k(\lambda-55)_0 + s(\lambda-55) \times [R(55) - R(55)_0]
\label{eqTABVALUES1}
\end{equation}
where $R(55)_0 = 3.02$ and $k(\lambda-55)_0$ is the extinction for the case $R(55) = R(55)_0$.  
If $R$ is expressed in $R(V)$ notation, then Equation (\ref{eqTABVALUES1}) transforms to:
\begin{equation}
\scriptsize
k(\lambda-55)_{R(V)} = k(\lambda-55)_{0} + s(\lambda-55) \times [R(V) - 3.10] \times 0.99
\label{eqTABVALUES2}
\end{equation}

Four examples of extinction curves constructed from Table \ref{tabAVERAGE} are 
shown in Figure \ref{figRCURVES}, for values of $R(V)$ = 2.5, 3.1, 4.5, and 6.0.  
The $R$-dependent extinction relationship is available in the dust\_extinction python 
package\footnote{https://dust-extinction.readthedocs.io/} as the ``F20'' model. 

\subsection{Reddening Slopes for Optical Photometry}\label{secPHOTOM}

Given the advent of large ground-based surveys and the overall prevalence of the data, optical 
photometry remains a powerful astronomical tool.  Knowledge of the effects 
of interstellar extinction on the various photometric systems is important for disentangling 
the intrinsic and extrinsic (i.e., interstellar) contributions to photometric measurements.

\begin{deluxetable*}{rccccccccccccc}
\tablenum{4}
\label{tabUBVuvby}
\tabletypesize{\scriptsize}
\tablewidth{0pc} 
\tablecaption{UBV and uvby Extinction Indices}
\tablehead{ 
\colhead{$T_{eff}$}                   & 
\colhead{}                            &
\colhead{$\frac{E(B-V)}{E(44-55)}$}   &
\colhead{$\frac{E(U-B)}{E(B-V)}$}     &
\colhead{$\frac{E(b-y)}{E(B-V)}$}     &
\colhead{$\frac{E(m_1)}{E(b-y)}$}     &
\colhead{$\frac{E(c_1)}{E(b-y)}$}     &
\colhead{}                            &
\colhead{}                            &
\colhead{$\frac{E(B-V)}{E(44-55)}$}   &
\colhead{$\frac{E(U-B)}{E(B-V)}$}     &
\colhead{$\frac{E(b-y)}{E(B-V)}$}     &
\colhead{$\frac{E(m_1)}{E(b-y)}$}     &
\colhead{$\frac{E(c_1)}{E(b-y)}$}     }
\startdata
& & \multicolumn{12}{c}{$R(V) = 2.5$} \\ \cline{3-14}
& & \multicolumn{5}{c}{E(44-55) = 0.50} & & & \multicolumn{5}{c}{E(44-55) = 1.00} \\ \cline{3-7} \cline{10-14}
$  7000$K &  & $ 0.945$ & $ 0.806$ & $ 0.752$ & $-0.299$ & $ 0.285$ & & & $ 0.933$ & $ 0.830$ & $ 0.762$ & $-0.299$ & $ 0.279$ \\
$  8000$K &  & $ 0.955$ & $ 0.760$ & $ 0.746$ & $-0.300$ & $ 0.289$ & & & $ 0.943$ & $ 0.782$ & $ 0.755$ & $-0.301$ & $ 0.283$ \\
$  9000$K &  & $ 0.964$ & $ 0.730$ & $ 0.738$ & $-0.300$ & $ 0.290$ & & & $ 0.953$ & $ 0.750$ & $ 0.747$ & $-0.300$ & $ 0.283$ \\
$ 10000$K &  & $ 0.972$ & $ 0.718$ & $ 0.732$ & $-0.297$ & $ 0.290$ & & & $ 0.961$ & $ 0.739$ & $ 0.740$ & $-0.298$ & $ 0.283$ \\
$ 12500$K &  & $ 0.981$ & $ 0.736$ & $ 0.725$ & $-0.295$ & $ 0.293$ & & & $ 0.970$ & $ 0.755$ & $ 0.733$ & $-0.295$ & $ 0.287$ \\
$ 15000$K &  & $ 0.985$ & $ 0.752$ & $ 0.722$ & $-0.294$ & $ 0.296$ & & & $ 0.974$ & $ 0.771$ & $ 0.730$ & $-0.295$ & $ 0.290$ \\
$ 20000$K &  & $ 0.990$ & $ 0.765$ & $ 0.718$ & $-0.293$ & $ 0.299$ & & & $ 0.979$ & $ 0.784$ & $ 0.726$ & $-0.294$ & $ 0.293$ \\
$ 25000$K &  & $ 0.993$ & $ 0.773$ & $ 0.716$ & $-0.292$ & $ 0.300$ & & & $ 0.982$ & $ 0.791$ & $ 0.724$ & $-0.292$ & $ 0.294$ \\
$ 30000$K &  & $ 0.996$ & $ 0.775$ & $ 0.713$ & $-0.291$ & $ 0.301$ & & & $ 0.985$ & $ 0.794$ & $ 0.722$ & $-0.291$ & $ 0.295$ \\
$ 35000$K &  & $ 0.997$ & $ 0.778$ & $ 0.712$ & $-0.290$ & $ 0.301$ & & & $ 0.986$ & $ 0.796$ & $ 0.721$ & $-0.291$ & $ 0.295$ \\
& & \multicolumn{12}{c}{$R(V) = 3.1$} \\ \cline{3-14}
& & \multicolumn{5}{c}{E(44-55) = 0.50} & & & \multicolumn{5}{c}{E(44-55) = 1.00} \\ \cline{3-7} \cline{10-14}
$  7000$K &  & $ 0.951$ & $ 0.793$ & $ 0.748$ & $-0.301$ & $ 0.263$ & & & $ 0.940$ & $ 0.815$ & $ 0.756$ & $-0.301$ & $ 0.258$ \\
$  8000$K &  & $ 0.959$ & $ 0.748$ & $ 0.742$ & $-0.302$ & $ 0.266$ & & & $ 0.949$ & $ 0.769$ & $ 0.750$ & $-0.303$ & $ 0.261$ \\
$  9000$K &  & $ 0.968$ & $ 0.718$ & $ 0.735$ & $-0.301$ & $ 0.267$ & & & $ 0.959$ & $ 0.738$ & $ 0.743$ & $-0.302$ & $ 0.262$ \\
$ 10000$K &  & $ 0.976$ & $ 0.707$ & $ 0.729$ & $-0.299$ & $ 0.267$ & & & $ 0.966$ & $ 0.727$ & $ 0.736$ & $-0.300$ & $ 0.261$ \\
$ 12500$K &  & $ 0.985$ & $ 0.724$ & $ 0.722$ & $-0.297$ & $ 0.270$ & & & $ 0.975$ & $ 0.743$ & $ 0.730$ & $-0.297$ & $ 0.264$ \\
$ 15000$K &  & $ 0.989$ & $ 0.739$ & $ 0.719$ & $-0.296$ & $ 0.273$ & & & $ 0.979$ & $ 0.758$ & $ 0.727$ & $-0.296$ & $ 0.267$ \\
$ 20000$K &  & $ 0.994$ & $ 0.752$ & $ 0.716$ & $-0.295$ & $ 0.276$ & & & $ 0.984$ & $ 0.770$ & $ 0.723$ & $-0.295$ & $ 0.270$ \\
$ 25000$K &  & $ 0.997$ & $ 0.759$ & $ 0.713$ & $-0.293$ & $ 0.276$ & & & $ 0.987$ & $ 0.777$ & $ 0.720$ & $-0.294$ & $ 0.271$ \\
$ 30000$K &  & $ 0.999$ & $ 0.761$ & $ 0.711$ & $-0.293$ & $ 0.277$ & & & $ 0.990$ & $ 0.779$ & $ 0.718$ & $-0.293$ & $ 0.271$ \\
$ 35000$K &  & $ 1.001$ & $ 0.764$ & $ 0.710$ & $-0.292$ & $ 0.277$ & & & $ 0.991$ & $ 0.782$ & $ 0.717$ & $-0.292$ & $ 0.272$ \\
& & \multicolumn{12}{c}{$R(V) = 4.5$} \\ \cline{3-14}
& & \multicolumn{5}{c}{E(44-55) = 0.50} & & & \multicolumn{5}{c}{E(44-55) = 1.00} \\ \cline{3-7} \cline{10-14}
$  7000$K &  & $ 0.964$ & $ 0.761$ & $ 0.739$ & $-0.305$ & $ 0.211$ & & & $ 0.958$ & $ 0.780$ & $ 0.744$ & $-0.306$ & $ 0.207$ \\
$  8000$K &  & $ 0.971$ & $ 0.720$ & $ 0.734$ & $-0.306$ & $ 0.214$ & & & $ 0.965$ & $ 0.738$ & $ 0.739$ & $-0.307$ & $ 0.210$ \\
$  9000$K &  & $ 0.979$ & $ 0.692$ & $ 0.728$ & $-0.305$ & $ 0.214$ & & & $ 0.973$ & $ 0.709$ & $ 0.733$ & $-0.306$ & $ 0.211$ \\
$ 10000$K &  & $ 0.986$ & $ 0.682$ & $ 0.722$ & $-0.303$ & $ 0.214$ & & & $ 0.980$ & $ 0.699$ & $ 0.727$ & $-0.304$ & $ 0.210$ \\
$ 12500$K &  & $ 0.994$ & $ 0.696$ & $ 0.716$ & $-0.300$ & $ 0.216$ & & & $ 0.989$ & $ 0.713$ & $ 0.720$ & $-0.301$ & $ 0.212$ \\
$ 15000$K &  & $ 0.998$ & $ 0.710$ & $ 0.713$ & $-0.300$ & $ 0.218$ & & & $ 0.992$ & $ 0.726$ & $ 0.718$ & $-0.301$ & $ 0.214$ \\
$ 20000$K &  & $ 1.002$ & $ 0.721$ & $ 0.710$ & $-0.298$ & $ 0.220$ & & & $ 0.997$ & $ 0.737$ & $ 0.714$ & $-0.299$ & $ 0.217$ \\
$ 25000$K &  & $ 1.005$ & $ 0.728$ & $ 0.708$ & $-0.297$ & $ 0.221$ & & & $ 0.999$ & $ 0.744$ & $ 0.712$ & $-0.298$ & $ 0.217$ \\
$ 30000$K &  & $ 1.007$ & $ 0.730$ & $ 0.706$ & $-0.296$ & $ 0.221$ & & & $ 1.002$ & $ 0.746$ & $ 0.710$ & $-0.297$ & $ 0.217$ \\
$ 35000$K &  & $ 1.008$ & $ 0.732$ & $ 0.705$ & $-0.296$ & $ 0.222$ & & & $ 1.003$ & $ 0.748$ & $ 0.710$ & $-0.297$ & $ 0.218$ \\
& & \multicolumn{12}{c}{$R(V) = 6.0$} \\ \cline{3-14}
& & \multicolumn{5}{c}{E(44-55) = 0.50} & & & \multicolumn{5}{c}{E(44-55) = 1.00} \\ \cline{3-7} \cline{10-14}
$  7000$K &  & $ 0.978$ & $ 0.728$ & $ 0.729$ & $-0.309$ & $ 0.155$ & & & $ 0.979$ & $ 0.743$ & $ 0.729$ & $-0.310$ & $ 0.153$ \\
$  8000$K &  & $ 0.983$ & $ 0.690$ & $ 0.725$ & $-0.310$ & $ 0.158$ & & & $ 0.983$ & $ 0.705$ & $ 0.726$ & $-0.312$ & $ 0.155$ \\
$  9000$K &  & $ 0.990$ & $ 0.664$ & $ 0.720$ & $-0.309$ & $ 0.158$ & & & $ 0.990$ & $ 0.678$ & $ 0.721$ & $-0.311$ & $ 0.155$ \\
$ 10000$K &  & $ 0.997$ & $ 0.654$ & $ 0.715$ & $-0.307$ & $ 0.156$ & & & $ 0.997$ & $ 0.668$ & $ 0.716$ & $-0.309$ & $ 0.154$ \\
$ 12500$K &  & $ 1.004$ & $ 0.667$ & $ 0.709$ & $-0.305$ & $ 0.158$ & & & $ 1.004$ & $ 0.681$ & $ 0.710$ & $-0.306$ & $ 0.155$ \\
$ 15000$K &  & $ 1.008$ & $ 0.679$ & $ 0.707$ & $-0.304$ & $ 0.159$ & & & $ 1.008$ & $ 0.692$ & $ 0.708$ & $-0.305$ & $ 0.157$ \\
$ 20000$K &  & $ 1.012$ & $ 0.688$ & $ 0.704$ & $-0.303$ & $ 0.161$ & & & $ 1.011$ & $ 0.702$ & $ 0.705$ & $-0.304$ & $ 0.159$ \\
$ 25000$K &  & $ 1.014$ & $ 0.694$ & $ 0.702$ & $-0.301$ & $ 0.161$ & & & $ 1.014$ & $ 0.708$ & $ 0.703$ & $-0.303$ & $ 0.159$ \\
$ 30000$K &  & $ 1.016$ & $ 0.695$ & $ 0.700$ & $-0.301$ & $ 0.161$ & & & $ 1.016$ & $ 0.709$ & $ 0.701$ & $-0.302$ & $ 0.159$ \\
$ 35000$K &  & $ 1.017$ & $ 0.697$ & $ 0.700$ & $-0.300$ & $ 0.161$ & & & $ 1.017$ & $ 0.711$ & $ 0.701$ & $-0.301$ & $ 0.159$ \\
\enddata
\end{deluxetable*}

\begin{deluxetable*}{rccccccccccc}
\label{tabugriz}
\tablenum{5}
\tabletypesize{\scriptsize}
\tablenum{5} 
\tablewidth{0pc} 
\tablecaption{ugriz Extinction Indices}
\tablehead{ 
\colhead{$T_{eff}$}                     & 
\colhead{}                              &
\colhead{$\frac{E(u-g)}{E(44-55)}$}     &
\colhead{$\frac{E(g-r)}{E(44-55)}$}     &
\colhead{$\frac{E(r-i)}{E(44-55)}$}     &
\colhead{$\frac{E(i-z)}{E(44-55)}$}     &
\colhead{}                              &
\colhead{}                              &
\colhead{$\frac{E(u-g)}{E(44-55)}$}     &
\colhead{$\frac{E(g-r)}{E(44-55)}$}     &
\colhead{$\frac{E(r-i)}{E(44-55)}$}     &
\colhead{$\frac{E(i-z)}{E(44-55)}$}     }
\startdata
& & \multicolumn{10}{c}{$R(V) = 2.5$} \\ \cline{3-12}
& & \multicolumn{4}{c}{E(44-55) = 0.50} & & & \multicolumn{4}{c}{E(44-55) = 1.00} \\ \cline{3-6} \cline{9-12}
$  7000$K &  & $ 1.101$ & $ 1.000$ & $ 0.562$ & $ 0.433$ & & & $ 1.114$ & $ 0.979$ & $ 0.560$ & $ 0.433$ \\
$  8000$K &  & $ 1.072$ & $ 1.014$ & $ 0.565$ & $ 0.436$ & & & $ 1.085$ & $ 0.992$ & $ 0.563$ & $ 0.436$ \\
$  9000$K &  & $ 1.054$ & $ 1.028$ & $ 0.567$ & $ 0.438$ & & & $ 1.067$ & $ 1.006$ & $ 0.565$ & $ 0.438$ \\
$ 10000$K &  & $ 1.049$ & $ 1.036$ & $ 0.567$ & $ 0.438$ & & & $ 1.063$ & $ 1.014$ & $ 0.565$ & $ 0.438$ \\
$ 12500$K &  & $ 1.068$ & $ 1.045$ & $ 0.567$ & $ 0.437$ & & & $ 1.082$ & $ 1.022$ & $ 0.565$ & $ 0.436$ \\
$ 15000$K &  & $ 1.082$ & $ 1.049$ & $ 0.567$ & $ 0.435$ & & & $ 1.096$ & $ 1.026$ & $ 0.565$ & $ 0.435$ \\
$ 20000$K &  & $ 1.094$ & $ 1.055$ & $ 0.567$ & $ 0.434$ & & & $ 1.108$ & $ 1.032$ & $ 0.565$ & $ 0.434$ \\
$ 25000$K &  & $ 1.101$ & $ 1.058$ & $ 0.567$ & $ 0.434$ & & & $ 1.116$ & $ 1.035$ & $ 0.565$ & $ 0.433$ \\
$ 30000$K &  & $ 1.104$ & $ 1.062$ & $ 0.567$ & $ 0.433$ & & & $ 1.118$ & $ 1.038$ & $ 0.565$ & $ 0.433$ \\
$ 35000$K &  & $ 1.106$ & $ 1.065$ & $ 0.566$ & $ 0.433$ & & & $ 1.121$ & $ 1.041$ & $ 0.564$ & $ 0.433$ \\
& & \multicolumn{10}{c}{$R(V) = 3.1$} \\ \cline{3-12}
& & \multicolumn{4}{c}{E(44-55) = 0.50} & & & \multicolumn{4}{c}{E(44-55) = 1.00} \\ \cline{3-6} \cline{9-12}
$  7000$K &  & $ 1.087$ & $ 1.037$ & $ 0.622$ & $ 0.502$ & & & $ 1.101$ & $ 1.018$ & $ 0.619$ & $ 0.503$ \\
$  8000$K &  & $ 1.059$ & $ 1.051$ & $ 0.625$ & $ 0.506$ & & & $ 1.073$ & $ 1.031$ & $ 0.623$ & $ 0.506$ \\
$  9000$K &  & $ 1.041$ & $ 1.064$ & $ 0.627$ & $ 0.508$ & & & $ 1.055$ & $ 1.044$ & $ 0.624$ & $ 0.508$ \\
$ 10000$K &  & $ 1.036$ & $ 1.072$ & $ 0.627$ & $ 0.508$ & & & $ 1.050$ & $ 1.051$ & $ 0.625$ & $ 0.508$ \\
$ 12500$K &  & $ 1.053$ & $ 1.081$ & $ 0.627$ & $ 0.506$ & & & $ 1.068$ & $ 1.060$ & $ 0.624$ & $ 0.506$ \\
$ 15000$K &  & $ 1.067$ & $ 1.085$ & $ 0.627$ & $ 0.504$ & & & $ 1.082$ & $ 1.064$ & $ 0.624$ & $ 0.504$ \\
$ 20000$K &  & $ 1.078$ & $ 1.091$ & $ 0.627$ & $ 0.503$ & & & $ 1.094$ & $ 1.069$ & $ 0.624$ & $ 0.503$ \\
$ 25000$K &  & $ 1.085$ & $ 1.094$ & $ 0.627$ & $ 0.502$ & & & $ 1.100$ & $ 1.072$ & $ 0.625$ & $ 0.502$ \\
$ 30000$K &  & $ 1.087$ & $ 1.097$ & $ 0.627$ & $ 0.502$ & & & $ 1.103$ & $ 1.075$ & $ 0.625$ & $ 0.502$ \\
$ 35000$K &  & $ 1.090$ & $ 1.100$ & $ 0.626$ & $ 0.502$ & & & $ 1.105$ & $ 1.078$ & $ 0.623$ & $ 0.502$ \\
& & \multicolumn{10}{c}{$R(V) = 4.5$} \\ \cline{3-12}
& & \multicolumn{4}{c}{E(44-55) = 0.50} & & & \multicolumn{4}{c}{E(44-55) = 1.00} \\ \cline{3-6} \cline{9-12}
$  7000$K &  & $ 1.055$ & $ 1.125$ & $ 0.760$ & $ 0.664$ & & & $ 1.071$ & $ 1.110$ & $ 0.758$ & $ 0.666$ \\
$  8000$K &  & $ 1.028$ & $ 1.137$ & $ 0.764$ & $ 0.668$ & & & $ 1.043$ & $ 1.121$ & $ 0.761$ & $ 0.670$ \\
$  9000$K &  & $ 1.010$ & $ 1.149$ & $ 0.766$ & $ 0.671$ & & & $ 1.026$ & $ 1.133$ & $ 0.763$ & $ 0.673$ \\
$ 10000$K &  & $ 1.005$ & $ 1.157$ & $ 0.766$ & $ 0.671$ & & & $ 1.021$ & $ 1.141$ & $ 0.763$ & $ 0.673$ \\
$ 12500$K &  & $ 1.020$ & $ 1.165$ & $ 0.766$ & $ 0.668$ & & & $ 1.036$ & $ 1.149$ & $ 0.763$ & $ 0.670$ \\
$ 15000$K &  & $ 1.032$ & $ 1.169$ & $ 0.766$ & $ 0.666$ & & & $ 1.049$ & $ 1.152$ & $ 0.763$ & $ 0.667$ \\
$ 20000$K &  & $ 1.042$ & $ 1.174$ & $ 0.766$ & $ 0.664$ & & & $ 1.059$ & $ 1.157$ & $ 0.763$ & $ 0.665$ \\
$ 25000$K &  & $ 1.048$ & $ 1.177$ & $ 0.766$ & $ 0.662$ & & & $ 1.065$ & $ 1.160$ & $ 0.763$ & $ 0.664$ \\
$ 30000$K &  & $ 1.049$ & $ 1.180$ & $ 0.766$ & $ 0.661$ & & & $ 1.067$ & $ 1.163$ & $ 0.763$ & $ 0.663$ \\
$ 35000$K &  & $ 1.051$ & $ 1.184$ & $ 0.764$ & $ 0.662$ & & & $ 1.069$ & $ 1.166$ & $ 0.761$ & $ 0.663$ \\
& & \multicolumn{10}{c}{$R(V) = 6.0$} \\ \cline{3-12}
& & \multicolumn{4}{c}{E(44-55) = 0.50} & & & \multicolumn{4}{c}{E(44-55) = 1.00} \\ \cline{3-6} \cline{9-12}
$  7000$K &  & $ 1.021$ & $ 1.220$ & $ 0.909$ & $ 0.838$ & & & $ 1.038$ & $ 1.211$ & $ 0.906$ & $ 0.843$ \\
$  8000$K &  & $ 0.994$ & $ 1.230$ & $ 0.913$ & $ 0.844$ & & & $ 1.011$ & $ 1.220$ & $ 0.910$ & $ 0.849$ \\
$  9000$K &  & $ 0.977$ & $ 1.241$ & $ 0.915$ & $ 0.847$ & & & $ 0.994$ & $ 1.231$ & $ 0.912$ & $ 0.852$ \\
$ 10000$K &  & $ 0.971$ & $ 1.248$ & $ 0.916$ & $ 0.846$ & & & $ 0.989$ & $ 1.238$ & $ 0.913$ & $ 0.851$ \\
$ 12500$K &  & $ 0.984$ & $ 1.256$ & $ 0.915$ & $ 0.842$ & & & $ 1.002$ & $ 1.245$ & $ 0.912$ & $ 0.847$ \\
$ 15000$K &  & $ 0.995$ & $ 1.260$ & $ 0.915$ & $ 0.839$ & & & $ 1.013$ & $ 1.249$ & $ 0.912$ & $ 0.844$ \\
$ 20000$K &  & $ 1.003$ & $ 1.265$ & $ 0.915$ & $ 0.837$ & & & $ 1.021$ & $ 1.253$ & $ 0.912$ & $ 0.841$ \\
$ 25000$K &  & $ 1.007$ & $ 1.267$ & $ 0.915$ & $ 0.835$ & & & $ 1.026$ & $ 1.256$ & $ 0.912$ & $ 0.839$ \\
$ 30000$K &  & $ 1.008$ & $ 1.270$ & $ 0.915$ & $ 0.834$ & & & $ 1.027$ & $ 1.259$ & $ 0.912$ & $ 0.838$ \\
$ 35000$K &  & $ 1.010$ & $ 1.274$ & $ 0.913$ & $ 0.834$ & & & $ 1.029$ & $ 1.262$ & $ 0.910$ & $ 0.839$ \\
\enddata
\end{deluxetable*}

To illustrate and quantify the effects of interstellar extinction on optical SEDs, we have computed the 
photometric reddening slopes predicted by our new optical extinction curve for three common photometric 
systems: Johnson \ubv, Str\"omgren \uvby, and Sloan Digital Sky Survey (SDSS) \ugriz.  The slopes were 
determined by performing synthetic photometry on intrinsic and reddened stellar energy distributions, 
using the {\it ATLAS9} atmosphere models, and then forming the various differences and ratios.  The 
results are presented in Tables \ref{tabUBVuvby} and \ref{tabugriz}.  The measurements were made for 
two values of reddening [\efourfive\/ =  0.50 and 1.0], four values of $R(V)$ (2.5, 3.1, 4.5, and 6.0), 
and ten values of \teff\/ (from 7000 K to 35000 K).  The synthetic Johnson and Str\"omgren photometry 
was performed as described by \citet{Fitzpatrick05b}.  We modified the calibration of the synthetic 
values, however, to take advantage of the new \hst\/ G430L and G750L observations obtained for this program. 
In brief, we computed synthetic, uncalibrated Johnson and Str\"omgren photometric indices ($B-V$, $U-B$, 
$b-y$, $m_1$ and $c_1$) using the \hst\/ spectrophotometry for each of our 72 reddened stars, plus three 
additional unreddened stars HD~38666, HD~34816, and HD~214680 from the \hst\/ archives.  We then 
determined the linear transformation between these synthetic values and the observed photometric indices.  
The results are similar to those reported in \citet{Fitzpatrick05b} and the scatter of the observed 
values about the synthetic results is consistent with the expected observational error.  The synthetic 
\ugriz\/ photometry was performed as described in \citet{Casagrande14}, with filter profiles 
from \citet{Doi10}.

The \ebv/\efourfive\/ column in Table \ref{tabUBVuvby} illustrates the uncertainty in measuring the 
``amount'' of extinction using broadband photometry.  For a fixed extinction [as measured by \efourfive], 
the observed values of \ebv\/ can vary significantly depending on the stellar \teff, reflecting variations 
in the effective wavelengths of the broad $B$ and $V$ filters.  For the same reason, the values of 
\ebv/\efourfive\/ also depend on the overall level of extinction.  The Johnson and Str\"omgren reddening 
slopes implied by the $R(V)=3.1$ curve are -- with the exception of 
{\it E(c$_1$)}/\ebv\/ -- consistent with the accepted values \citep[see][]{Fitzpatrick99review}, 
including the \ebv-dependence 
of \eub/\ebv\/ \citep[e.g.,][]{Fitzgerald70}.  The accepted {\it E(c$_1$)}/\ebv\/ ratio ($\simeq 0.20$) is from 
\citet{Crawford75} and based on an ``eye fit'' to photometry of $\sim$50 O-type stars.  However, it is clear 
from Crawford's Figure 4 that a steeper ratio is needed to explain the most heavily reddened O stars in 
the sample.  In addition, Crawford notes that a preliminary study of B-type stars yields a larger value 
($\simeq 0.24$).  We suggest that the results in Table \ref{tabUBVuvby} present the most reliable estimates 
for the Str\"omgren reddening slopes.  Note also that, while the intermediate-bandwidth Str\"omgren slopes 
are relatively insensitive to effective wavelength shifts [i.e., minimal dependence on \teff\/ or 
\efourfive], the $E(c_1)$ ratio is strongly dependent on $R(V)$ due to the short wavelength $u$ filter.  
The same sensitivity is also seen in the \eub\/ slope, potentially allowing both indices to be used as 
diagnostics of $R(V)$.

Qualitatively similar effects, in terms of \teff\/ and \efourfive\/ sensitivity, can be seen in the 
\ugriz\/ results in Table \ref{tabugriz}.  Note that the reddening slopes involving the longer wavelength 
{\it riz} filters ($\lambda_{\rm eff}$ = 6231, 7625 and 9134~\AA, respectively) are particularly 
sensitive to the values of $R(V)$.  Given that there is less curve-to-curve scatter relative to the 
$R$-correlation at these longer wavelengths (see Figure \ref{figTABFORM}), as compared to that for the 
Str\"omgren $u$ or Johnson $U$ filters, these filters could provide a very strong diagnostic of $R(V)$.

\subsection{Balmer Decrement}\label{secBALMER}

The formation rate of massive stars in external galaxies can be inferred 
from the luminosities of emission lines -- notably, the Balmer lines -- arising in the galaxies' H II 
regions \citep[e.g.,][]{Kennicutt83,Gallagher84}.  This requires that the line emission first be corrected 
for the attenuating effects of interstellar extinction.  A common technique is to compare the observed 
luminosity ratio of two lines, e.g., $H\alpha/H\beta$, to the expected intrinsic ratio and interpret any 
difference as a result of the wavelength-dependent effects of reddening.  This is the ``Balmer Decrement'' 
method and allows an estimate of \ebv, which further allows an estimate of the total extinction at the 
line wavelengths \citep[e.g.,][]{Calzetti94}. This process requires an understanding of the intrinsic 
emission mechanism and, apropos to this paper, a well-determined interstellar extinction law.

Using the results in Table \ref{tabAVERAGE} and following the derivation in, for example, \S 3 
of \citet[][]{Dominguez13}, we can 
compute the implied reddening based on any pair of lines for which the intrinsic ratio is known. Two examples are:  
\begin{equation}
\scriptsize
E(44-55) = \frac{2.5}{1.16+0.09\times(R_{55}-3.02)}\times  \rm{log}_{10}\left[\frac{(H\alpha/H\beta)_{obs}}{(H\alpha/H\beta)_{int}}\right]
\label{eqHalpha}
\end{equation}
and 
\begin{equation}
\scriptsize
E(44-55) =-5.43\times \rm{log}_{10}\left[\frac{(H\gamma/H\beta)_{obs}}{(H\gamma/H\beta)_{int}}\right]   \:  .
\label{eqHgamma}
\end{equation}
In both cases, the subscripts ``obs'' and ``int'' refer to the observed line ratios and the expected 
intrinsic ratios, respectively.  
Equations (\ref{eqHalpha}) and (\ref{eqHgamma}) show that the observed ratio is slightly sensitive to the 
value of $R$ for the H$\alpha$ case, and not at all for the H$\gamma$ case.  The overall weakness of 
the $R$-dependence is because the emission lines lie within or 
close to the region where the curves are normalized and, therefore, highly constrained.

For the case of the average Milky Way value of $R(55)$ = 3.02 (i.e., $R(V) = 3.10$), the constant term in 
Equation (\ref{eqHalpha}) becomes 
2.15. To find \ebv, this value is reduced by a factor of $\sim$0.99 (see Table \ref{tabUBVuvby}), yielding 
a scale factor of 2.13. This 
can be compared with the value of 1.97 as computed by \citet{Dominguez13} using the attenuation law 
of \citet{Calzetti00}, which characterizes regions of starburst activity.  Given the difference in the 
environments involved (Milky Way average vs. starburst), this level of disagreement might be 
considered unsurprising.  However, a closer look shows that it actually reveals a significant problem; namely
that the \citet{Calzetti00} law is not rigorously normalized to the $BV$ system.  This is a common 
problem among most current representations of optical extinction and will be graphically illustrated below 
in \S5.1.  In brief, the \citet{Calzetti00} law is too steep in the optical and produces \ebv\/ values about 
6\% larger than intended.  For example, if the curve were scaled to \ebv\/ = 1.0 and applied to an SED, 
synthetic photometry would reveal that the $(B-V)$ color of the SED actually increased by about 1.06 mag.  
A simple rescaling of the law in the optical, to force consistency with the intended normalization, 
eliminates most of the difference with our result for the H$\alpha$/H$\beta$ ratio.  

For emission lines near the normalization region (4400--5500~\AA), the results for the average Milky Way 
curve, e.g., in Equations 
(\ref{eqHalpha}) and (\ref{eqHgamma}), are likely to be universally applicable.  For emission lines 
shortward of the normalization region, 
however, the Decrement method critically depends on the detailed nature of the extinction law (i.e., 
large-$R$, small-$R$, Milky Way 
average, starburst, etc.), since it is here that the wide range in normalized extinction curve shapes 
becomes evident.

\section{Discussion}
\subsection{Comparison to Previous Work}\label{secCOMPARE}

\begin{figure*}
\figurenum{9}
\epsscale{1.1}
\plottwo{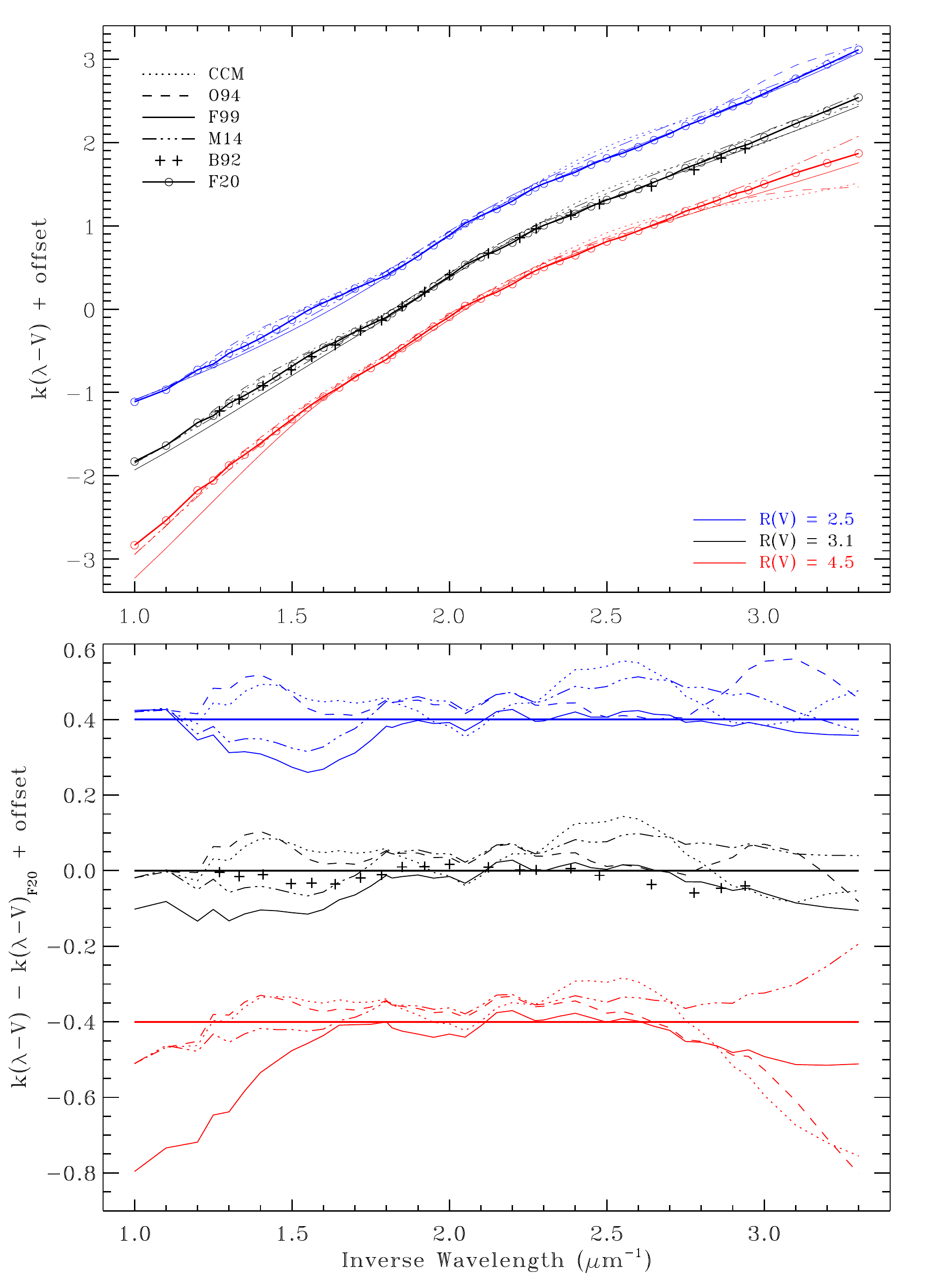}{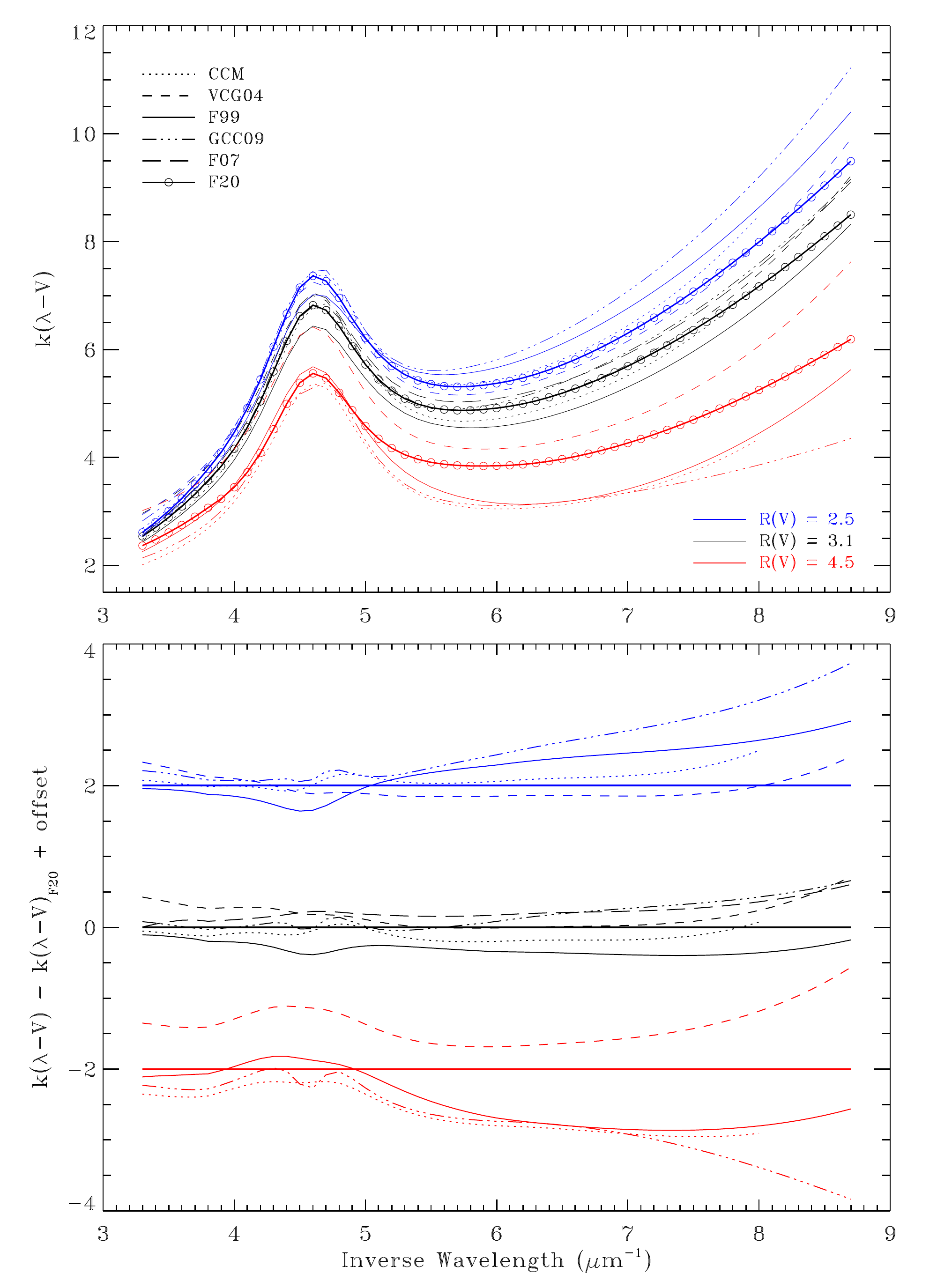}
\caption{A comparison between our results and previously published estimates of extinction in 
the optical/NIR (left side) and UV (right side).  The extinction curves themselves are shown in the upper 
half of each panel. The difference between these curves and our new results are shown in the lower half 
of each panel.  Results for several values of $R(V)$ are presented, and identified by different colors 
(see the lower right portion of each upper panel).  
The various published studies are distinguished by differing line styles and symbols, as indicated in the 
top left portion of each upper panel.  Vertical offsets are utilized to increase clarity.  The key to 
the various studies, in alphabetical order, is:
B92 = \citealt{Bastiaansen92}; CCM = \citealt{Cardelli89}; F07 = \citealt{Fitzpatrick07}; F20 = this paper; 
F99 = \citealt{Fitzpatrick99review}; GCC09 = \citealt{Gordon09FUSE}; M14 = \citealt{MaizApellaniz14}; 
O94 = \citealt{ODonnell94}; VCG04 = \citealt{Valencic04}.}
\label{figCOMPARE}
\end{figure*}

Our new results are compared with past work in Figure \ref{figCOMPARE}.  The left and right sides of the figure 
show the optical/NIR and UV spectral regions, respectively. The top panels on both sides show the extinction curves 
themselves, while the bottom panels highlight the differences between our results and those of the other investigations. 
Vertical offsets are used for clarity. Three representative values of $R(V)$ are illustrated (distinguished by color) for 
most of the comparison studies, which -- with the exception of \citet[][``B92'']{Bastiaansen92} in the optical and
\citet[][``F07'']{Fitzpatrick07} in the UV -- all present families of R-dependent curves.  \citetalias{Bastiaansen92}'s and 
\citetalias{Fitzpatrick07}'s results are explicitly measures 
of mean diffuse medium extinction properties and are grouped with the $R(V)=3.1$ curves. The native units for all the comparison studies 
are \klam, and so we converted our \kfive\/ curves to \klam\/ for this comparison, using Equation \ref{eq:klamtransform}.

In the optical/NIR, the figure shows that the agreement among the various curves is at the level of $\sim$0.1 within and slightly 
beyond the normalization region (i.e., $\sim$1.6--2.7~\invmic).  The work of \citetalias{Bastiaansen92} shows the best agreement with 
our results (for the case $R(V) = 3.1$).  \citetalias{Bastiaansen92}'s curve was produced using narrowband multi-color photometry, yielding a low-resolution 
estimate of the true wavelength dependence of the optical curve.  The other studies, with the partial exception 
of \citet[][``F99'']{Fitzpatrick99review}, were not intended to yield detailed representations 
of the optical extinction law.  Rather, they 
are generally ``connect-the-dots'' curves guided by the extinction ratios and effective wavelengths from broadband photometry. As such,
they only crudely represent the actual shape of the curve, leading to the sizable deviations
seen even within the normalization region.  In general, these curves do not yield the 
intended values of \ebv, $R(V)$ and other photometric extinction indices when applied to stellar SEDs, as already noted in the 
previous section for the \citet{Calzetti00} starburst extinction curve. (See the discussion in \citetalias{Fitzpatrick99review}).  

The discrepancies between the comparison curves and ours are
larger outside the normalization region.  None of the earlier studies utilize measurements in the gap 
between the optical and UV, and the curves  in these regions are extrapolations or interpolations between 
fundamentally different datasets.  Note the large discrepancy in the \citetalias{Fitzpatrick99review} curve 
at long wavelengths and large $R(V)$. \citetalias{Fitzpatrick99review} used a simple interpolation between the optical (at $\sim$6000~\AA) and 
the NIR $JHK$ region and this clearly does not follow the true shape of the curve as revealed by the new spectrophotometry.

The righthand panel of Figure \ref{figCOMPARE} shows good agreement among all the studies on the properties of 
mean Milky Way extinction in the UV (i.e., for 
the case $R(V) = 3.1$).  This is not surprising since there is considerable overlap in the sightlines used 
in the various studies, all having been drawn 
from the database of \iue\/ satellite observations. The differences that do exist arise from a number of 
potential sources, including (1) the specific subset of sightlines used, (2) the use of updated photometry (particularly 
the \tmass\/ database), 
(3) differing techniques used to produce the curves (e.g., pair method vs. extinction-without-standards), 
and (4) the technique used for deriving the mean curve (e.g., via the $R(V)$-dependence or from a simple average).  
The differences among the various studies are much 
larger as $R(V)$ departs from the average Milky Way value.  In the most general terms, our results show a weaker 
overall $R(V)$-dependence than most of the studies illustrated in Figure \ref{figCOMPARE}.  Given that the 
$R(V)$-dependences are strongly driven by a relatively small number of sightlines at extreme $R(V)$, the specific 
sample selections are likely to have a major effect on the results. The other sources of uncertainty listed above 
for the mean curve are also probable contributors.  

To address the sensitivity of our results to the sample selection, we reexamined the 328 star sample from \citetalias{Fitzpatrick07}. They 
derived a mean extinction curve from a simple average of all curves with $2.4 \leq R(V) \leq 3.6$, but did not 
quantify the $R(V)$-dependence.  We used the same methodology as described in \S \ref{secAVERAGE} to measure the UV 
$R(V)$-dependence of the 260 stars from \citetalias{Fitzpatrick07} for which $\ebv > 0.30$.  The results are shown by the blue 
dotted curves in the three panels of Figure \ref{figMEANCURVE}, converted to the \kfive\/ normalization. As can be seen, the 
\citetalias{Fitzpatrick07} sample yielded very similar results to those derived here. Specifically, the $R(55) = 3.02$ curve and the sample 
standard deviation curve are nearly identical to the current results, and virtually indistinguishable in Figure \ref{figMEANCURVE} (top and 
bottom panels), while the $R(V)$-dependence 
is similar in shape and differs by only $\sim$10\%  from the current result (middle panel of 
Figure \ref{figMEANCURVE}).  We conclude that the agreement between \citetalias{Fitzpatrick07} and our much smaller current sample indicates that the new 
results are not subject to a strong sample bias and are generally applicable -- at least to the extent that the current body of UV 
Milky Way extinction measurements can be considered representative of the Milky Way.  

\subsection{Variations beyond $R$-dependence}

Our focus in this paper has been on the trends in the shapes of interstellar extinction curves that can be 
related to changes in $R$.  While this accounts for the broadest trends seen among extinction curves 
(see Figure \ref{figRCURVES}), it by no means reproduces all the observed sightline-to-sightline variations. The 
$\sigma [k(\lambda-55)]$ curve in Figure \ref{figMEANCURVE} reveals that there is significant scatter around the mean $R$-dependence, 
which is typically much larger than the individual curve uncertainties (with the likely exception of the longest 
wavelength G750L observations).  This cosmic scatter reveals specific variations among the grain populations that are not 
distinguished (or distinguishable) by $R$.  Notably, the strength of the 2175 \AA\/ bump and the level of the curve at the
shortest UV wavelengths are only lightly tied to $R$. The level of the far-UV curve is particularly interesting.  There are several 
sightlines (e.g., towards HD~204827, HD~210072, and HD~210121) that are well known for having very steep UV 
extinction curves, seemingly accompanied by weaker-than-average 2175 \AA\/ bumps 
(see \citetalias{Fitzpatrick07}).  These are {\it not} reproduced by our $R$-dependent curves.  In fact, these sightlines are among those
that were ``clipped'' from the $R(55)$-dependence because they do not follow the general trend, as can be seen in the bottom 
panels of Figure \ref{figSLOPES}.  These relatively rare, steep far-UV/weak bump sightlines appear to be associated in general 
with low values of $R$, but not in an easily predictable way.  They may represent a separate $R$-dependent sequence of curves 
or perhaps demonstrate an extreme sensitivity of the curve to grain population details in environments where a generally small value 
of $R$ pertains. In either case, a detailed study of the grain environments along such sightlines may be necessary to understand 
the formation of such curves.  Extinction studies in external galaxies, such as the Small Magellanic Cloud, where steep/weak-bumped 
curves appear to be the norm \citep[e.g.,][]{Gordon03SMC}, may also help shed light on the relationship between grain population characteristics 
and extinction curve properties.

An additional level of curve variability outside the scope of this paper is seen in the fine structure revealed by our measurements of 
optical extinction. As noted in \S \ref{secCURVES} and illustrated in the curves of Figure 4, the optical region contains distinct extinction 
features known as the Very Broad Structure.  A detailed examination of these features is presented
by \citet[][]{Massa20}. This study reveals significant sightline-to-sightline variations among these features, 
which is correlated with other aspects of the extinction curve -- but not with $R$.  See \citet[][]{Massa20} for the complete discussion. 

\section{Summary}

We have produced the first spectrophotometrically-derived measurement of the average Milky Way 
interstellar extinction curve over the wavelength range 1150-to-10000~\AA\/ and determined its general 
dependence on the ratio of total-to-selective extinction in the optical, $R$, over the range $2.5 \leq R \leq 6.0$.  
This curve resolves structures in the optical spectral region down to a level of $\sim$100-200~\AA\/ and fully 
resolves all structure in the UV region.

The curve is based on a unique, homogeneous dataset consisting of \iue\/ and 
\hst/STIS spectrophotometry of normal, early-type stars. These data were all calibrated in 
a mutually consistent way and fit to model atmospheres to produce a uniform, 
self-consistent set of extinction curves which span 1150-to-10000~\AA\/ at high 
resolution and extended to 2.2~\mic\/ with broad band $JHK$ photometry.  
These curves provide the first systematic measurements of extinction in the transition 
region between past ground-based studies and 
space-based UV measurements.  The use of optical spectrophotometry allowed us to eliminate the uncertainties 
introduced by normalizing extinction curves by broadband optical photometry and we adopted a
monochromatic normalization scheme, based on the extinction at 4400 and 5500~\AA.  
The new curves were used to evaluate the dependence of extinction on $R$ over the full wavelength range and we ultimately 
parametrized this dependence as a simple linear function of $R$, essentially a first-order 
Taylor expansion. The Average Milky Way Extinction Curve we present is that which corresponds to the case $R(V) = 3.10$.  
We present this curve in tabular form, along with the linear $R$ dependence and a measure of the curve-to-curve variations 
about this mean dependence. 

Because the coverage between 
1150 and 10000~\AA\/ is free of gaps and at relatively high resolution, it 
is possible to use the tabular data to derive the behavior of extinction for 
any given photometric system, and several examples are presented.  The new, high resolution curves also allow the study of extinction features on intermediate 
wavelength scales in the optical.  While such features have been seen before, 
the new curves show them with considerably more detail and higher signal-to-noise 
than in the past.  The exact shape and sightline-to-sightline variability of this structure is discussed more thoroughly by \cite{Massa20}.

\acknowledgments
E.F. and D.M. acknowledge support by Grant \# HST-GO-13760 provided by NASA through the Space 
Telescope Science Institute, which is operated by the Association of 
Universities for Research in Astronomy, Inc., under NASA contract NAS5-26555.  Some of the data 
presented in this paper were obtained from the Mikulski Archive for Space 
Telescopes (MAST). STScI is operated by the Association of Universities for Research in 
Astronomy, Inc., under NASA contract NAS5-26555.

\bibliographystyle{aasjournal}
\bibliography{optextI}

\appendix

\section{STIS Data Reduction}
\label{appendix_stis}

The G430L and G750L spectra are processed with the Instrument Definition Team
(IDT) pipeline software written in IDL by D. J. Lindler in 1996--1997. This IDL
data reduction is more flexible and offers the following advantages over the
STScI pipeline results that are available from the
MAST\footnote{http://archive.stsci.edu/hst/search.php} archive:

\begin{enumerate}
\item Small shifts in pointing or instrumental flexure between the images of a
cosmic ray split (CR-split) observation can often cause an erroneous reduction
of net signal, because pixels with the lower signal are retained and the higher
pixel value is rejected in the case that the CR-split image pair is not
perfectly aligned and pixel pair values differ by more than their statistical
noise estimates. Figure~\ref{crej} shows a standard display of the IDT
processing, which makes obvious the deficiency in the reduction with the default
of \textit{mult\_noise=0} parameter for combining the CR-split pair of images
that have a small shift. This excess of false CR flags is completely
eliminated by a relaxed \textit{mult\_noise=0.04} value, which does not
significantly increase the number of unflagged actual cosmic-ray hits because of
our relatively short exposure times. Figure~\ref{crerr} shows the error in
extracted spectral signal for \textit{mult\_noise=0} in black and the
corresponding error in the STScI pipeline error of the ocmv0j010\_sx1.fits
reduction in red. Errors in spectral shape of 2\% are common and have two
narrower band error features in the STScI standard product. For our observed 98
SNAP program stars, 12 G430L and 7 G750L observations required relaxed values
for \textit{mult\_noise} in order to extract the proper spectral signal. For slightly 
saturated STIS spectral observations, the CR-reject algorithm also
fails with a display similar to Figure~\ref{crej}. However, changing
\textit{mult\_noise} has little effect. Because saturated data at Gain=4 do
not lose charge and the total signal is still present in the raw images, the
solution for extracting the proper spectrum is to omit any CR-reject attempts
and average the extracted signal from the pair of *flt.fits images. This
approach works well in the common case of a bright star and short exposures
where cosmic ray hits are minimal.

\begin{figure}
\figurenum{10}
\epsscale{0.75}
\plotone{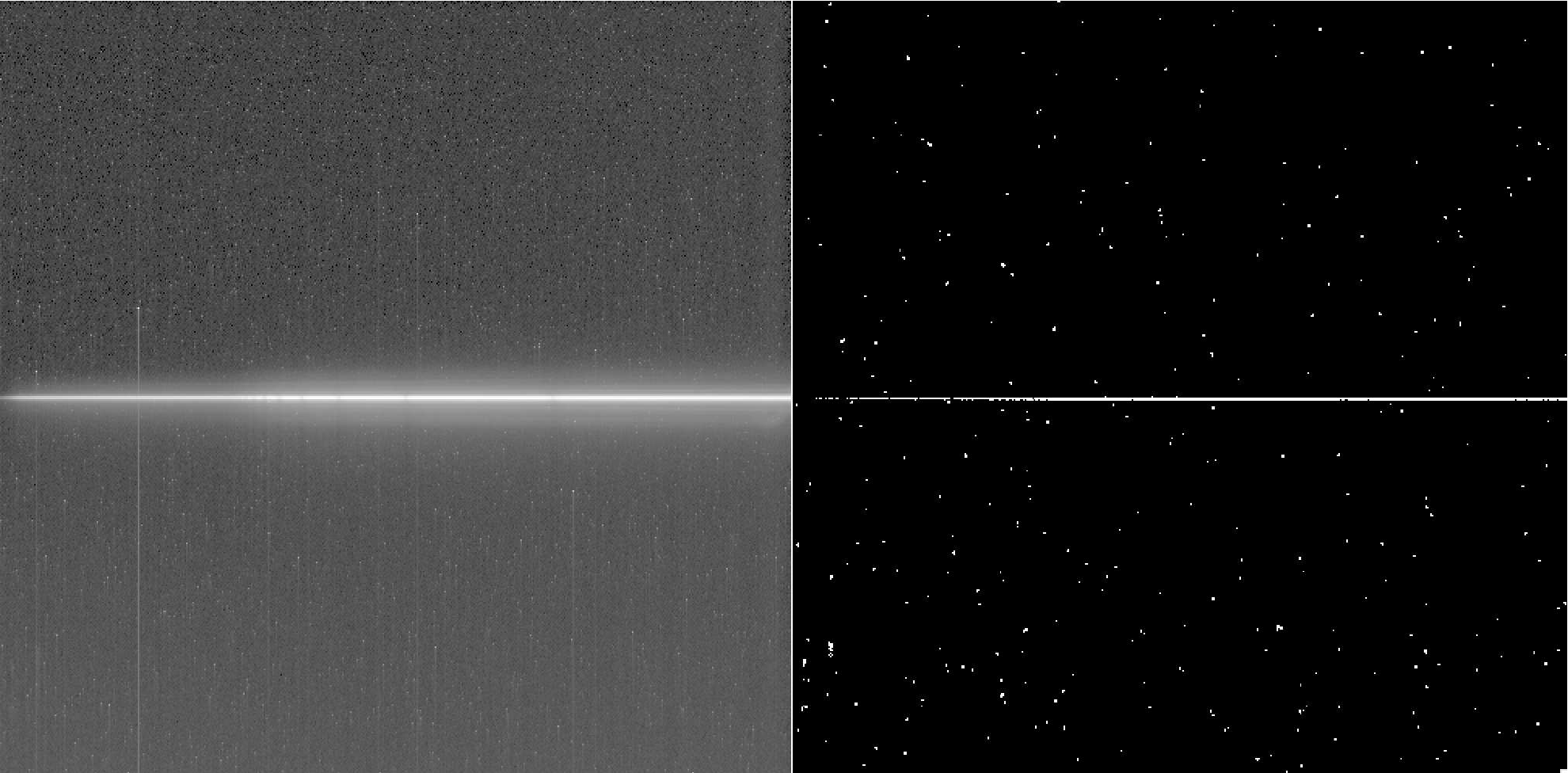}
\caption{1024x1024 pixel STIS CCD image of the G430L CR-split=2 observation of HD146285
(ocmv0j010) on the left and the corresponding image on the right, which flags the 
location of each falsely detected cosmic ray hit for the default \textit{mult\_noise=0}. 
Because of a pointing shift between the pair of CR-split images, there is an easily 
recognized excess of flagged pixels at the location of the spectrum. With a relaxed 
value of \textit{mult\_noise=0.04},  
there is no excess of CR-flagged pixels at the spectral location and the correct
combined image of the CR-split pair is produced.
\label{crej}}
\end{figure}

\begin{figure}
\figurenum{11}
\epsscale{0.75}
\plotone{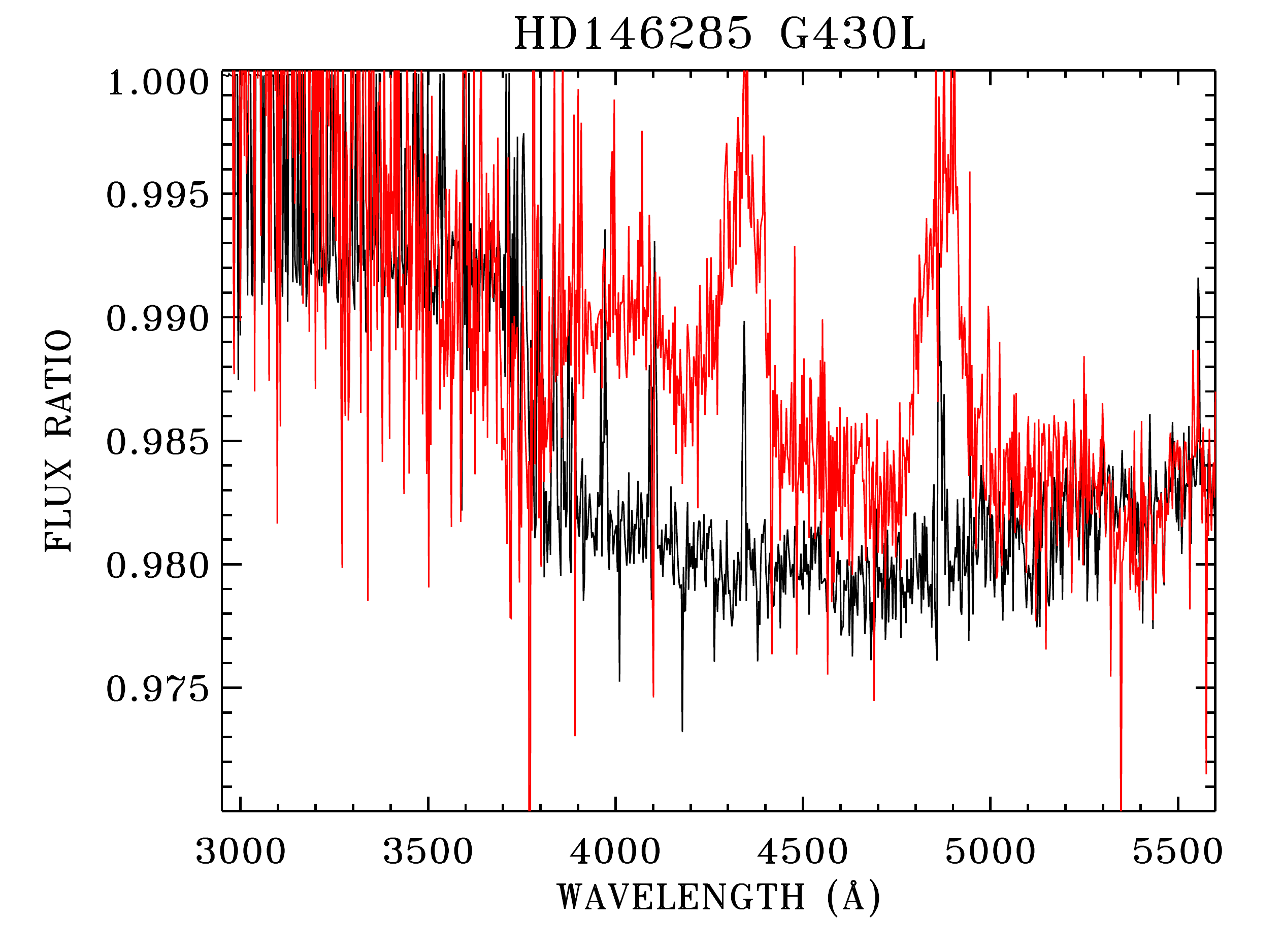}
\caption{Ratio (black) of extracted spectra for \textit{mult\_noise=0} to 
\textit{mult\_noise=0.04} for the CR-split=2 observation ocmv0j010 and the ratio (red) of 
the STScI pipeline reduction ocmv0j010\_sx1.fits to the same denominator with the IDL 
parameter \textit{mult\_noise=0.04}. Errors in spectral shape of up to $\sim$2\% exist 
for the two incorrect spectral extractions.
\label{crerr}}
\end{figure}

\begin{figure}
\figurenum{12}
\epsscale{0.75}
\plotone{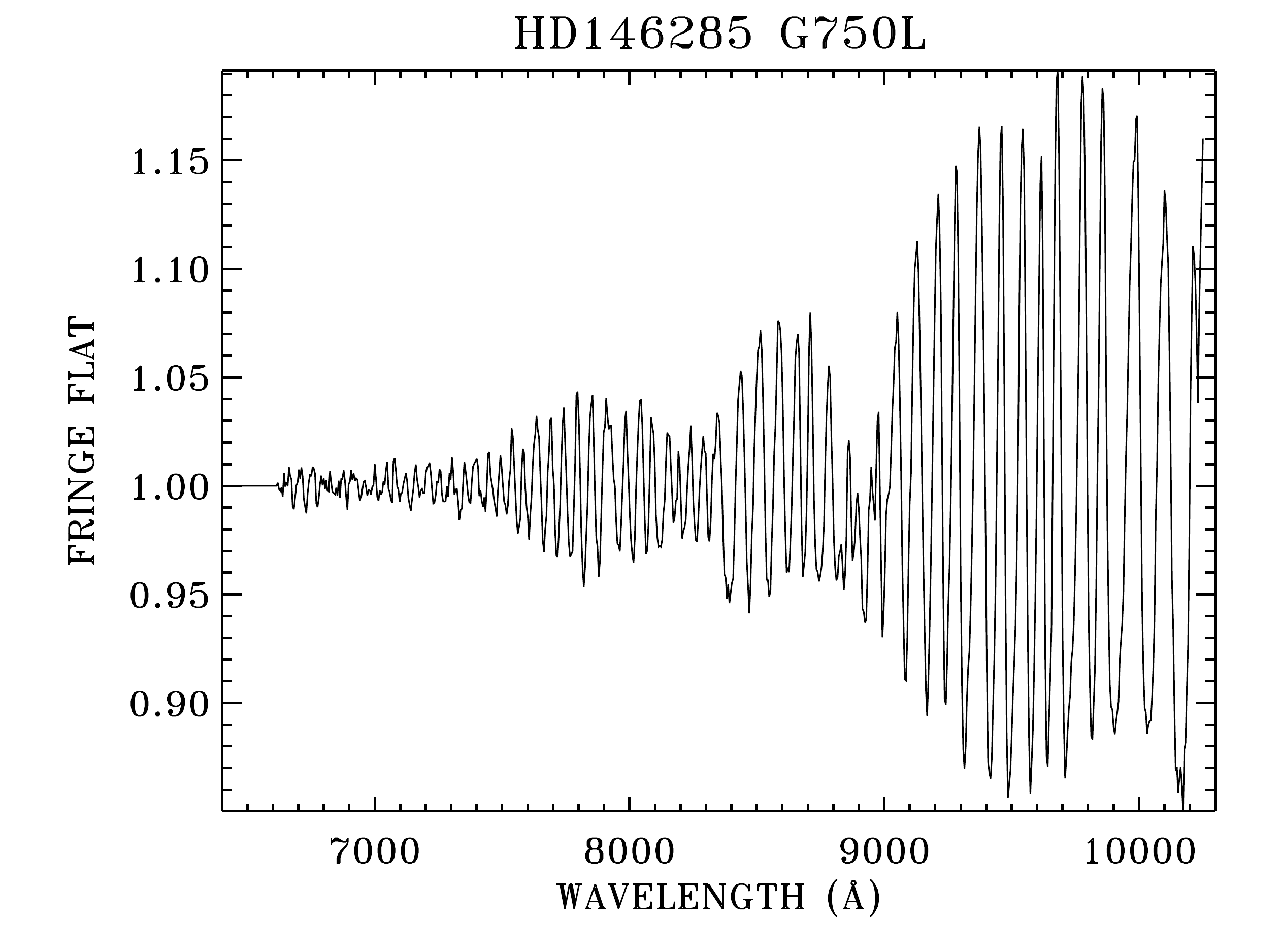}
\caption{Fringe flat correction for the G750L spectrum ocmv0j020 of HD146285.
\label{fringe}}
\end{figure}

\item For G750L, especially at the longer wavelengths, the spectral width has
increased significantly in recent years and a wider extraction height (EXTRSIZE)
of 11 pixels, instead of the STScI pipeline EXTRSIZE=7 pixels, is required to
maintain photometric repeatability of G750L observations \citep{Bohlin15}. 
The EXTRSIZE=7 spectra show errors in spectral shape of up to 3\% in
ratio of the long to the short wavelengths of G750L. The wider extraction width 
encompasses more low signal pixel values, which decreases the spectral S/N slightly.

\item Pixels that are flagged as hot are not used, and those values are reset to
the value interpolated in the spectral direction of the affected row of the
image before extraction of the spectrum. Very occasionally, a hot pixel  flag
can occur at the center of a spectral line, resulting in a small reduction in
equivalent width. For this paper, our interest is in the continuum and not in
precise equivalent width measures.

\item To remove the CCD fringing in G750L, tungsten lamp spectra through the
0.03\arcsec x 0.09\arcsec slit must be used to align this artificial star with
the position of the actual star. The following spline nodes for fitting these
tungsten flats are a prerequisite for the G750L sensitivity calibration: 5240,
5265, 5300, 5318, 5340, 5380, 5420, 5460, 5500, 5580, 5700, 5800, 5840, 6000,
6200, 6300, 6450, 6550, 6600, 6620, 6700, 6800, 6900, 7000, 7150, 7400, 7600,
7800, 8000, 8300, 8500, 9100, 9500, 9800, and 10200~\AA. These wavelengths are
selected to track the features of the tungsten lamp spectrum, while maintaining
the proper stiffness and avoiding $\sim$1\% bumps due to the $\sim$20\% fringe
structure at the longer wavelengths. The nodes below 6600~\AA\ are not crucial
because the fringe flat is set to unity below 6620~\AA. These short-slit
tungsten flats are applied to all G750L observations after normalization to
unity, reduction of the fringe amplitude by 11\% to compensate for narrower
PSF of the artificial star, and cross-correlating with the stellar spectrum 
to remove small residual offsets in the dispersion direction. An example of a 
fringe flat for ocmv0j020 appears in Figure~\ref{fringe}.
\end{enumerate}

In addition to the above data extraction advantages, our post-processing
includes the CTE corrections of \citet[][]{Goudfrooij06} and an up-to-date
correction for the changes in STIS sensitivity with time and with temperature.
The observations of the three primary flux standards G191B2B, GD153, and GD71,
as corrected for CTE and the latest measures of sensitivity change with time,
produce a current absolute flux calibration \citep[{\it CALSPEC}; see][]{Bohlin14}. 

\section{Conversions between extinction relationship equations}
\label{appendix_ccm}

Interstellar extinction curves are commonly expressed using the ``$k$\,'' notation, in the form
\begin{equation}
k(\lambda-V) \equiv \frac{E(\lambda-V)}{E(B-V)} \equiv \frac{A(\lambda)-A(V)}{A(B)-A(V)}  \; ,
\label{eq:app1}
\end{equation}
where $A(\lambda)$ is the total extinction at wavelength $\lambda$ and $B$ and $V$ refer
to measurements made using the Johnson $B$ and $V$ filters.  A related quantity, commonly known
as $R$, is the ratio of total-to-selective extinction in the optical, given by
\begin{equation}
R(V) \equiv \frac{A(V)}{E(B-V)} \; .
\label{eq:app2}
\end{equation}

In this paper, we adopt the analogous forms 
\begin{equation}
k(\lambda-55) \equiv \frac{E(\lambda-55)}{E(44-55)} \equiv \frac{A(\lambda)-A(55)}{A(44)-A(55)}  \; ,
\label{eq:app3}
\end{equation}
and 
\begin{equation}
R(55) \equiv \frac{A(55)}{E(44-55)} \; ,
\label{eq:app4}
\end{equation}
in which measurements with the Johnson $B$ and $V$ filters are replaced by monochromatic measurements at 
4400 and 5500 \AA, respectively. Our data show that the two normalizations are related by
\begin{equation}
k(\lambda-55) = \alpha \; k(\lambda-V) + \beta  \; ,
\label{eq:app5}
\end{equation}
which implies
\begin{equation}
R(55) = \alpha \; R(V) - \beta  \; . 
\label{eq:app6}
\end{equation}
The precise values of $\alpha$ and $\beta$ depend on 
the shapes of the stellar energy distributions (reddened and unreddened), due to the broadband nature 
of the $B$ and $V$ filters. For our particular sample, which on average consists of middle B stars with 
$E(B-V) \simeq 0.5$, we find that $\alpha = 0.990$ and $\beta = 0.049$.  With these coefficients, the 
generally accepted Milky Way mean value of $R(V)_0 = 3.10$ corresponds to $R(55)_0 = 3.02$. 

As described in the text, we find the following functional relationship between the $k(\lambda-55)$ 
extinction curves and $R(55)$
\begin{equation}
k(\lambda-55) = s(\lambda-55) \; [R(55)-R(55)_0] + k(\lambda-55)_0 \; , 
\label{eq:app7}
\end{equation}
which is essentially a first order Taylor expansion in $R(55)$ at each wavelength. In this equation, 
$s(\lambda-55)$ is the observed slope between $R(55)$ and $k(\lambda-55)$, 
$R(55)_0$ is the average Milky Way value of $R(55)$, and $k(\lambda-55)_0$ is the extinction curve 
corresponding to $R(55)_0$. The complete set of $k(\lambda-55)_0$ and $s(\lambda-55)$ values, which 
allow a detailed extinction curve to be constructed over the range 
1150 to 10000 \AA, are given in Table \ref{tabAVERAGE}.

In their early investigation of the $R$-dependence of extinction, \citetalias{Cardelli89} adopted a different 
extinction curve normalization, i.e., $A(\lambda)/A(V)$. From Equations \ref{eq:app1} and \ref{eq:app2}, 
it can be seen that this is related to $k(\lambda-V)$ by
\begin{equation}
\frac{A(\lambda)}{A(V)} = \frac{k(\lambda - V)}{R(V)} + 1  \: .
\label{eq:CCM1}
\end{equation}
\citetalias{Cardelli89} parameterized the $R$-dependence of $A(\lambda)/A(V)$ as
\begin{equation}
\frac{A(\lambda)}{A(V)} = a(\lambda) +b(\lambda) R(V)^{-1}  \: .
\label{eq:CCM2}
\end{equation}
At first look, this appears to be a very different functional relationship between $R$ and the 
shape of the extinction curve than we have found (i.e., Eq. \ref{eq:app7}). However, when Equation 
\ref{eq:CCM1} is substituted 
into Equation \ref{eq:CCM2}, the following result is obtained:
\begin{equation}
k(\lambda-V) = [a(\lambda) -1] R(V) + b(\lambda)  \: .
\label{eq:CCM3}
\end{equation}
Thus it is seen that, like us, \citetalias{Cardelli89} found their extinction curves to be linearly related to $R$, when expressed 
in the ``$k$\,'' notation.  Moreover, the relationship between \citetalias{Cardelli89}'s $a(\lambda)$ and $b(\lambda)$ and our 
coefficients $s(\lambda-55)$ and $k(\lambda-55)_0$ can be found by comparing Equations \ref{eq:app7} and \ref{eq:CCM3} 
and making the appropriate substitutions from Equations \ref{eq:app5} and \ref{eq:app6}.  The result is
\begin{eqnarray}
a(\lambda)  & = & s(\lambda-55) + 1  \nonumber \\
b(\lambda)  & = & -s(\lambda-55)\left \lbrack \frac{R(55)_0 + \beta}{\alpha} \right \rbrack + \left\lbrack \frac{k(\lambda-55)_0 - \beta}{\alpha}\right\rbrack  \: .
\end{eqnarray}
In principal, identical results would be achieved regardless of whether our ``$k$'' normalization or 
\citetalias{Cardelli89}'s $A(\lambda)$ normalization were used.  One must keep in mind,
however, that these mathematical transformations only apply in the absense of errors.  Least squares solutions 
resulting from the two different formulations could yield different results.  

\end{document}